\documentclass[twocolumn,nobibnotes,nofootinbib,superscriptaddress,preprintnumbers,aps,prd,10pt]{revtex4-1}
\pdfoutput=1

\usepackage{graphicx}
\usepackage{latexsym}
\usepackage{amsfonts}
\usepackage{amssymb}
\usepackage{color}
\usepackage{amsmath}
\usepackage{slashed}
\usepackage{subcaption}
\usepackage{dcolumn}
\usepackage{feynmf}
\DeclareGraphicsExtensions{.pdf,.png,.jpeg}

\definecolor{orange}{rgb}{1,0.5,0}
		
\newcommand{\gsim}{\gtrsim}
\newcommand{\lsim}{\lesssim}
\newcommand{\ra}{\rightarrow}

\newcommand{\Mtwo}{M_{\tilde{g}2}}
\newcommand{\Mone}{M_{\tilde{g}1}}
\newcommand{\Msq}{M_{\tilde{q}}}
\newcommand{\Mbino}{M_{\tilde{B}}}
\newcommand{\Mwino}{M_{\tilde{W}}}
\newcommand{\g}{\tilde{g}}

\begin{document}	

\title{Mixed Gauginos Sending Mixed Messages to the LHC}

\author{Graham D. Kribs}
\affiliation{School of Natural Sciences, Institute for Advanced Study,
             Princeton, NJ 08540}
\affiliation{Department of Physics, University of Oregon, Eugene, OR 97403}

\author{Nirmal Raj}
\affiliation{Department of Physics, University of Oregon, Eugene, OR 97403}

\begin{abstract}

Supersymmetric models with a Dirac gluino have been shown to be 
considerably less constrained from LHC searches than models
with a Majorana gluino.  In this paper, we generalize this discussion
to models with a ``Mixed Gluino'', that acquires both Dirac and Majorana masses,
as well as models in which electroweak gauginos contribute to squark production.
Our primary interest is the degree of suppression of the cross section for
first generation squarks compared with a gluino with a pure Majorana mass.
We find that not all Majorana masses are alike -- a Majorana mass
for the gluino can \emph{suppress} the squark production cross section,
whereas a Majorana mass for the adjoint fermion partner leads to an
increased cross section, compared with a pure Dirac gluino.  In the presence of
electroweak gauginos, squark production can increase by at most
a factor of a few above the pure Dirac gluino case when the electroweak gauginos
have purely Majorana masses near to the squark masses.
This unusual set of gaugino masses is interesting since 
the usual Higgs quartic coupling would not be suppressed.
When the electroweak gaugino masses are much lighter than the squarks, 
there is a \emph{negligible} change to the squark production cross section.
We explain both of these cases in detail by considering the various
subprocesses for squark production at the LHC\@.
Continued searches for squark production with cross sections much
smaller than expected in the MSSM are absolutely warranted, and
we suggest new simplified models to fully characterize the
collider constraints.

\end{abstract}

\maketitle

\section{Introduction}
\label{sec:intro}

The strongest constraints on weak scale supersymmetry from the LHC
are on first generation squarks and the
gluino~\cite{ATLASHeavyMSSM:2013,CMS2013multijet}.
First generation
squark production proceeds through $pp \ra \tilde{q}\tilde{q}$
that is dominated by $t$-channel exchange of a gluino that acquires 
a Majorana mass (``Majorana gluino'') using valence quarks from the proton.
Not surprisingly, the largest contributions
come from sub-processes involving a chirality flip in the $t$-channel gluino
exchange diagram which is a comparatively unsuppressed dimension-5 interaction.
The bounds on first generation squarks, typically combined with the
second generation in a simplified model involving $\Msq$ and
$M_{\tilde{g}}$, are currently $\Msq > 1.8$~TeV
for $M_{\tilde{g}} \simeq \Msq$ \cite{ATLASHeavyMSSM:2013}.

In a previous paper \cite{Kribs:2012gx},
one of us (GDK) with Adam Martin showed that
the presence of a gluino that acquires a Dirac mass
(referred to as a ``Dirac gluino'') -- instead of
a Majorana mass -- significantly weakens these collider constraints.
This was due to three reasons:  first, a Dirac gluino can be
significantly heavier than a Majorana gluino, with respect to
fine-tuning of the electroweak symmetry-breaking scale.
This is because a Dirac gluino yields one-loop \emph{finite}
contributions to squark masses \cite{Fox:2002bu}.
Second, no ``chirality-flipping'' Dirac gluino $t$-channel
exchange diagrams exist, and thus several subprocesses
for squark production simply vanish.  Third, the remaining
squark production subprocess amplitudes are suppressed by
$|p|/M_{\tilde{g}}^2$, where $|p|$ is the typical
momentum exchanged through the Dirac gluino.  For a heavier
Dirac gluino ($M_{\tilde{g}} \gsim 2$-$3$~TeV), 
this significantly suppresses $t$-channel
gluino exchange to the point where it is subdominant to
the gluino-independent squark--anti-squark 
production processes \cite{Kribs:2012gx}.

Dirac gaugino masses have been considered
long ago \cite{Fayet:1978qc,Polchinski:1982an,Hall:1990hq}
and have inspired more recent model building
\cite{Fox:2002bu,Nelson:2002ca,Chacko:2004mi,Carpenter:2005tz,Antoniadis:2005em,Nomura:2005rj,Antoniadis:2006uj,Kribs:2007ac,Amigo:2008rc,Benakli:2008pg,Blechman:2009if,Carpenter:2010as,Kribs:2010md,Abel:2011dc,Frugiuele:2011mh,Itoyama:2011zi,Unwin:2012fj,Abel:2013kha}
and phenomenology
\cite{Hisano:2006mv,Hsieh:2007wq,Blechman:2008gu,Kribs:2008hq,Choi:2008pi,Plehn:2008ae,Harnik:2008uu,Choi:2008ub,Kribs:2009zy,Belanger:2009wf,Benakli:2009mk,Kumar:2009sf,Chun:2009zx,Benakli:2010gi,Fok:2010vk,DeSimone:2010tf,Choi:2010gc,Choi:2010an,Chun:2010hz,Davies:2011mp,Benakli:2011kz,Kumar:2011np,Davies:2011js,Heikinheimo:2011fk,Fuks:2012im,Kribs:2012gx,GoncalvesNetto:2012nt,Fok:2012fb,Fok:2012me,Frugiuele:2012pe,Frugiuele:2012kp,Benakli:2012cy,Agrawal:2013kha,Hardy:2013ywa,Buckley:2013sca}.
As beautiful as Dirac gauginos may be, there are two objections
that are sometimes raised:
\begin{itemize}
\item Supersymmetry-breaking sectors do not generically have
$F$-terms much smaller than $D$-terms.  In the absence of
a specialized mediation sector that sequesters the $F$-term
contributions \cite{Chacko:2004mi}, we might expect
both Dirac and Majorana masses to be generated (for example, \cite{Kribs:2010md}).
Moreover, even if $F$-term mediation is sequestered,
gauginos do acquire Majorana masses through anomaly-mediation
\cite{Randall:1998uk,Giudice:1998xp}.
\item In the presence of a pure Dirac wino and bino, the usual
tree-level $D$-term quartic coupling for the Higgs potential is not
generated \cite{Fox:2002bu}.  This requires additional couplings
to regenerate the quartic coupling.  While there are mechanisms to
generate a quartic in models with a pure Dirac gaugino mass
(see \cite{Fok:2012fb} in the context of $R$-symmetric supersymmetry),
it is obviously of interest to understand the impact of electroweak
gauginos acquiring Majorana masses on squark production cross section limits.
\end{itemize}
In this context, we consider two generalizations of Ref.~\cite{Kribs:2012gx}:
(i) models with a ``Mixed Gluino'' that acquires both Dirac and Majorana masses, and
(ii) models in which the electroweak gauginos acquire purely Majorana masses,
and contribute to squark production.
As we will see, both cases have surprising outcomes.

Our primary interest is to compare squark production cross sections
with mixed gauginos against the pure Dirac and pure Majorana cases.
Mixed gauginos were also considered in \cite{Choi:2008pi}, where the
main emphasis was on distinguishing the different types of gaugino masses
well before the strong bounds on colored superpartner production
were set by the LHC collaborations.  Our interest in this paper is
largely orthogonal, examining in detail the modifications to
squark production when the gaugino is \emph{heavy}.
We used MadGraph5 \cite{Alwall:2011uj} to simulate squark production
at leading order (LO) for the LHC operating at a center-of-mass energy
of $8$~TeV and $14$~TeV using the CTEQ6L1 parton distribution
functions (PDFs).
We did not, however, incorporate next-to-leading-order (NLO) corrections
in our cross sections, for several reasons:  first, in some cases there is
a large range of scales between the squark mass and the gaugino mass,
and unfortunately existing codes (Prospino~2.1 \cite{Beenakker:1996ch}
and \cite{NLLFast})
are not designed to handle this.  Second, to the best
of our knowledge, the NLO corrections for a Dirac gaugino as well as a
mixed gaugino have not been computed.  This is an important outstanding
problem, but it is not the primary interest of this paper.
In much of the results presented below, we consider \emph{ratios}
of production cross sections, where most of the large NLO corrections
are expected to cancel.  We do show some LO cross sections as a
function of gaugino mass, to better explain our results, however, in
these cases we are generally interested in the trend as a function of
the gaugino mass rather than the precise cross section values.
The full NLO calculation would be interesting to compute, but it is
beyond the scope of this paper.

\section{Mixed Gauginos}
\label{sec:theoryformulae}

``Mixed Gauginos'' are, by definition, the Majorana mass eigenstates
of gauginos that acquire both a Dirac mass with an adjoint fermion
partner as well as Majorana masses for the gluino, the adjoint fermion,
or both.  This occurs when
the supersymmetry-breaking hidden sector contains superfields
that acquire both $F$-type and $D$-type supersymmetry-breaking vevs.
Let us first write the operators that lead to these contributions
to the gaugino masses, using the spurions $X \equiv F \theta^2$ and
$W'_\alpha \equiv D' \theta_\alpha$.
A Majorana mass arises from the usual operator
\begin{equation}
c_m \int d^2\theta \frac{X}{M} W^{\alpha}W_{\alpha}
\label{Ftermbreak}
\end{equation}
and a Dirac mass from \cite{Fox:2002bu}
\begin{equation}
c_d \int d^2\theta \sqrt{2} \frac{W'_{\alpha}}{M} W^{\alpha}_j A_j \, ,
\label{Dtermbreak}
\end{equation}
where $M$ is the mediation scale and $A_j$ is a chiral superfield
in the adjoint representation of the relevant gauge group of
the Standard Model. 
Whether a gaugino acquires a Dirac mass obviously depends on the existence
of a chiral adjoint to pair up with.
There are additional operators that can contribute to gaugino masses.
The chiral adjoint can acquire a Majorana mass through
\begin{equation}
c_{m'} \int d^4\theta \frac{1}{2} \frac{X^\dagger}{M} \mathrm{tr} A_j A_j
 + \text{h.c.} \, ,
\label{eq:adjmajorana}
\end{equation}
familiar from the Giudice-Masiero mechanism for generating $\mu$
in the MSSM\@.  Here we are assuming that the adjoint fermion only
acquires mass after supersymmetry breaking, i.e., there is no
``bare'' contribution to its mass in the superpotential.

Scalar masses can be generated by contact interactions
\begin{equation*}
\int d^4\theta \frac{X^{\dagger}X}{M^2} Q^{\dagger}Q \, ,
\end{equation*}
at the messenger scale,
as well as the ``soft'' and ``supersoft'' contributions from
Majorana and Dirac gauginos, respectively.  In this paper,
we neglect flavor mixing among the squarks, since the existence
of sizable Majorana masses means we do not have $R$-symmetry to
protect us against flavor-changing neutral currents \cite{Kribs:2007ac}.

Renormalization group evolution from the messenger scale
to the weak scale affects the relative size of the Dirac and
Majorana masses.  Let us first define the Dirac mass,
the Majorana gaugino mass, and the Majorana adjoint mass as
\begin{eqnarray}
M_d  &=& c_d D'/M \nonumber \\
M_m  &=& c_m F/M  \label{eq:massdefs} \\
M_m' &=& c_{m'} F^\dagger/M \, . \nonumber
\end{eqnarray}
All of these quantities are generated at the messenger scale
(possibly with additional hidden sector renormalization \cite{Cohen:2006qc}).
For a gauge group $i$ with beta function coefficient $b_i$
and quadratic Casimir of the adjoint $c_i$, the Dirac operator
receives significant RG effects (neglecting Yukawa couplings)
\cite{Fox:2002bu,Kribs:2010md}
\begin{equation}
M_d(\mu) = M_d(M) \times \left\{ \begin{array}{lcl}
           \left( \frac{\mu}{M} \right)^{-c_i \alpha_i/(2 \pi)}
             & \mathrm{for} & b_i = 0 \\
           \left( \frac{\alpha_i(\mu)}{\alpha_i(M)} \right)^{(b_i - 2 c_i)/(2 b_i)}
             & \mathrm{for} & b_i \not= 0 \, .
           \end{array} \right.
\end{equation}
We calculated the RG evolution of the Majorana adjoint mass to be
(again neglecting Yukawa couplings)
\begin{equation}
M_{m'}(\mu) = M_{m'}(M) \times \left\{ \begin{array}{lcl}
           \left( \frac{\mu}{M} \right)^{-c_i \alpha_i/\pi}
             & \mathrm{for} & b_i = 0 \\
           \left( \frac{\alpha_i(\mu)}{\alpha_i(M)} \right)^{- 2 c_i/b_i}
             & \mathrm{for} & b_i \not= 0 \\
           \end{array} \right.
\end{equation}
which can be obtained directly from the wavefunction renormalization
of the superpotential (and agrees with resuming the RG equation given
in Ref.~\cite{Jack:1999ud} without Yukawa couplings).
The size of the RG evolution can be substantial \cite{Kribs:2010md},
but depends heavily on several assumptions about the mediation
as well as the particle content of the model above the electroweak scale.
These ``ultraviolet'' (UV) issues will not be discussed further in this paper.

\section{Mixed Gluino}
\label{sec:MixedGluino}

Let us now specialize our discussion to a gluino that acquires a
Dirac and Majorana mass.  All of what we say below can also be
straightforwardly applied to the electroweak
gauginos.\footnote{There is an amusing subtlety involving charginos that
acquire ``Dirac'' masses (by this we mean charginos that acquire
Dirac masses by pairing up with additional fermions in the
triplet representation of $SU(2)_W$), that we relegate to
App.~\ref{sec:DbutnotD}.}
Using Eq.~(\ref{eq:massdefs}),
the resulting mass terms for the gaugino and
adjoint superfield are (in $2$-component language)
\begin{equation}
\mathcal{L}_{\tilde{g} \, \mathrm{mass}} \; = \;
\left( \begin{array}{cc}
       g & \psi
       \end{array} \right)
\left( \begin{array}{cc}
       M_m & M_d \\
       M_d & M_m'   	
       \end{array} \right)
\left( \begin{array}{c}
       g \\ \psi
       \end{array} \right)
+ \text{h.c.}
\label{eq:basicmass}
\end{equation}
where we have suppressed the $SU(3)_c$ color indices on the fields.
The relative size of the Dirac and Majorana contributions are set
by the coefficients of the operators (evaluated at the weak scale).
While we take the coefficients to be arbitrary, our main phenomenological
interest is the range $M_d \gg M_m,M_m'$ to $M_d \gsim M_m,M_m'$.

From Eq.~(\ref{eq:basicmass}), the $2$-component fermions $g$ and $\psi$
mix, giving us the mass eigenstates of the gluino
\begin{equation}
\left( \begin{array}{c} g_1 \\ g_2 \end{array} \right) =
  \left( \begin{array}{cc}    \cos \theta_{\g} & \sin \theta_{\g} \\
                            - \sin \theta_{\g} & \cos \theta_{\g}   	
  \end{array} \right)	
  \left( \begin{array}{c} \psi \\ g \end{array} \right) \, ,
\label{eq:eigstats}
\end{equation}
where the mixing angle is given by
\begin{equation}
\cos \theta_{\g} \; = \; \sqrt{\frac{1}{2}}
                  \left(1 + \frac{M_m - M_m'}{\sqrt{(M_m - M_m')^2
                                              + 4 M_d^2}}\right)^{1/2} \, .
\label{eq:MixAng}
\end{equation}
Diagonalizing the Lagrangian, Eq.~(\ref{eq:basicmass}),
gives the two eigenvalues that we write as $-\Mone$ and $\Mtwo$
respectively,
\begin{eqnarray}
\!\!\!\!\!\!\!\!\!\!
-\Mone &=& \frac{1}{2} \left( M_m + M_m'
                              - \sqrt{(M_m - M_m')^2 + 4 M_d^2} \right)
           \nonumber \\
\!\!\!\!\!\!\!\!\!\!
\Mtwo  &=& \frac{1}{2} \left( M_m + M_m'
                              + \sqrt{(M_m - M_m')^2 + 4 M_d^2} \right)
           \label{eq:EigMass}
\end{eqnarray}
We have chosen to define $\Mone$ to be the negative of the eigenvalue
of the mass matrix so that when $M_d^2 > M_m M_m'$, both $\Mone$
and $\Mtwo$ are positive.  We could have instead redefined the eigenstates
to absorb this sign, however this would lead to proliferation of
$i$'s in the following, that we prefer to avoid.

The two familiar limits of these equations are now evident:
For a pure Dirac gluino ($M_m = 0$), $\Mone = \Mtwo = M_d$,
the mixing angle $\theta_{\g} = \pi/4$, and then the gluino eigenstates
are $g_{1,2} = (g \pm \psi)/\sqrt{2}$.  For a pure Majorana gluino ($M_d = 0$),
the mixing angle $\theta_{\g} = 0$, which means the gluino and its
adjoint fermion partner do not mix, i.e., $g_1 = g$, $g_2 = \psi$.
Consequently, $\Mone = M_m$ and $\Mtwo = M_m'$.
	
The quark-gluino-squark interactions are given by
\begin{eqnarray}
\lefteqn{\mathcal{L}_{int} \; =} \nonumber \\
  & &{} -\sqrt{2} g_s \big( \tilde{u}^*_{L,i}\ t^a \ g_a \ u_{L,i}
                            + \tilde{d}^*_{L,i}\ t^a \ g_a \ d_{L,i}
        \nonumber \\
  & &{} - \tilde{u}^*_{R,i} \ t^a \ g_a u_{R,i}
        - \tilde{d}^*_{R,i} \ t^a \ g_a \ d_{R,i}  \big) + \text{h.c.}
\end{eqnarray}	
where the index $i$ runs over each quark generation and
the squark color indices have been suppressed.
Expanding using Eq.~(\ref{eq:eigstats}), this becomes
\begin{eqnarray}
\lefteqn{-\mathcal{L}_{int}/\sqrt{2} g_s \; =} \nonumber \\
& &{} + \tilde{u}^*_{L,i} \ t^a \ g_{1,a} \cos \theta_{\g} \ u_{L,i}
      + \tilde{u}^*_{L,i} \ t^a \ g_{2,a} \sin \theta_{\g} \ u_{L,i}
      \nonumber \\
& &{} + \tilde{d}^*_{L,i} \ t^a \ g_{1,a} \cos \theta_{\g} \ d_{L,i}
      + \tilde{d}^*_{L,i} \ t^a \ g_{2,a} \sin \theta_{\g} \ d_{L,i}
      \nonumber \\
& &{} - \tilde{u}^*_{R,i} \ t^a \ g_{1,a} \cos \theta_{\g} \ u_{R,i}
      - \tilde{u}^*_{R,i} \ t^a \ g_{2,a} \sin \theta_{\g} \ u_{R,i}
      \nonumber \\
& &{} - \tilde{d}^*_{R,i} \ t^a \ g_{1,a} \cos \theta_{\g} \ d_{R,i}
      - \tilde{d}^*_{R,i} \ t^a \ g_{2,a} \sin \theta_{\g} \ d_{R,i}
      \nonumber \\
& &{} + \ \text{h.c.}
\label{eq:qsqgmixed}
\end{eqnarray}
This is the form of the interaction Lagrangian most useful for
our phenomenological study.
		
In order to understand the implications of a mixed gluino arising
from both a Dirac and a Majorana mass, we first need to parameterize
the mixing in a way relevant to our collider study.
There are two distinct effects when simultaneously varying
$M_d$, $M_m$, and $M_m'$:  the coupling constants to the squarks and quarks
change, according to Eq.~(\ref{eq:qsqgmixed}), and the masses of
the gluino eigenstates change, according to Eq.~(\ref{eq:EigMass}).
This leads to changes in both the \emph{dynamics} (the coupling
constants) and the \emph{kinematics} (the gluino masses) of the
squark production cross sections.
We are interested in separating these effects, to the extent possible.

\subsection{Review of pure Dirac gluinos}
\label{sec:TotalSigmaQCD}

\begin{figure*}
\begin{subfigure}[t]{0.32\textwidth}
\includegraphics[width=5.3cm]{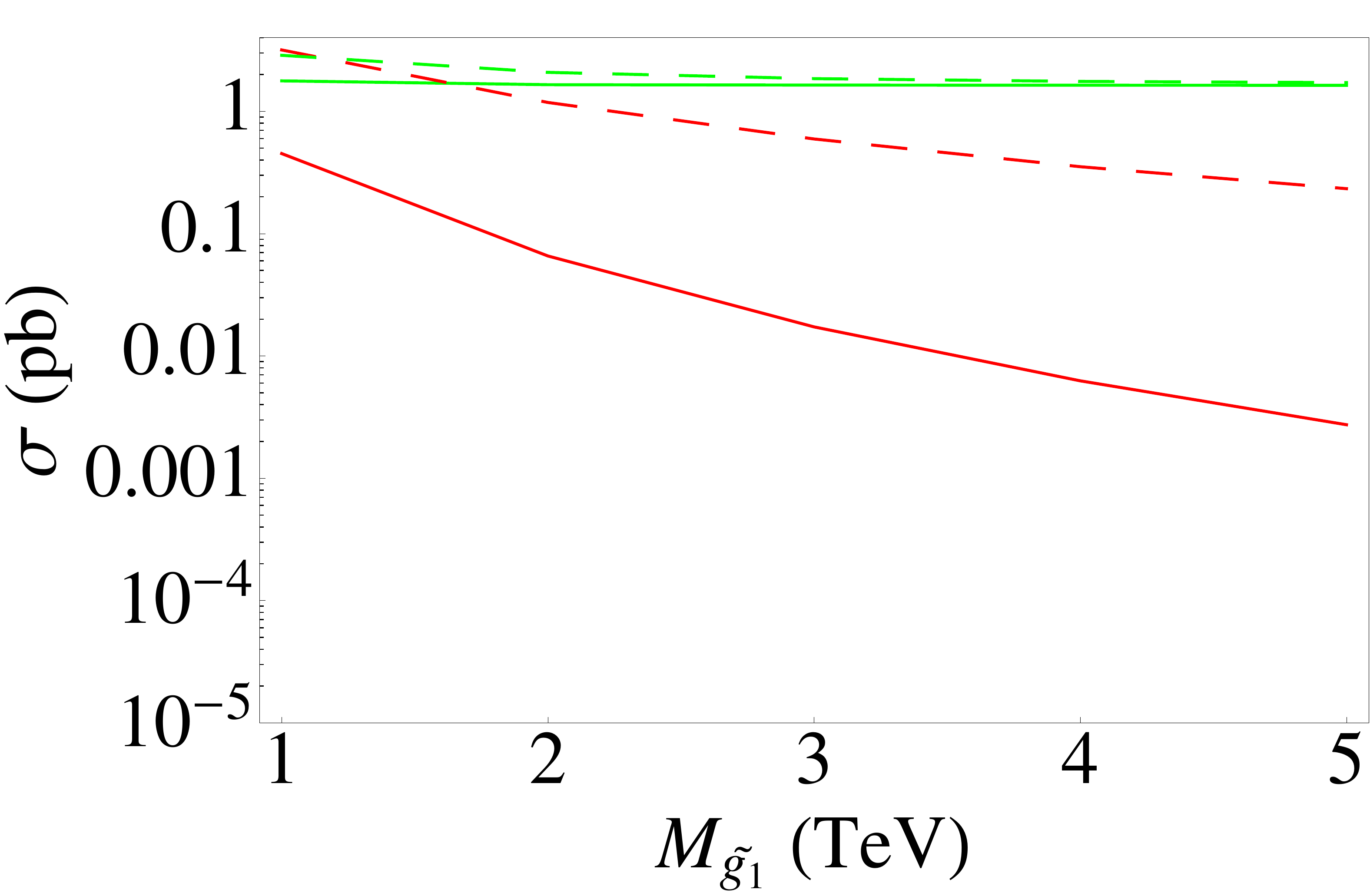}
\caption{$\Msq=400$~GeV}
\label{fig:DirGlu400}
\end{subfigure}
\hfill
\begin{subfigure}[t]{0.32\textwidth}
\includegraphics[width=5.3cm]{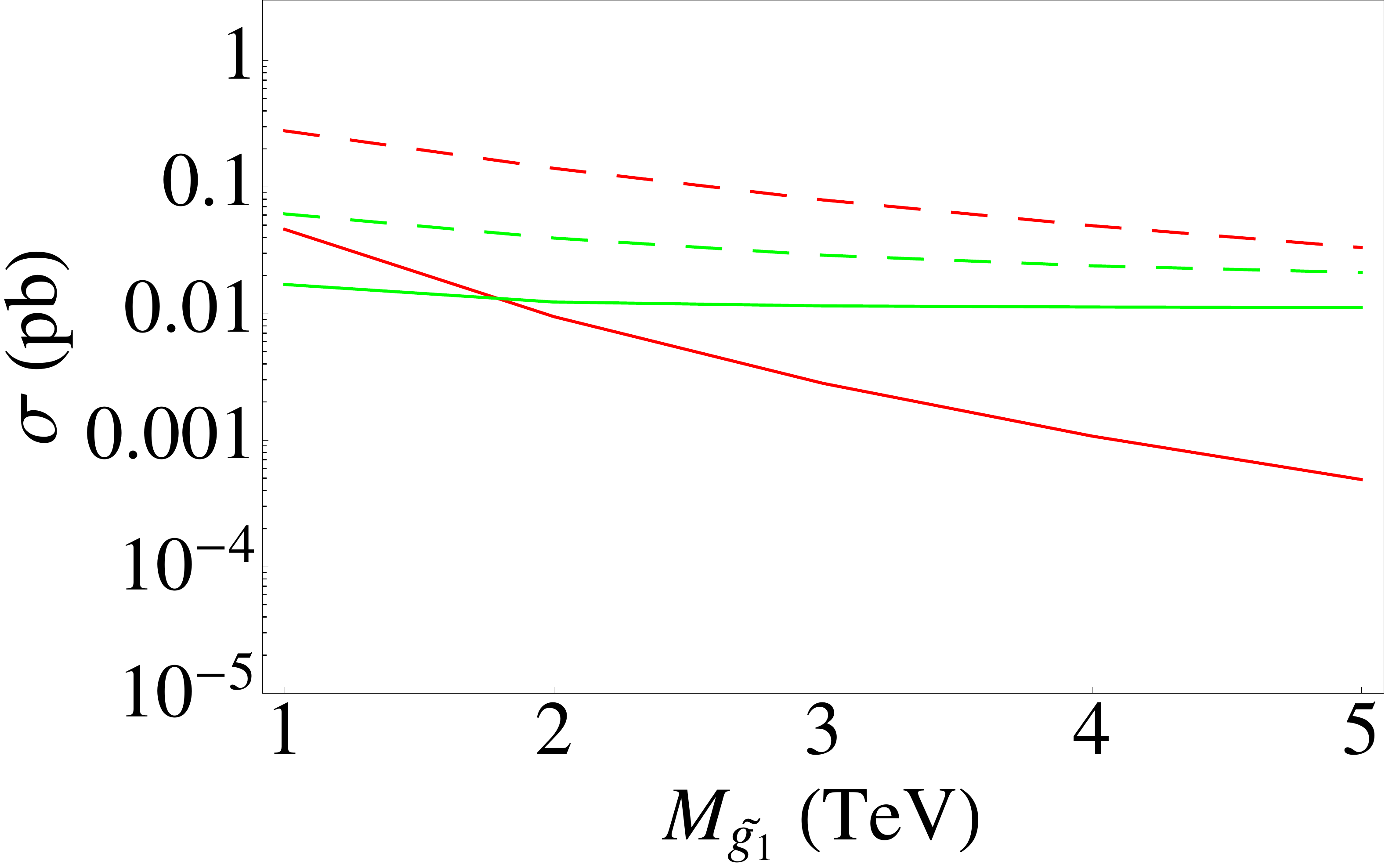}
\caption{$\Msq=800$~GeV}
\label{fig:DirGlu800}
\end{subfigure}
\hfill
\begin{subfigure}[t]{0.32\textwidth}
\includegraphics[width=5.3cm]{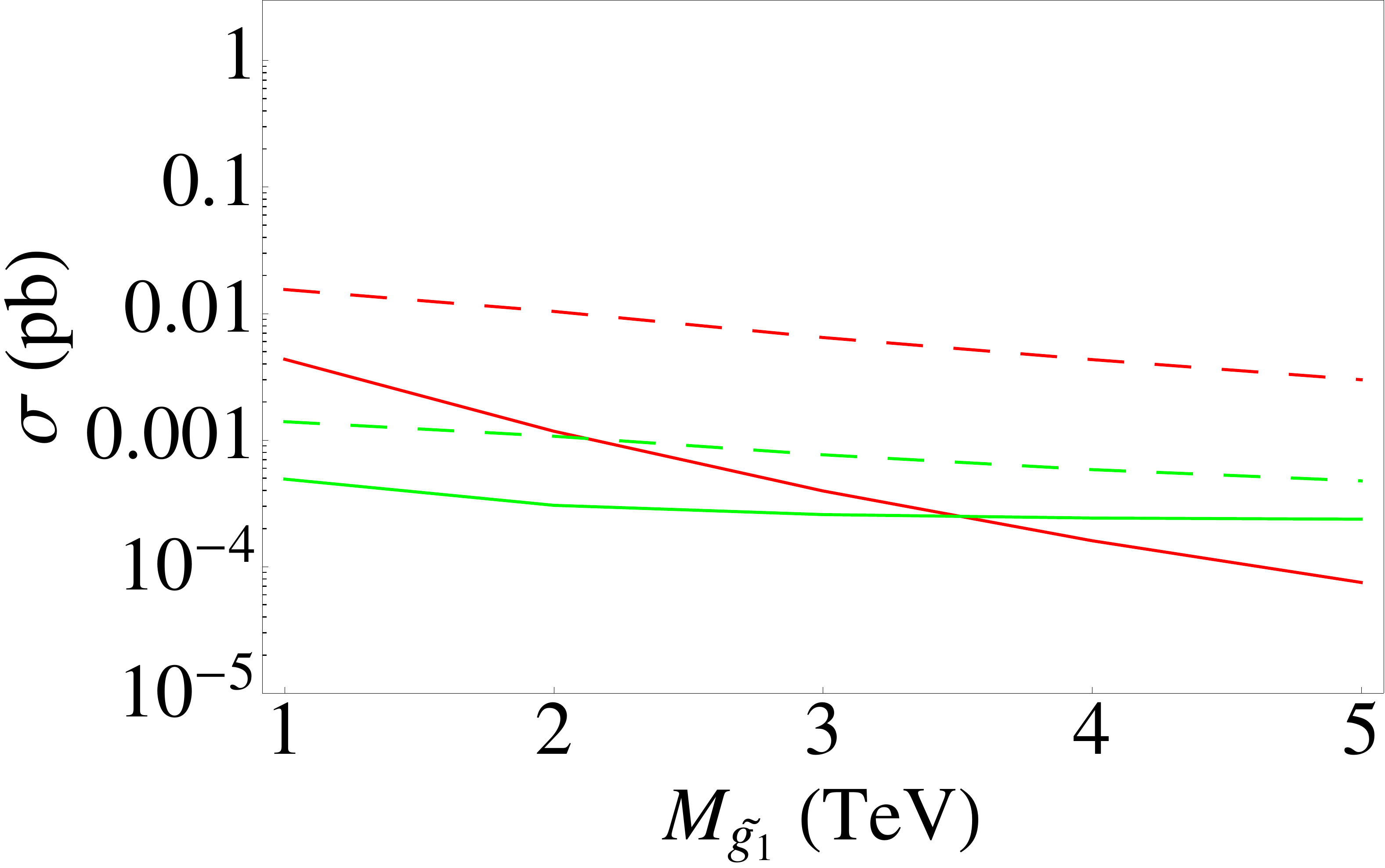}
\caption{$\Msq=1200$~GeV}
\label{fig:DirGlu1200}
\end{subfigure}
\caption{Comparing the squark pair production cross section (red)
against squark--anti-squark production cross section (green) summing over
the first two generations of squarks with masses of $400$, $800$ and
$1200$~GeV\@.  The solid lines denote the case in which the
Majorana masses vanish ($M_m = M_m' = 0$),
so the $x$-axis corresponds to a pure Dirac gluino mass.
At low squark masses, squark-anti-squark production through an $s$-channel
gluon that dominates over the $t$-channel gluino-mediated squark-squark production.
However, for $m_{\tilde{q}} = 800,1200$~GeV we find that squark-squark
production dominates up to $M_d \simeq 2,3.5$~TeV\@. The dotted lines depict the
behavior when the Dirac mass vanishes ($M_d = 0$), with the $x$-axis corresponding
to a pure Majorana mass. Only at a very low squark mass of $400$~GeV does
squark--anti-squark production dominate. For higher squark masses $800$ and
$1200$~GeV, squark pair production dominates for all gluino masses. This
is due to $t$-channel mediated \emph{same-handed}
squark production, which was absent in the case of a pure Dirac gluino.}
\label{fig:DirGluVariedDomin}
\end{figure*}

Before embarking on our study of mixed gluinos, we first want to review the
effects of a pure Dirac gluino on the various squark production processes.
The relevant squark production processes
include\footnote{The third combination,
antisquark-antisquark production, can be ignored since its rate is
highly suppressed by PDFs.}
$pp \ra \tilde{q}_{L,R} \tilde{q}_{L,R}$ and
$pp \ra \tilde{q}_{L,R} \tilde{q}^*_{L,R}$.
Fig.~\ref{fig:DirGluVariedDomin} shows the relative contributions of these
two production modes for different (pure Dirac) gluino masses, depicted by
the solid curves.
The dominant effects of $t$-channel gluino exchange impact just
the first generation of squarks.
However, since a common simplified model that ATLAS
and CMS use in quoting bounds is to sum over all squarks of the
first two generations assuming the flavors and chiralities are degenerate
in mass, we do this also.  The lightest supersymmetric particle (LSP)
is taken to be a neutral particle odd under $R$-parity.  The gravitino
is one possibility, though as we will see, a Majorana bino is another
distinct possibility.

At low squark masses, $400$~GeV (Fig.~\ref{fig:DirGlu400}), the production
cross section is heavily
dominated by squark-antisquark production with quarks or gluons in the
initial state. This is because squark pair production through
$t$-channel (Dirac) gluino exchange can only yield
$pp \ra \tilde{q}_L \tilde{q}_R$; the other processes $(LL,RR)$ are absent.
As the squark mass is increased, the modes
$\tilde{q}_{L/R}, \tilde{q}^*_{L/R}$ and $\tilde{q}_L, \tilde{q}_R$
become comparable to each other.  For $M_{\tilde{q}} = 800$~GeV,
this occurs for Dirac gluino masses near $\simeq 2$~TeV,
as shown in Fig.~\ref{fig:DirGlu800}.
In other words, the gluino $t$-channel exchange diagrams of squark pair production
are not as suppressed in this range. Considering even larger squark masses,
$M_{\tilde{q}} = 1200$~GeV,
we find squark pair production becomes comparable to squark--anti-squark
production for a (Dirac) gluino mass $\simeq 4$~TeV, shown in
Fig.~\ref{fig:DirGlu1200}.

The dashed lines in Fig.~\ref{fig:DirGluVariedDomin} depict the two
production modes for a pure Majorana gluino. At a low squark mass of
$400$~GeV, squark--anti-squark production dominates the cross section for gluino
masses greater than $\sim 2$~TeV, while for $\Msq = 800$~GeV and $\Msq = 1200$~GeV,
squark pair production dominates for all gluino masses shown in the figures.
This is because $t$-channel production of same-handed squark production
is the dominant production mode for these masses and energies
with a Majorana gluino.

\begin{figure*}[t]
\begin{subfigure}[t]{0.32\textwidth}
\includegraphics[width=5.1cm]{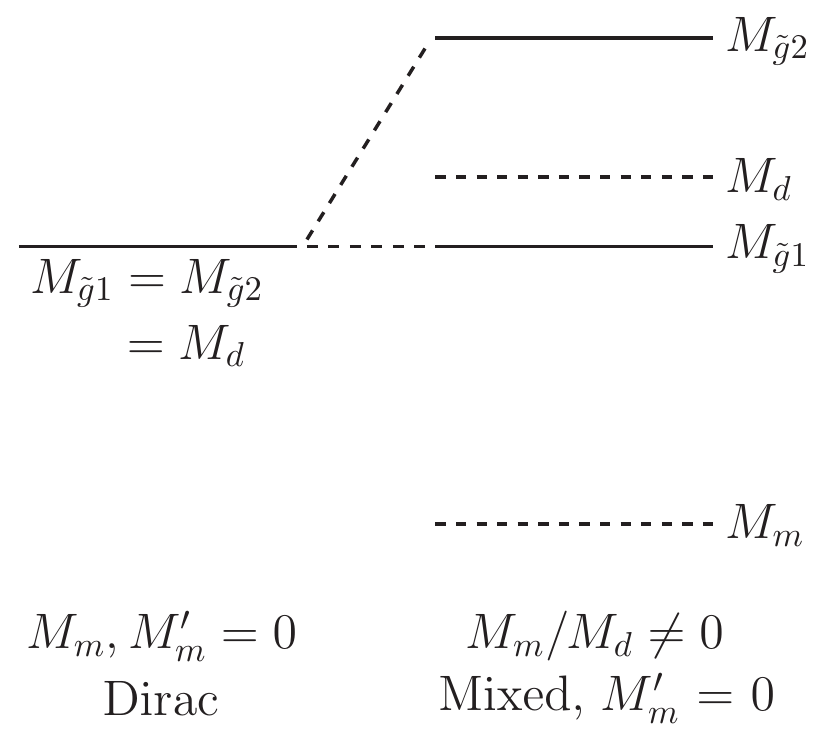}
\caption{}
\label{fig:Splits_M_mp_OFF}
\end{subfigure}
\hfill
\begin{subfigure}[t]{0.32\textwidth}
\includegraphics[width=5.1cm]{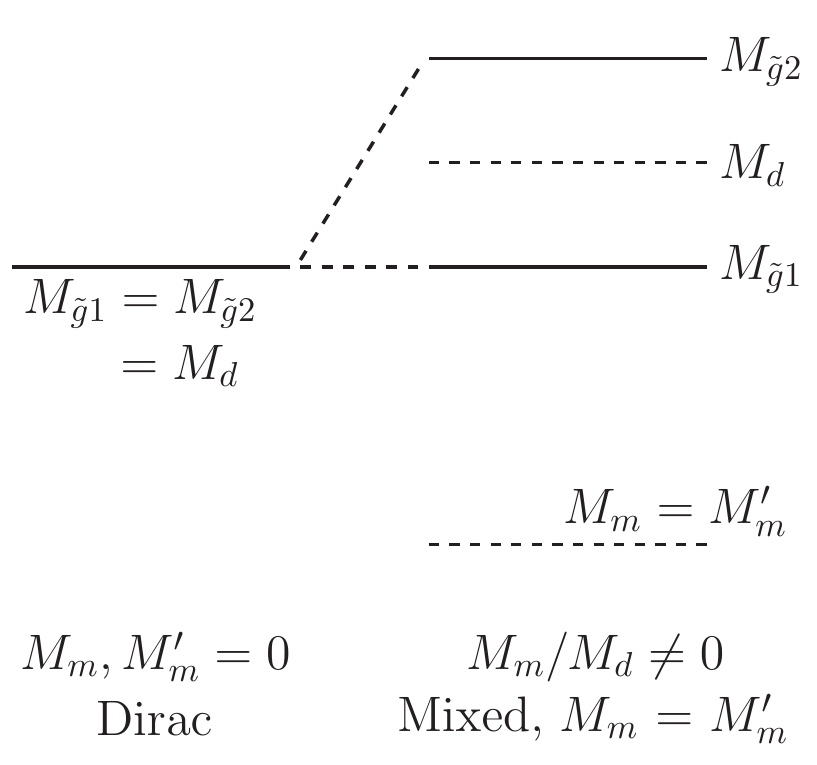}
\caption{}
\label{fig:Splits_Both_ON}
\end{subfigure}
\hfill
\begin{subfigure}[t]{0.32\textwidth}
\includegraphics[width=5.1cm]{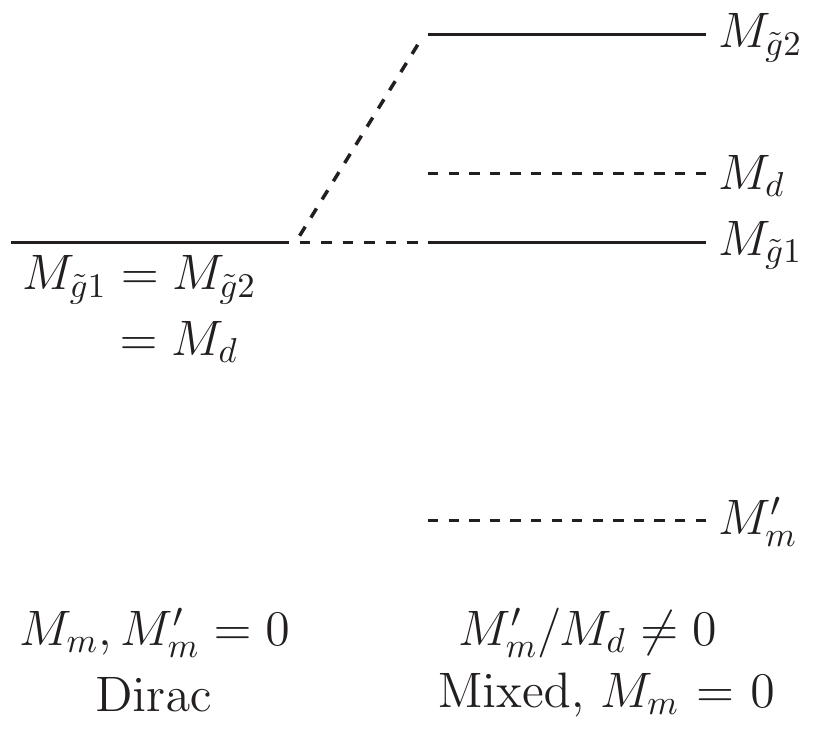}
\caption{}
\label{fig:Splits_M_m_OFF}
\end{subfigure}
\caption{The method we employ for adding Majorana masses
$M_m, M_m'$ to the supersoft Dirac mass $M_d$ of a gaugino.
The lower eigenvalue $\Mone$ is kept constant as $M_m/M_d$ or $M_m'/M_d$ is varied.}
\label{fig:splits}
\end{figure*}

\subsection{Case I:  $M_m' = 0$}

First, we consider the scenario $M_m' = 0$, $M_m \lsim M_d$.
In this Case, we can simplify the expressions for the masses and
mixing angle of the mixed gluino:
\begin{eqnarray}
-\Mone &=& \frac{1}{2}\left(M_m - \sqrt{M_m^2 + 4 M_d^2} \right) \label{eq:Monesimp} \\
\Mtwo  &=& \frac{1}{2}\left(M_m + \sqrt{M_m^2 + 4 M_d^2} \right) \label{eq:Mtwosimp} \\
\cos \theta_{\g} &=& \sqrt{\frac{\Mtwo}{\Mtwo + \Mone}} \; . \label{eq:costsimp}
\end{eqnarray}

Next, to separate the ``kinematics'' from the ``dynamics'',
we take the parameterization where we hold the mass eigenvalue
of the lightest gluino, $\Mone$, fixed, while varying the ratio
$x \equiv M_m/M_d$.  This gives two Majorana gluinos with masses 
$\Mone$ and $\Mtwo$ with mass difference given by $\Mtwo - \Mone = M_m$.
In the case $x < 1$, the mixing angle is in the range 
$1/\sqrt{2} < \cos\theta_{\g} \lsim 0.85$.  The mass spectrum
is illustrated in Fig.~\ref{fig:Splits_M_mp_OFF}.

To explore a wider range of mixing angles, $0.85 \lsim \cos\theta_{\g} \le 1$,
the parameter $x \gg 1$, that corresponds to $M_m \gg M_d$.  
In this regime, we get the usual see-saw formula,
familiar from neutrino physics, for the mass of the lightest gluino
eigenstate, $\simeq M_d^2/M_m$.  Here, however, the lighter mass eigenstate
decouples from squarks and quarks, while it is the heavier nearly pure
Majorana gluino eigenstate that maximally couples.
Without adjusting our basic premise -- hold the kinematics constant --
there is no way to enter this regime of parameters without taking
the Majorana mass for the gluino unnaturally large.

\begin{figure*}
\begin{centering}
\begin{subfigure}[t]{0.32\textwidth}
\caption{$\Msq=400$~GeV: ratios}
\includegraphics[width=6.0cm]{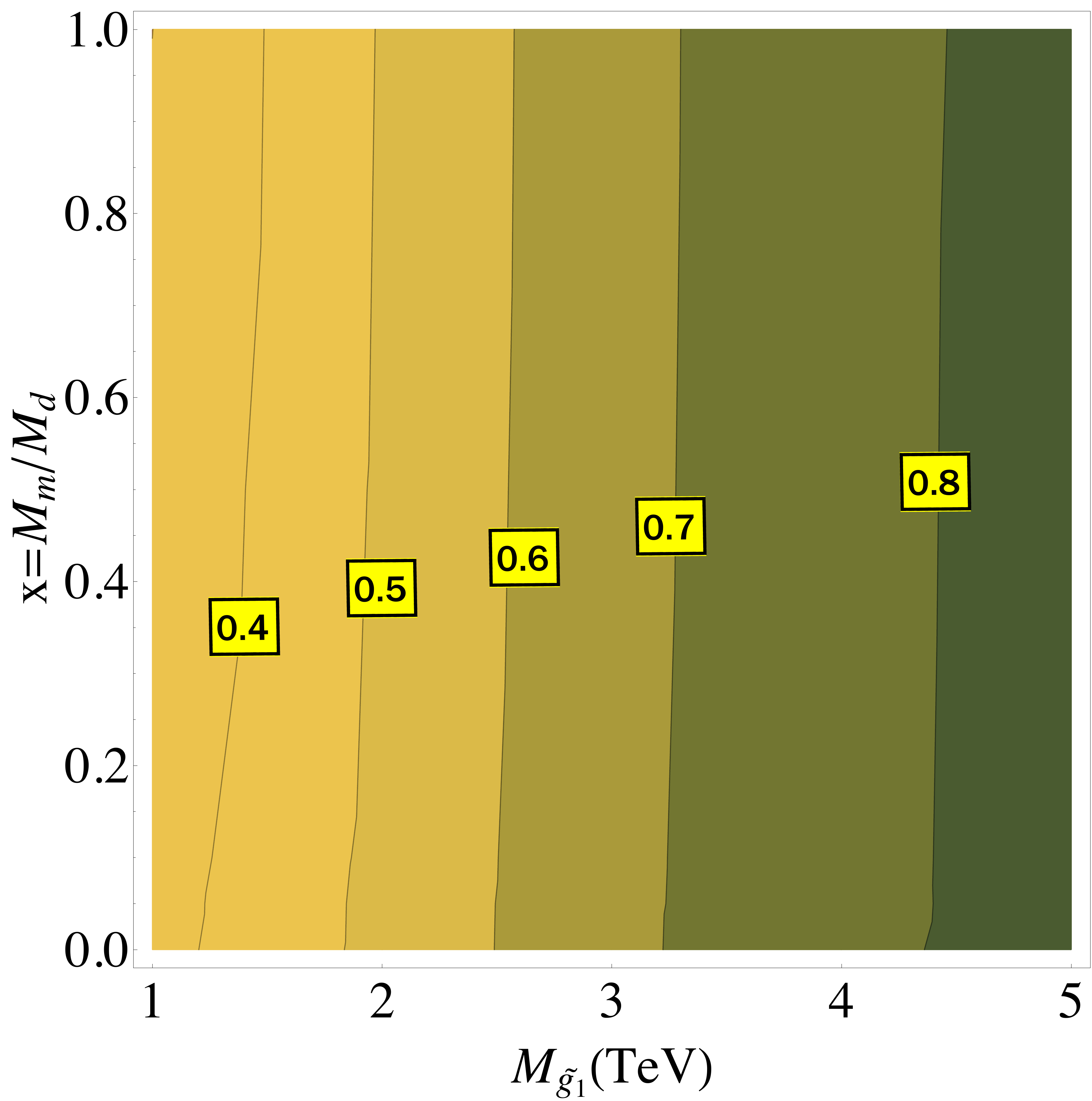}
\label{fig:MbyD400RatioContour}
\end{subfigure} \quad \quad \quad
\begin{subfigure}[t]{0.32\textwidth}
\caption{$\Msq=400$~GeV: cross sections}
\includegraphics[width=6.0cm]{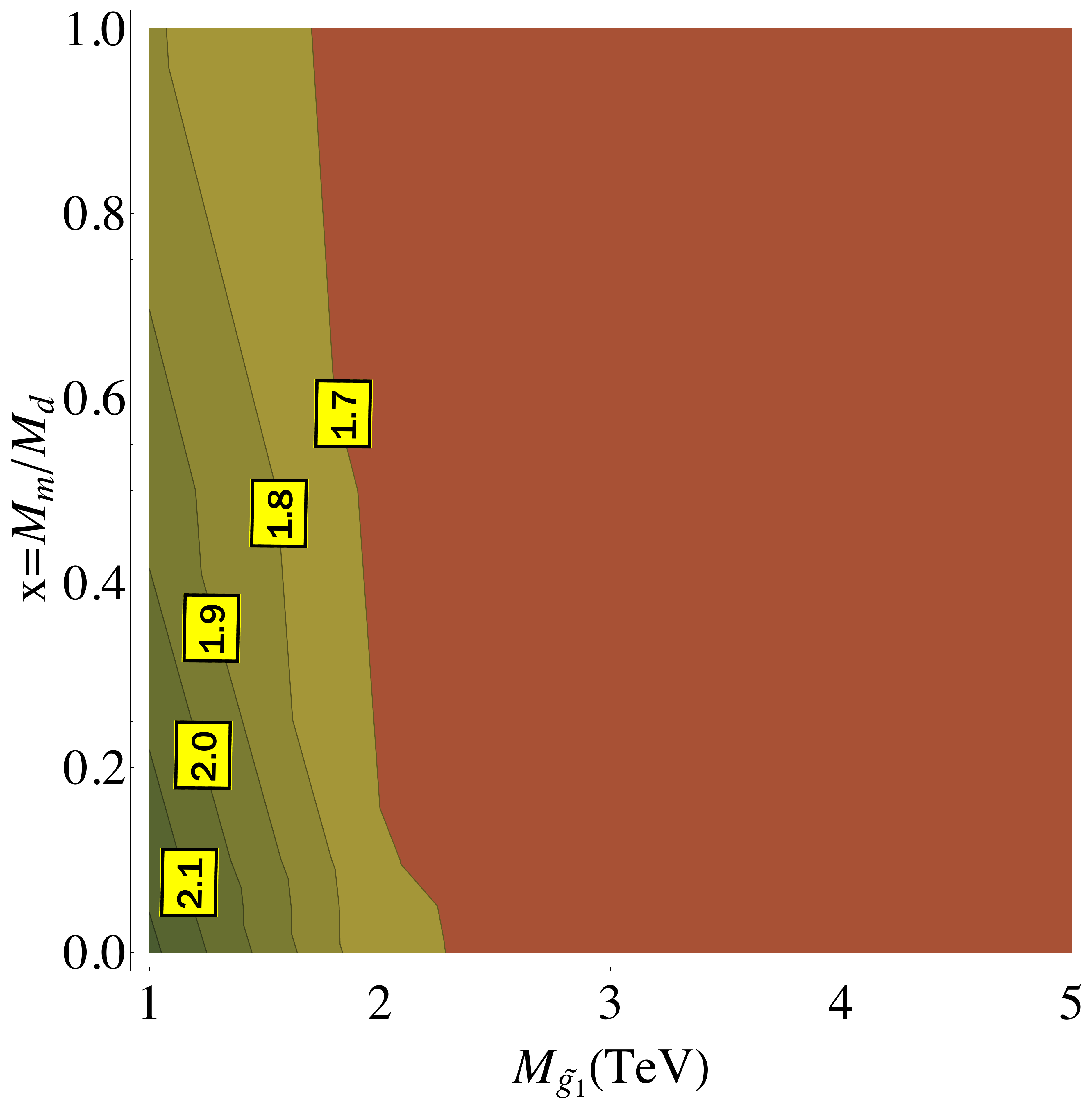}
\label{fig:MbyD400}
\end{subfigure}
\\

\begin{subfigure}[t]{0.32\textwidth}
\caption{$\Msq=800$~GeV: ratios}
\includegraphics[width=6.0cm]{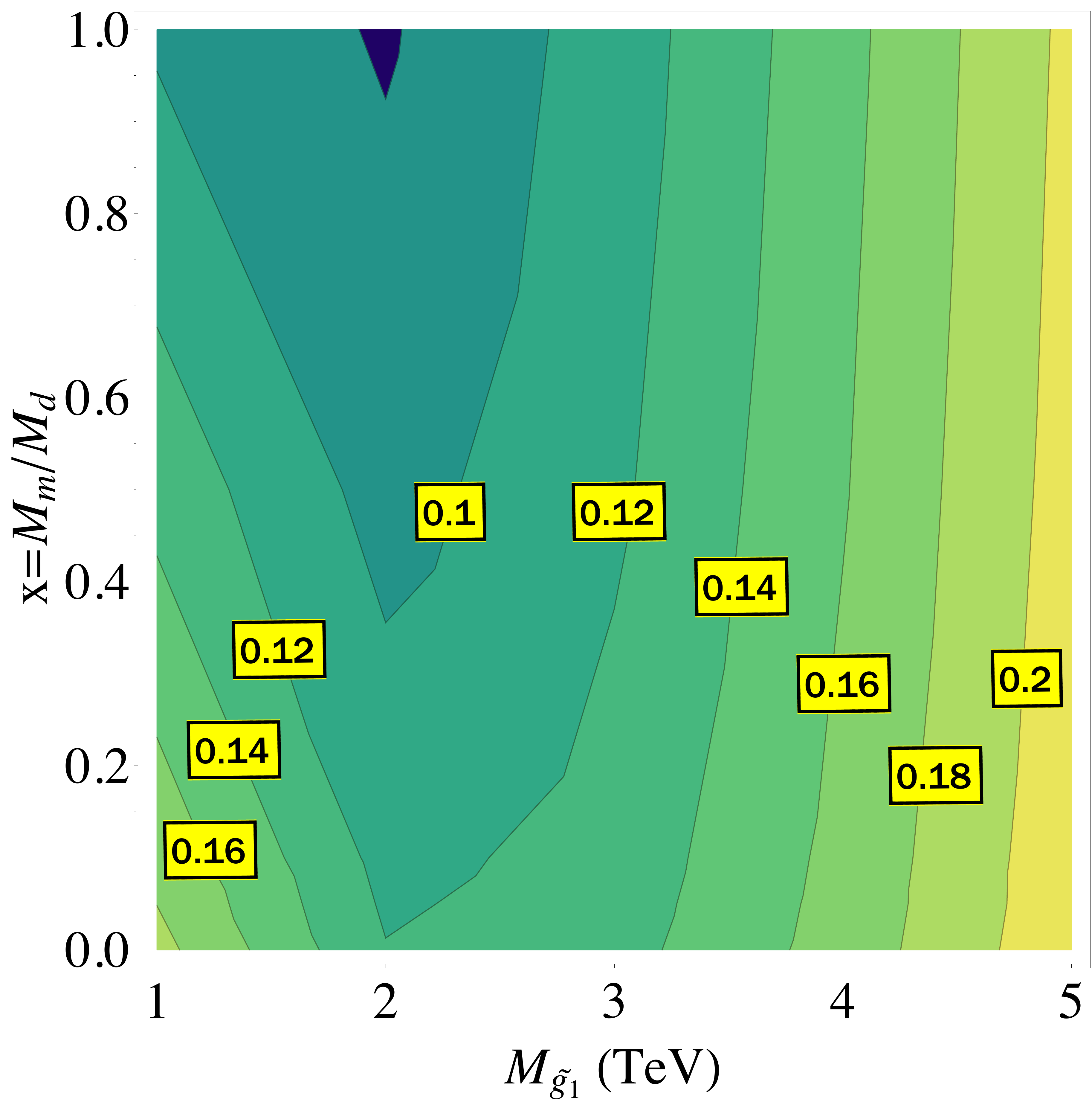}
\label{fig:MbyD800RatioContour}
\end{subfigure} \quad \quad \quad
\begin{subfigure}[t]{0.32\textwidth}
\caption{$\Msq=800$~GeV: cross sections}
\includegraphics[width=6.0cm]{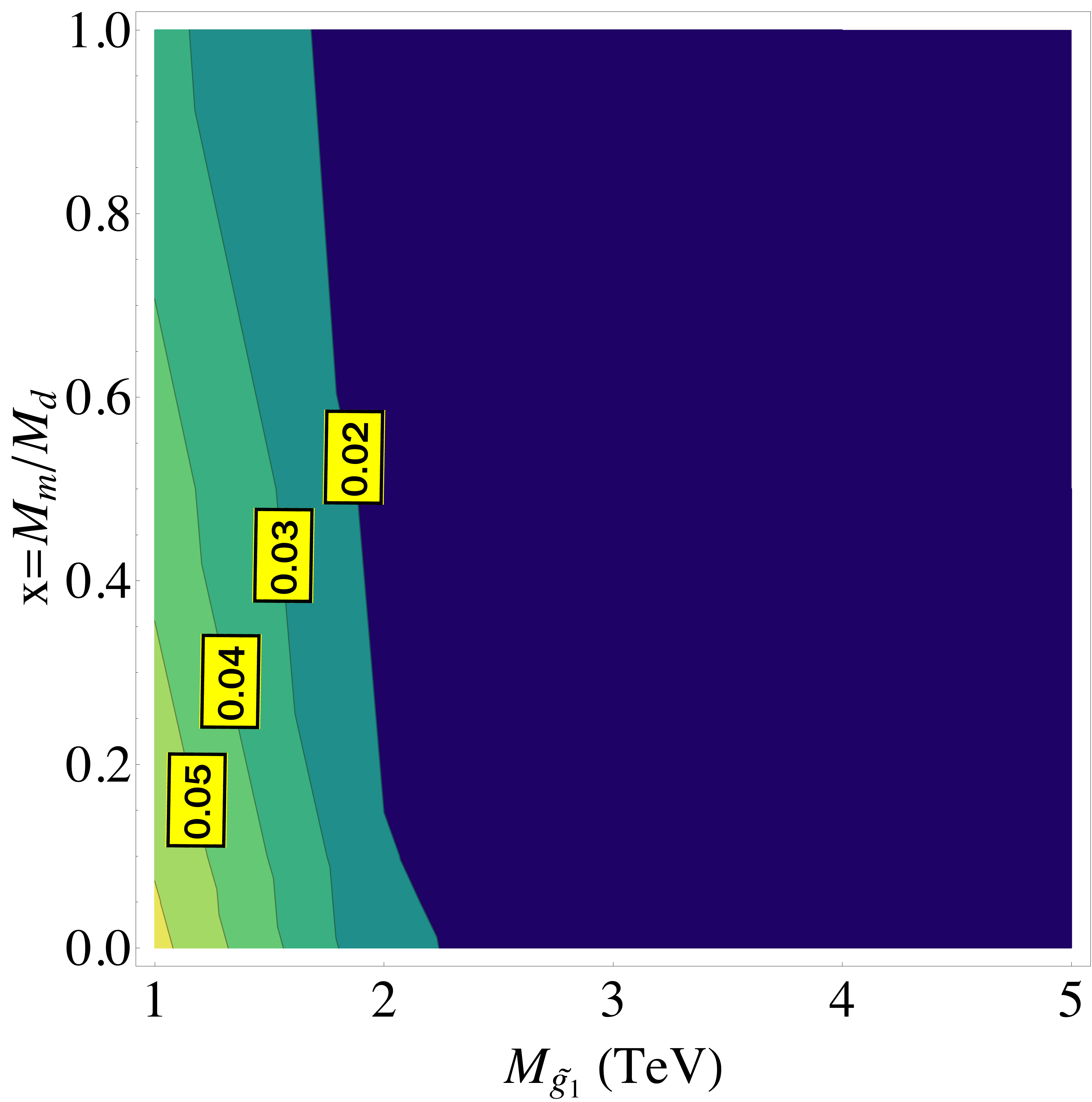}
\label{fig:MbyD800}
\end{subfigure}
\\

\begin{subfigure}[t]{0.32\textwidth}
\caption{$\Msq=1200$~GeV: ratios}
\includegraphics[width=6.0cm]{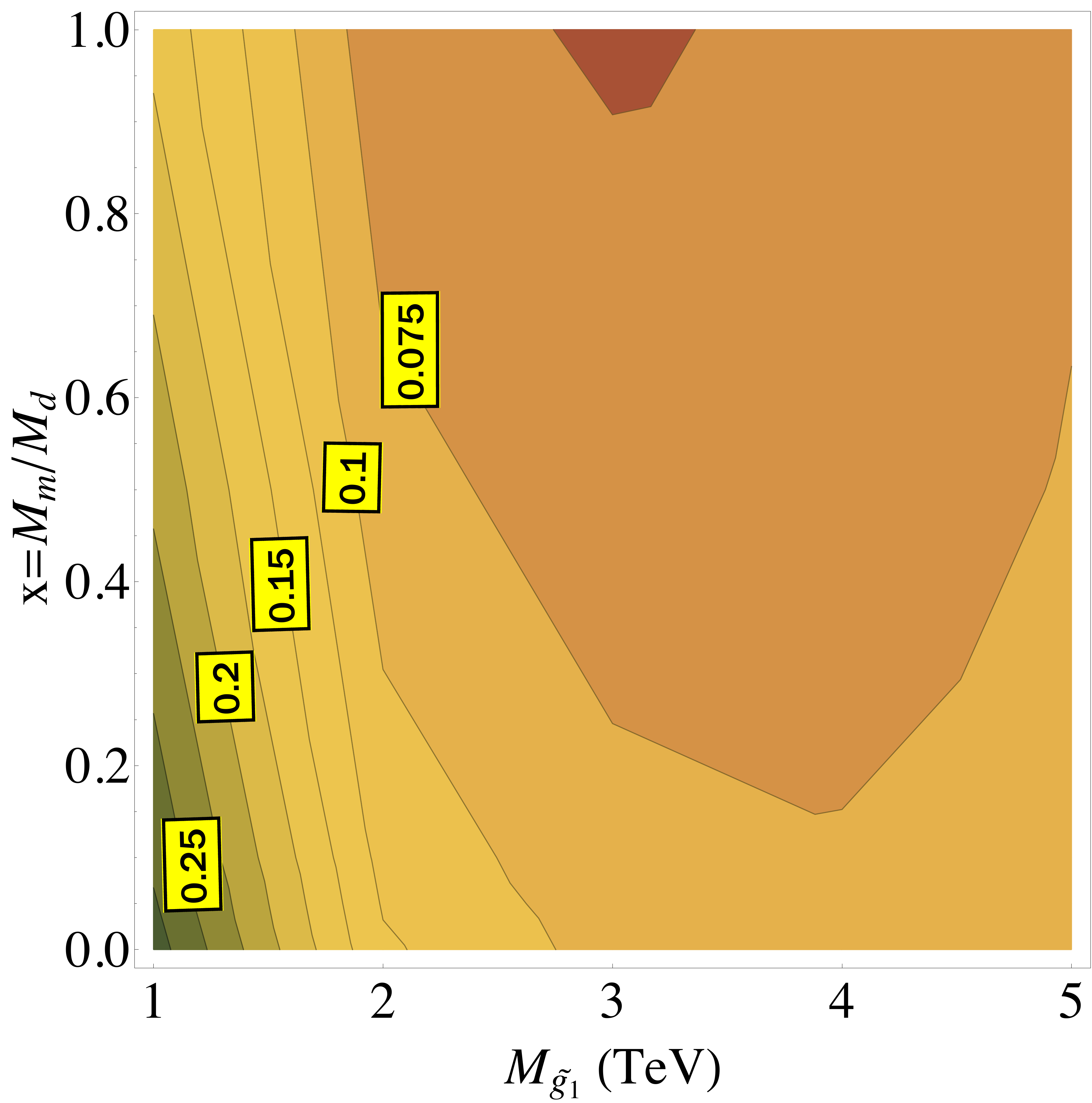}
\label{fig:MbyD1200RatioContour}
\end{subfigure} \quad \quad \quad
\begin{subfigure}[t]{0.32\textwidth}
\caption{$\Msq=1200$~GeV: cross sections}
\includegraphics[width=6.0cm]{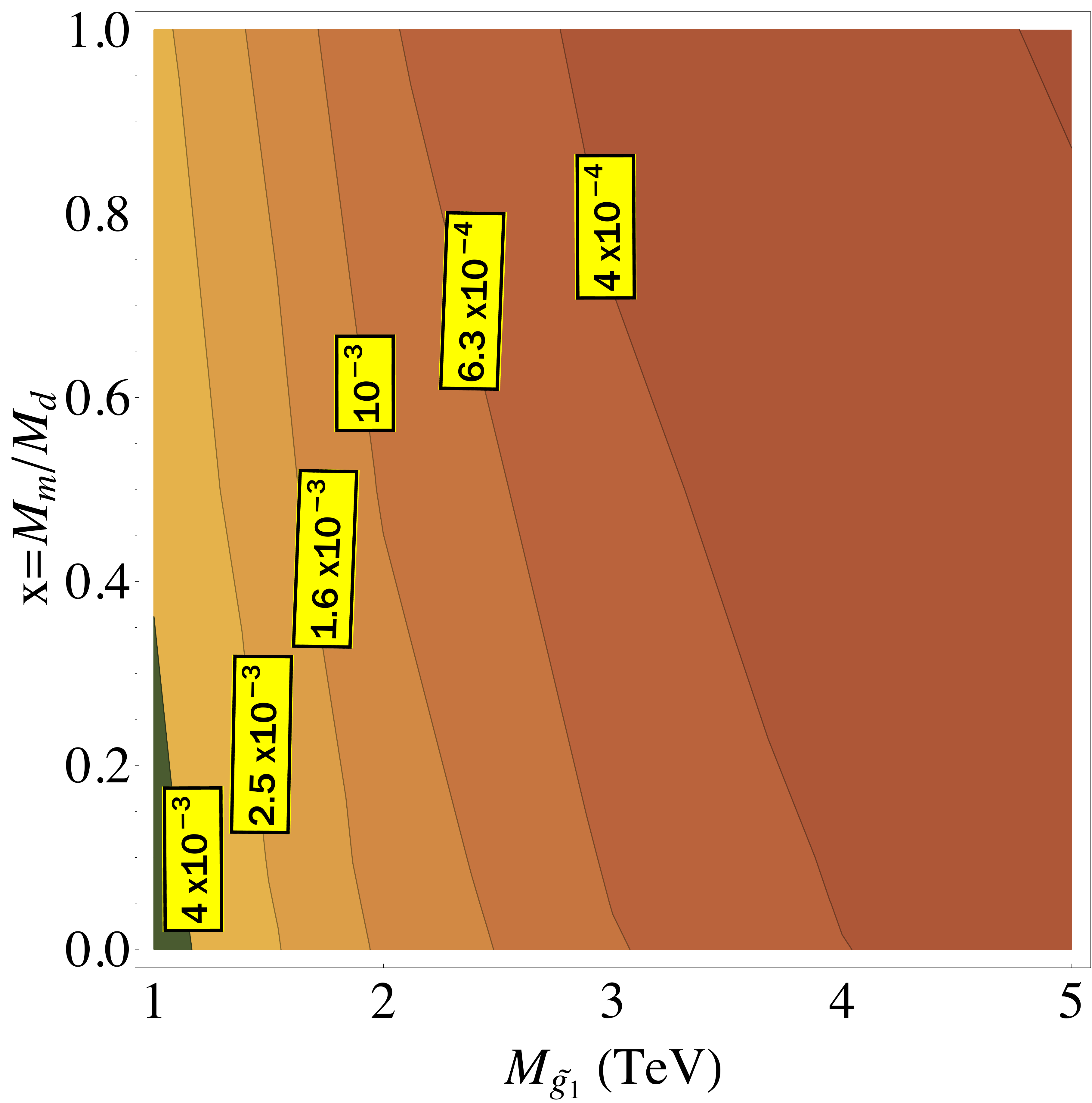}
\label{fig:MbyD1200}
\end{subfigure}

\caption{Plots illustrating Case I. LEFT: Contours of the ratio of the total production cross section
of the first two generations of squarks at LHC with $\sqrt{s} = 8$~TeV\@
in our model to the cross sections in MSSM\@.
RIGHT: Contours of the cross sections themselves (at leading order), in pb, at LHC
with $\sqrt{s} = 8$~TeV\@.
In these plots, we show the variation
as the lightest gaugino mass ($\Mone$) is varied simultaneous
with varying the relative size of the $M_m$ and $M_d$,
parameterized by $x = M_m/M_d$. The details of the critical features
are explained in the text.}
\label{MajbyDirSqXS}
\end{centering}
\end{figure*}

\begin{figure*}
\begin{subfigure}[t]{0.32\textwidth}
\includegraphics[width=5.3cm]{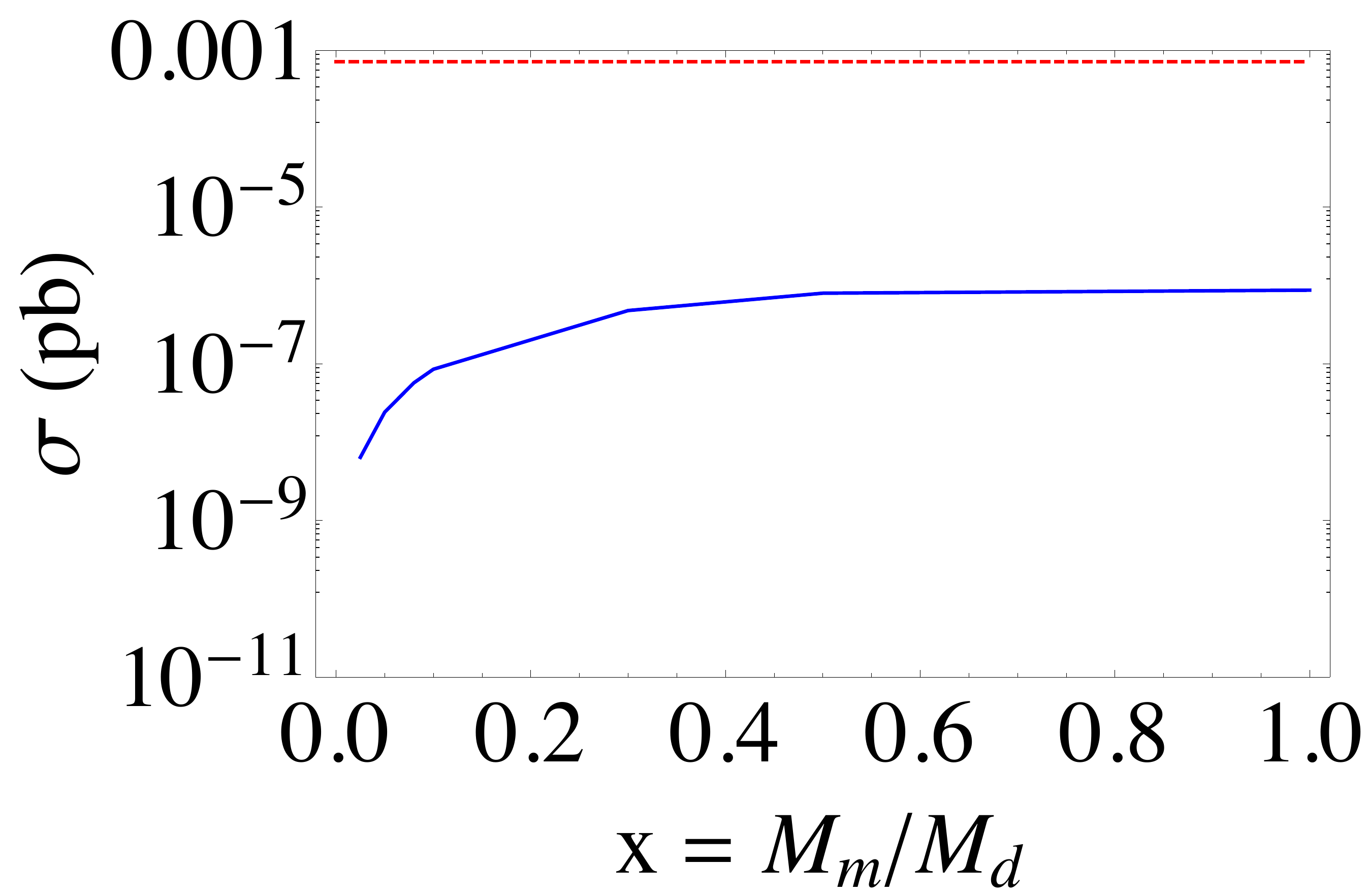}
\caption{$\tilde{u}_L \tilde{u}_L$}
\label{fig:uLuL}
\end{subfigure}
\begin{subfigure}[t]{0.32\textwidth}
\includegraphics[width=5.3cm]{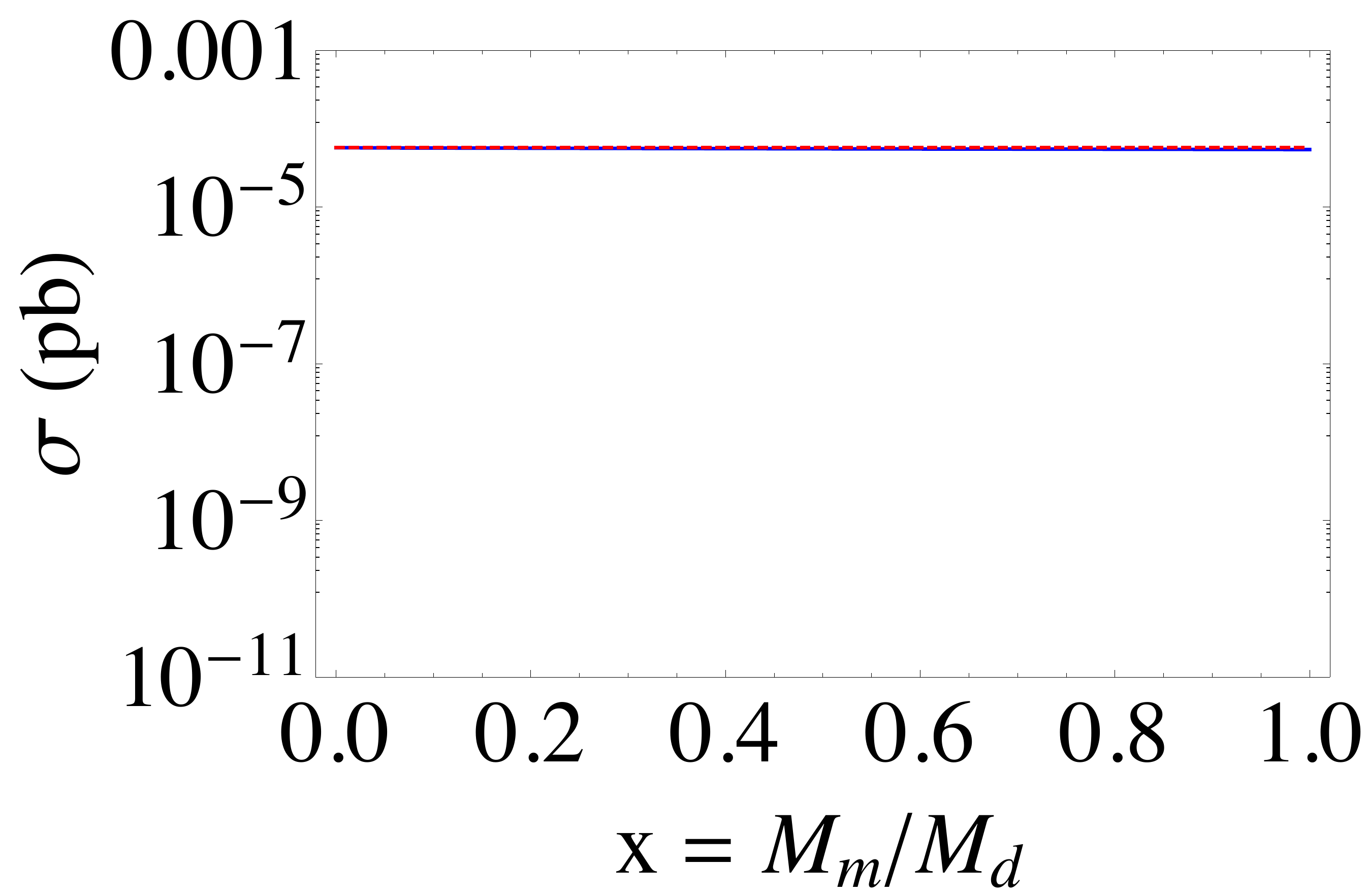}
\caption{$\tilde{u}_L \tilde{u}^*_L$}
\label{fig:uLuLStar}
\end{subfigure}
\begin{subfigure}[t]{0.32\textwidth}
\includegraphics[width=5.3cm]{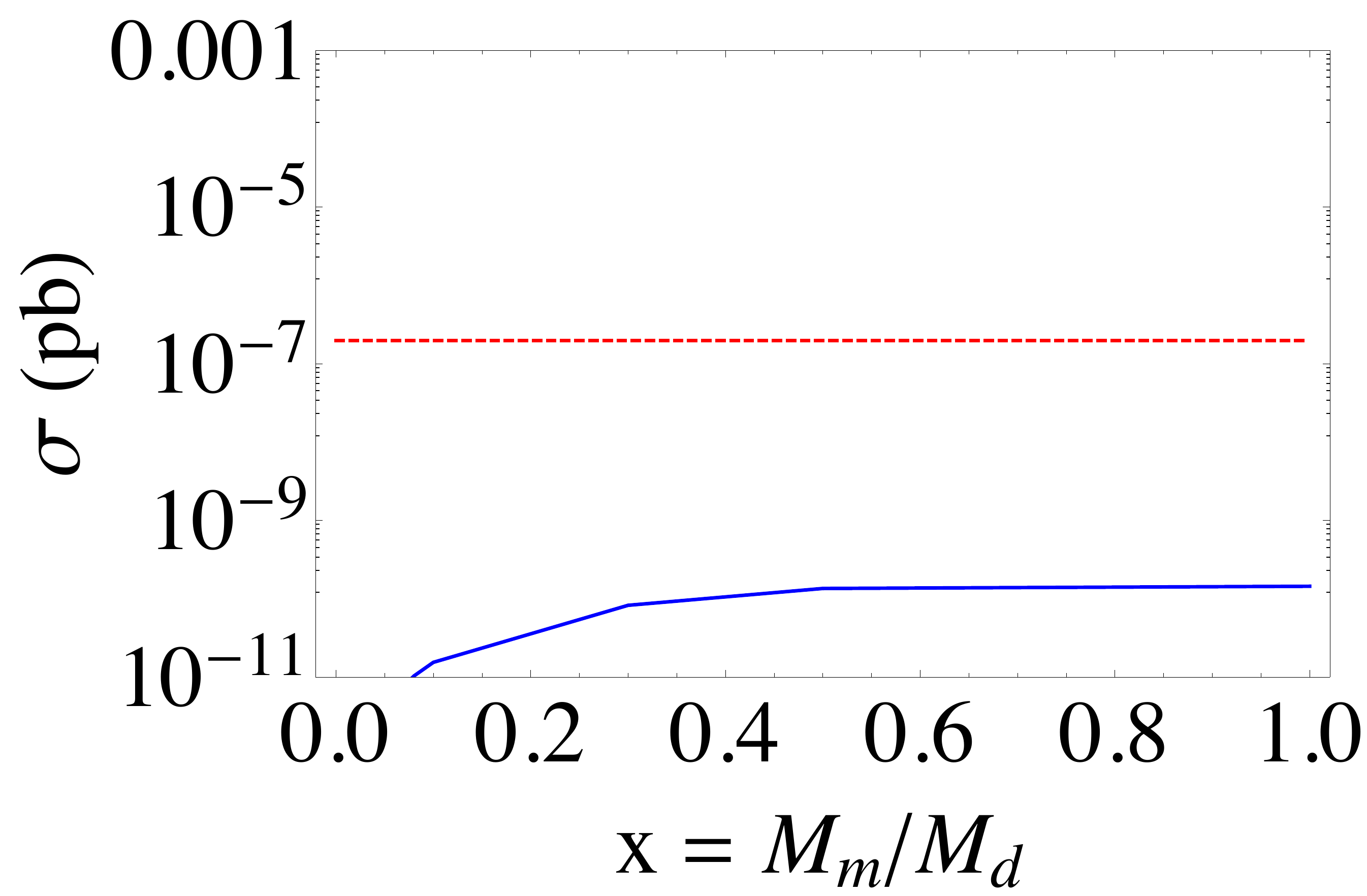}
\caption{$\tilde{u}^*_L \tilde{u}^*_L$}
\label{fig:uLStaruLstar}
\end{subfigure}
\begin{subfigure}[t]{0.32\textwidth}
\includegraphics[width=5.3cm]{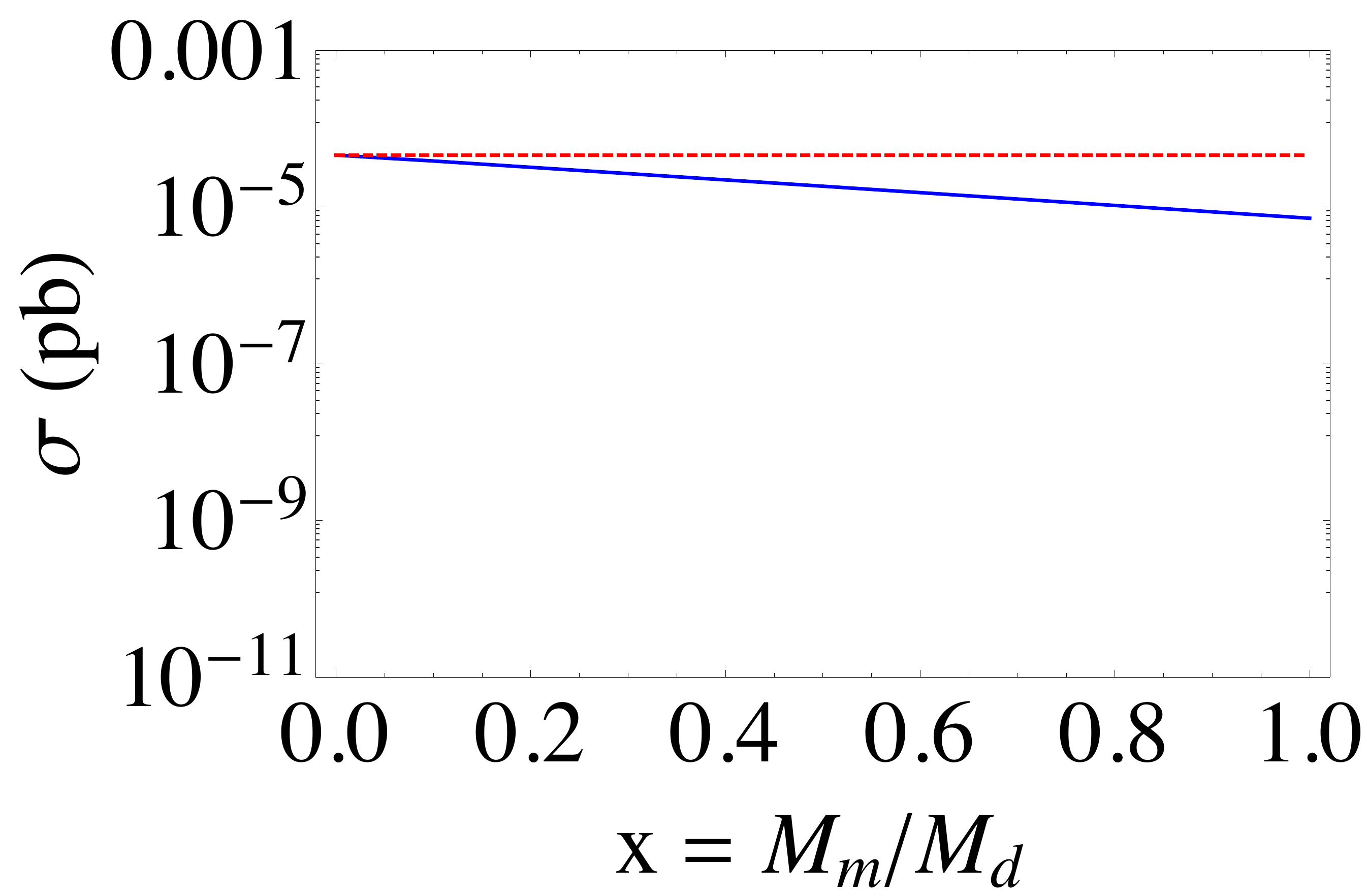}
\caption{$ \tilde{u}_L\tilde{u}_R $}
\label{fig:uLuR}
\end{subfigure}
\begin{subfigure}[t]{0.32\textwidth}
\includegraphics[width=5.3cm]{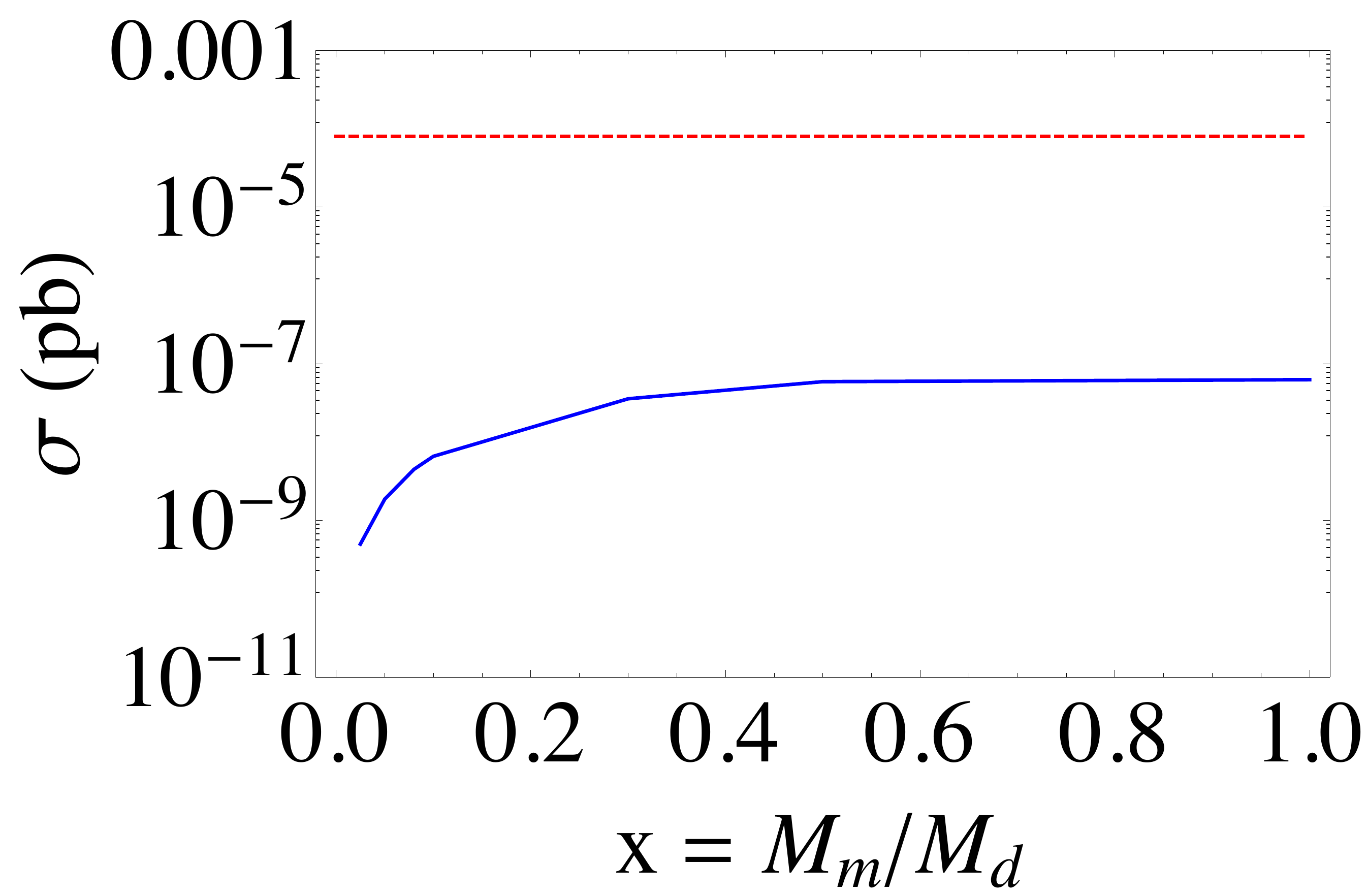}
\caption{$ \tilde{u}_L\tilde{u}^*_R $}
\label{fig:uLuRStar}
\end{subfigure}
\begin{subfigure}[t]{0.32\textwidth}
\includegraphics[width=5.3cm]{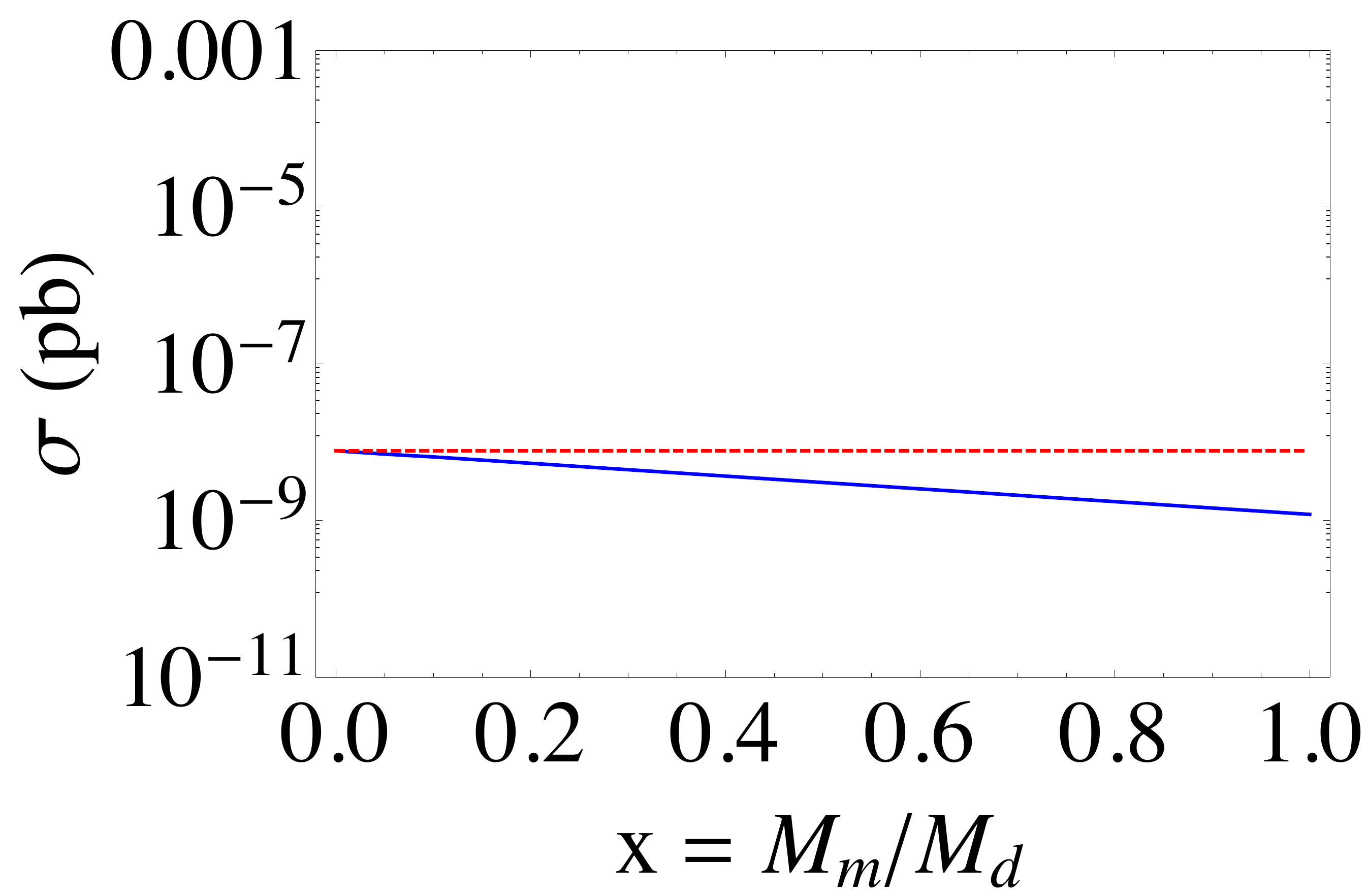}
\caption{$ \tilde{u}^*_L\tilde{u}^*_R $}
\label{fig:uLStaruRStar}
\end{subfigure}
\caption{Cross sections of the various unique modes that constitute
up squark production when $M_m'$ is set to zero.
The blue curves show these as a function $x = M_m/M_d$,
while the dashed red horizontal lines denote the corresponding cross section
for the case of a pure Majorana gluino of the same mass as $\Mone$.
Here the squark mass $M_{\tilde{u}}$ is $1200$~GeV and the mass of the
lighter gluino eigenstate $\Mone$ is $5$~TeV\@.}
\label{ResolvedXS}
\end{figure*}

\begin{figure*}
\begin{centering}
\begin{subfigure}[t]{0.32\textwidth}
\caption{$\Msq=400$~GeV: ratios}
\includegraphics[width=6.0cm]{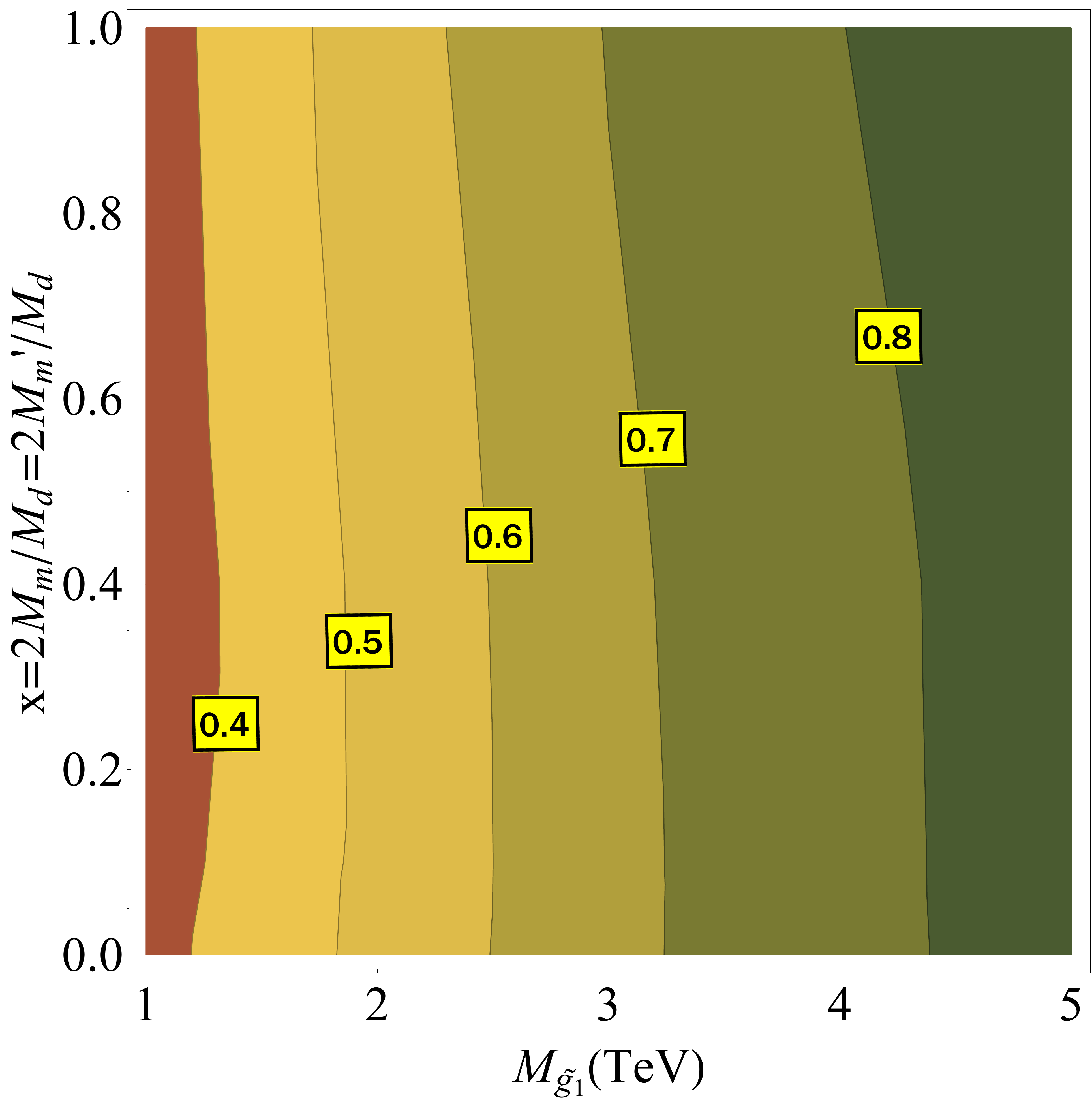}
\label{fig:MEqualbyD400RatioContour}
\end{subfigure} \quad \quad \quad
\begin{subfigure}[t]{0.32\textwidth}
\caption{$\Msq=400$~GeV: cross sections}
\includegraphics[width=6.0cm]{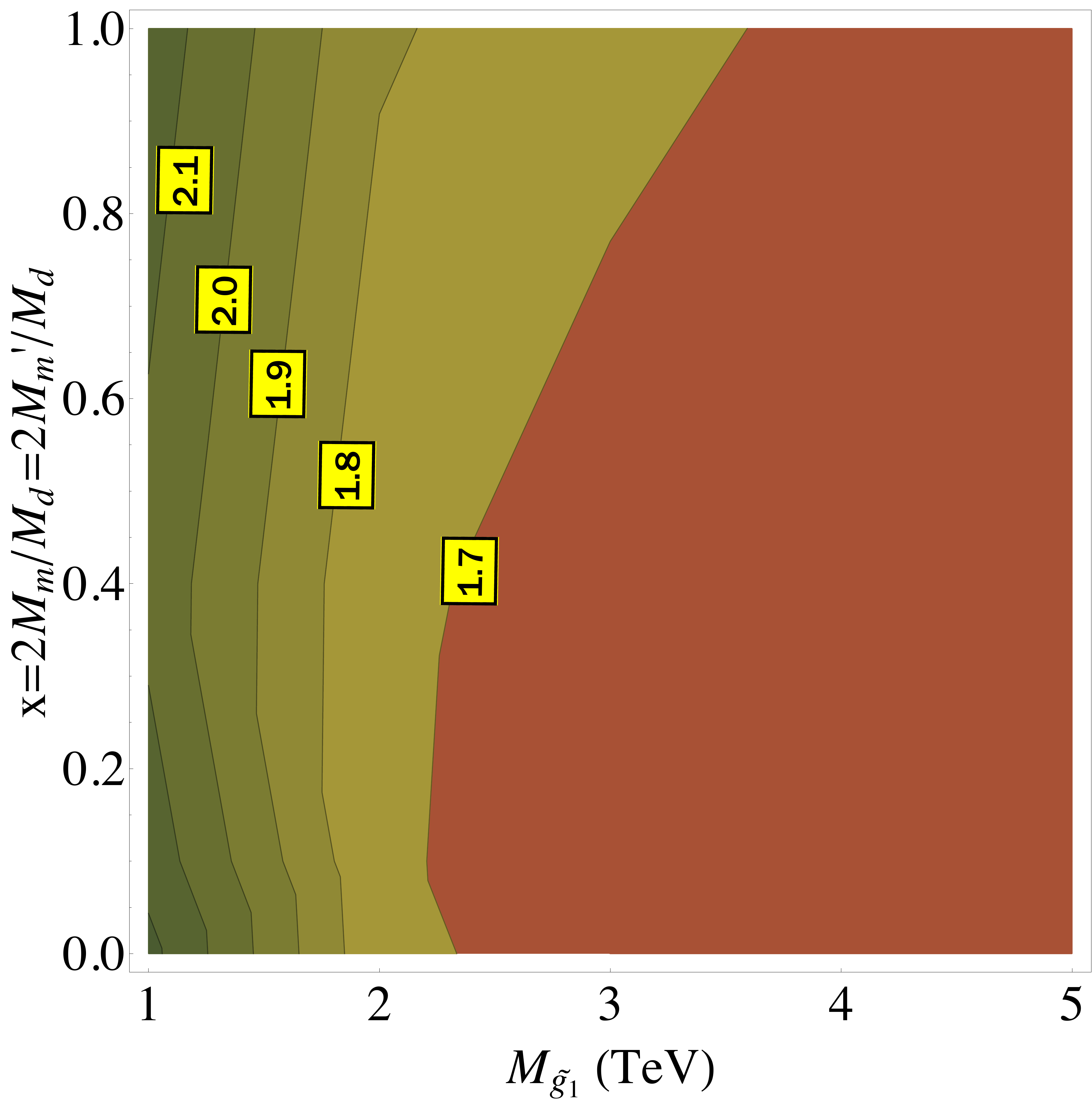}
\label{fig:MEqualbyD400}
\end{subfigure}
\\

\begin{subfigure}[t]{0.32\textwidth}
\caption{$\Msq=800$~GeV: ratios}
\includegraphics[width=6.0cm]{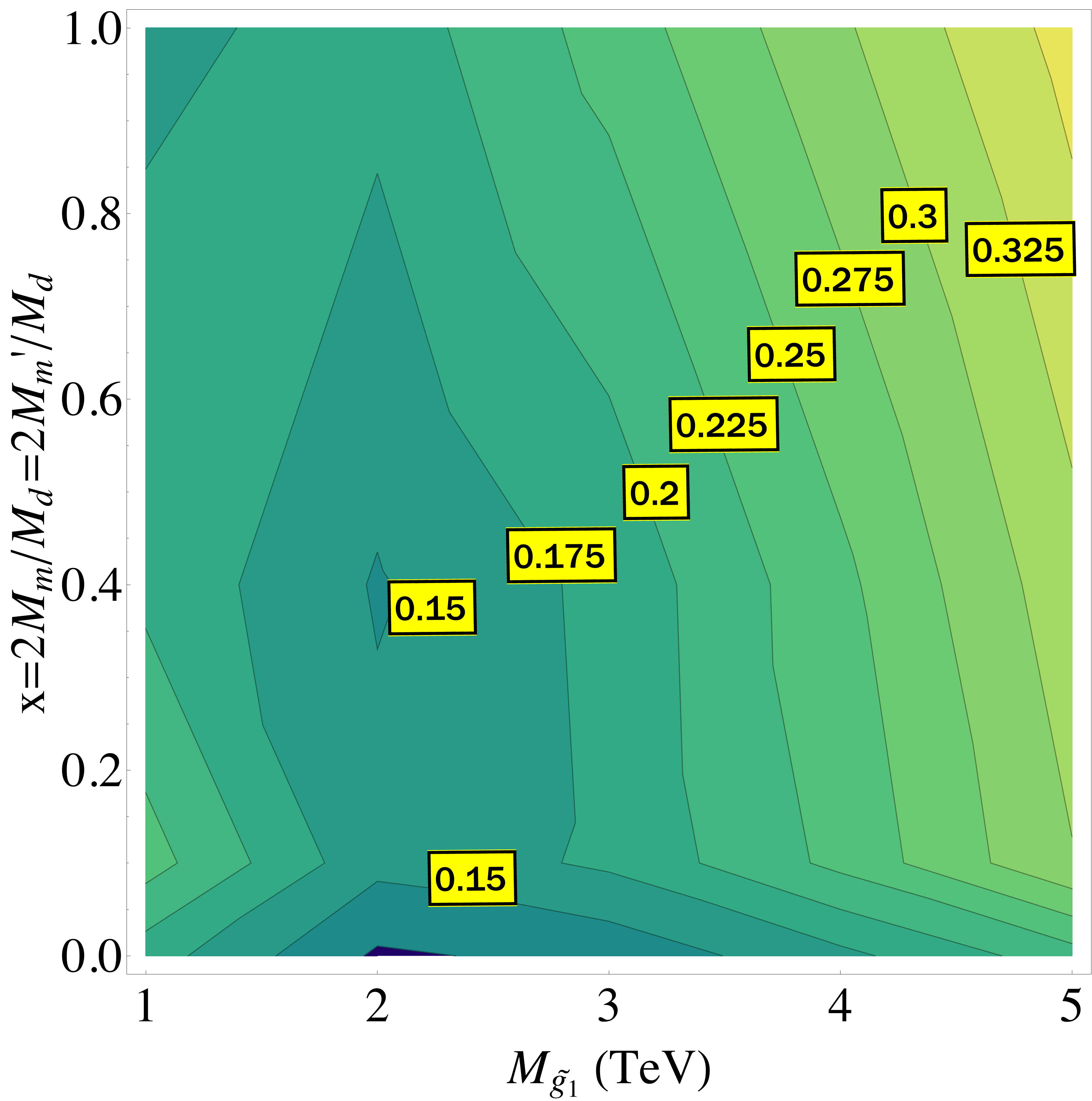}
\label{fig:MEqualbyD800RatioContour}
\end{subfigure} \quad \quad \quad
\begin{subfigure}[t]{0.32\textwidth}
\caption{$\Msq=800$~GeV: cross sections}
\includegraphics[width=6.0cm]{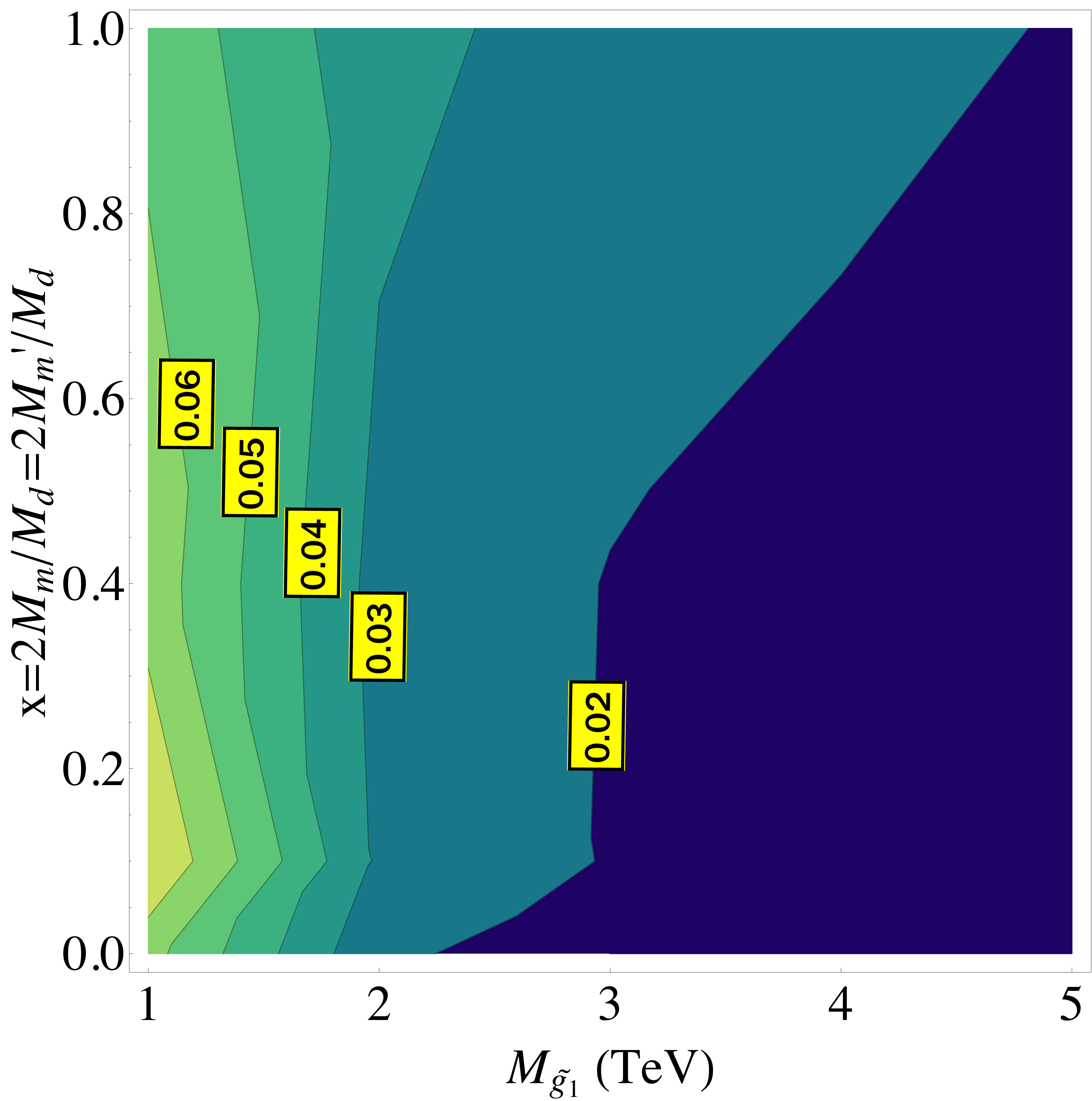}
\label{fig:MEqualbyD800}
\end{subfigure}
\\

\begin{subfigure}[t]{0.32\textwidth}
\caption{$\Msq=1200$~GeV: ratios}
\includegraphics[width=6.0cm]{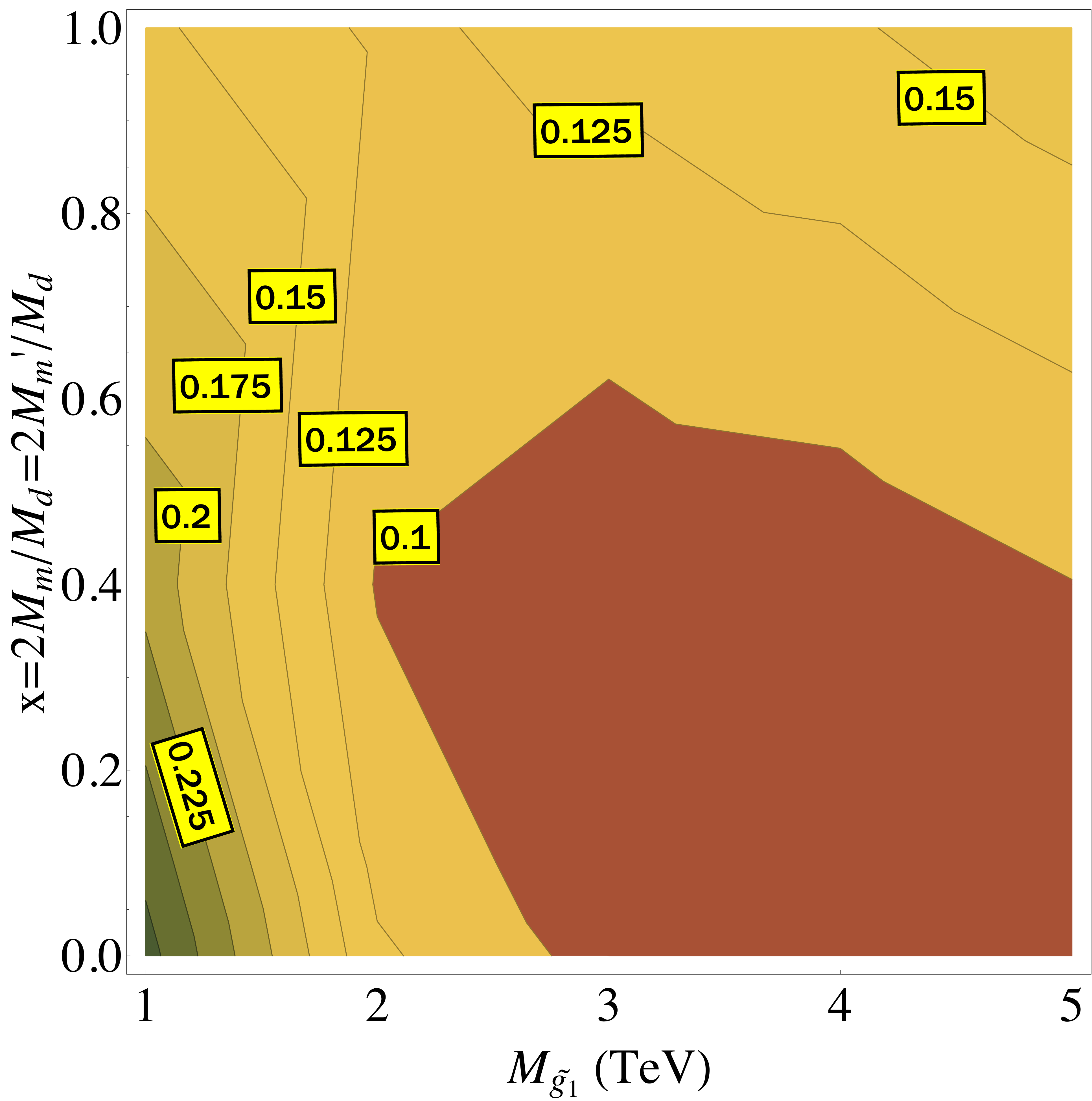}
\label{fig:MEqualbyD1200RatioContour}
\end{subfigure} \quad \quad \quad
\begin{subfigure}[t]{0.32\textwidth}
\caption{$\Msq=1200$~GeV: cross sections}
\includegraphics[width=6.0cm]{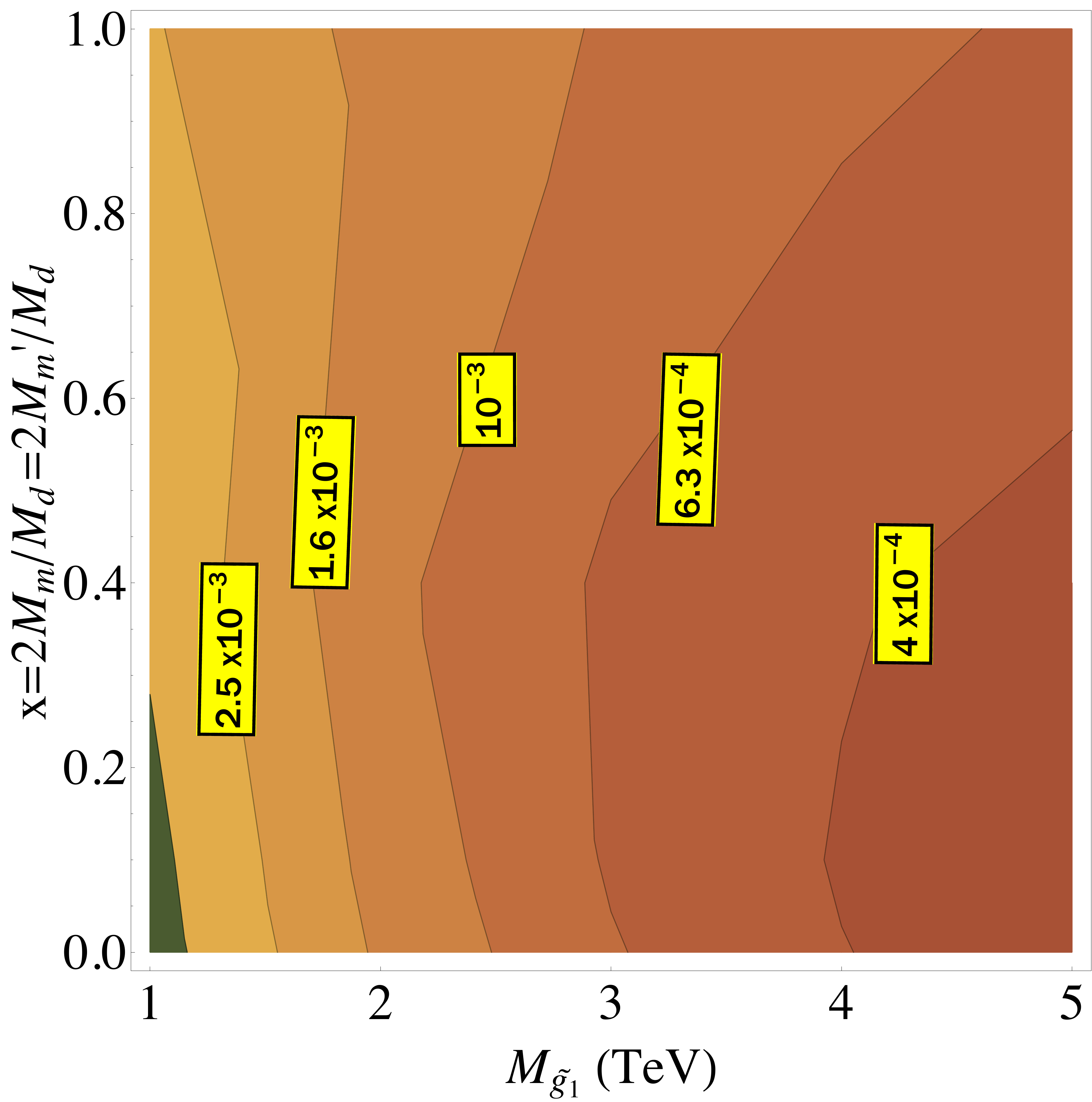}
\label{fig:MEqualbyD1200}
\end{subfigure}
\caption{Plots illustrating Case II. LEFT: Contours of the ratio of the total production cross section of the
first two generations of squarks at LHC with $\sqrt{s} = 8$~TeV\@ in our model
to the cross sections in MSSM\@.  RIGHT: Contours of the cross sections themselves (at leading order),
in pb, at LHC with $\sqrt{s} = 8$~TeV\@.
Here we have taken $M_m = M_m'$, and we show the variation
as the lightest gaugino mass ($\Mone$) is varied simultaneous
with varying $M_m = M_m'$ and $M_d$,
parameterized by $x = 2M_m/M_d = 2M_m'/M_d$. The critical features are
explained in the text.}
\label{MajEqualbyDirSqXS}
\end{centering}
\end{figure*}

\begin{figure*}
\begin{subfigure}[t]{0.32\textwidth}
\includegraphics[width=5.3cm]{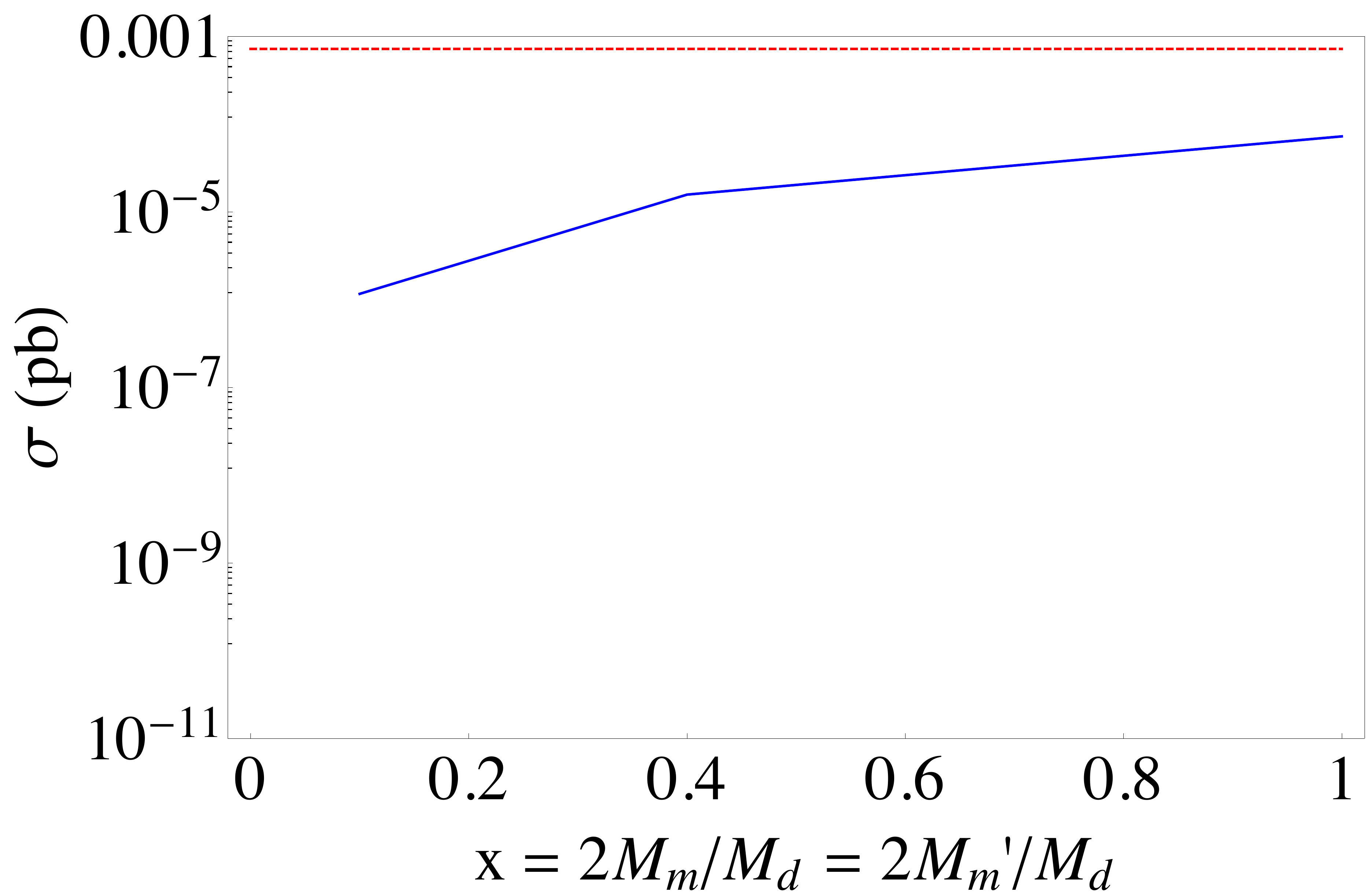}
\caption{$\tilde{u}_L \tilde{u}_L$}
\label{fig:uLuLEqual}
\end{subfigure}
\begin{subfigure}[t]{0.32\textwidth}
\includegraphics[width=5.3cm]{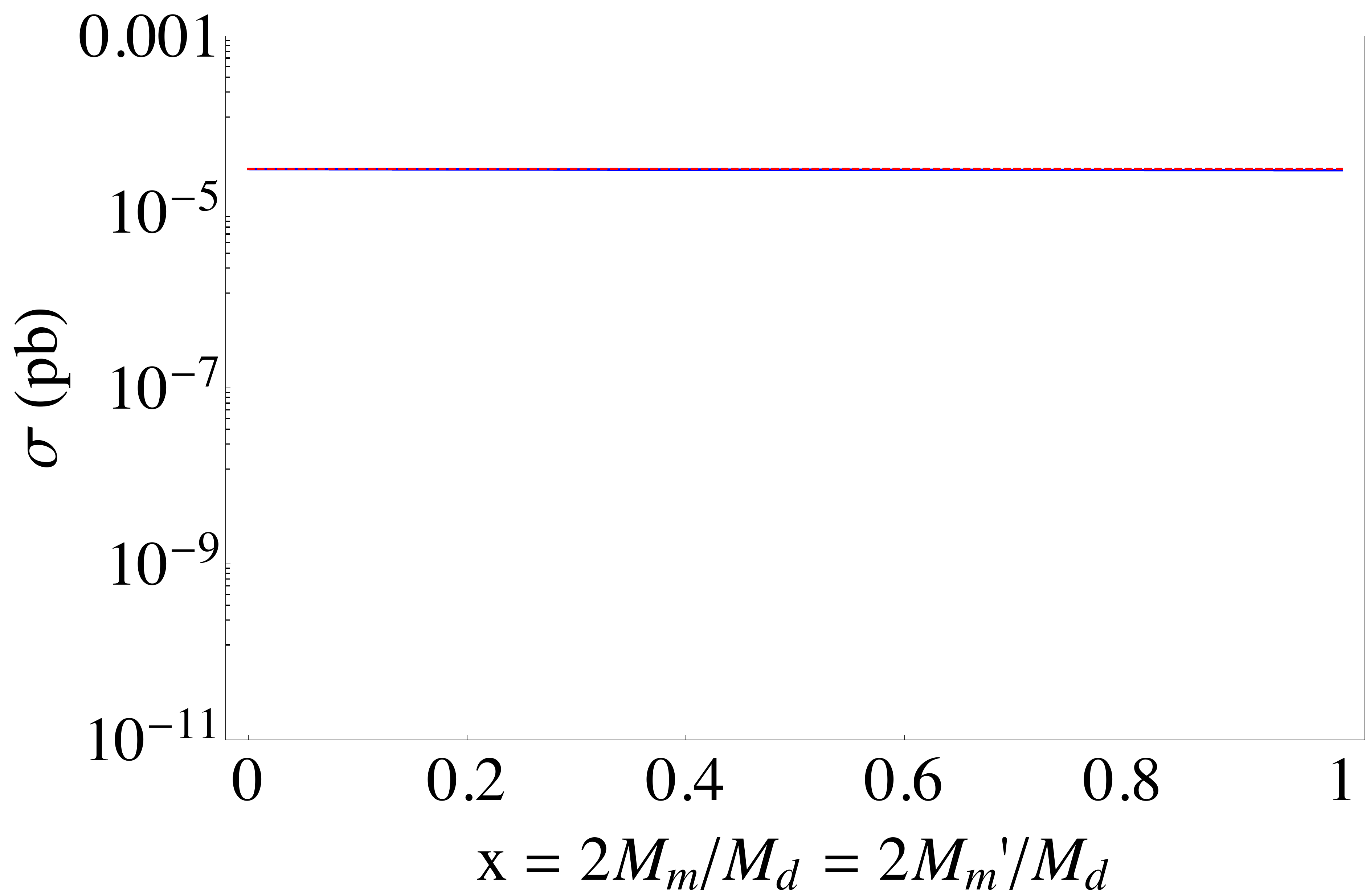}
\caption{$\tilde{u}_L \tilde{u}^*_L$}
\label{fig:uLuLStarEqual}
\end{subfigure}
\begin{subfigure}[t]{0.32\textwidth}
\includegraphics[width=5.3cm]{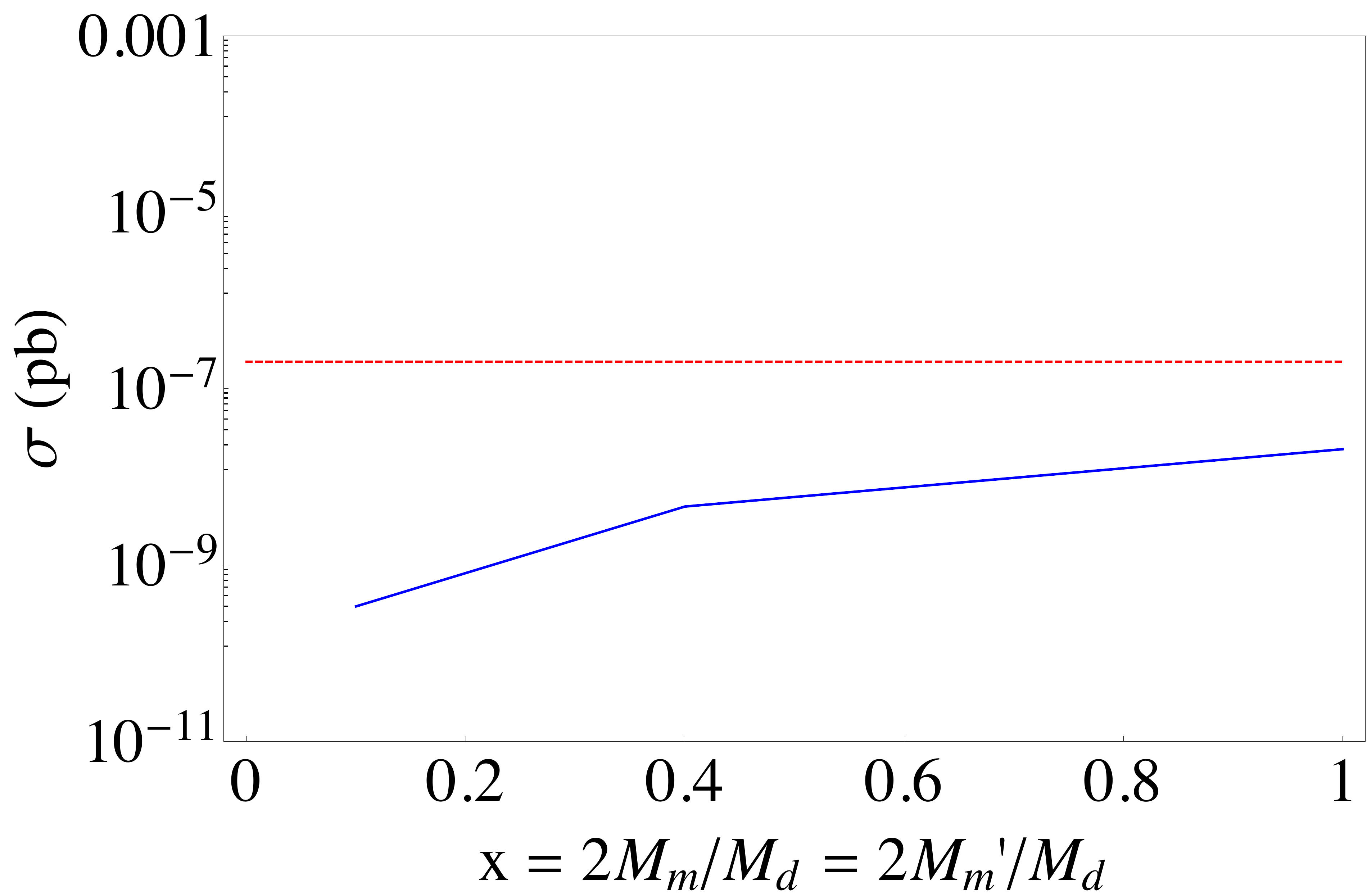}
\caption{$\tilde{u}^*_L \tilde{u}^*_L$}
\label{fig:uLStaruLstarEqual}
\end{subfigure}
\begin{subfigure}[t]{0.32\textwidth}
\includegraphics[width=5.3cm]{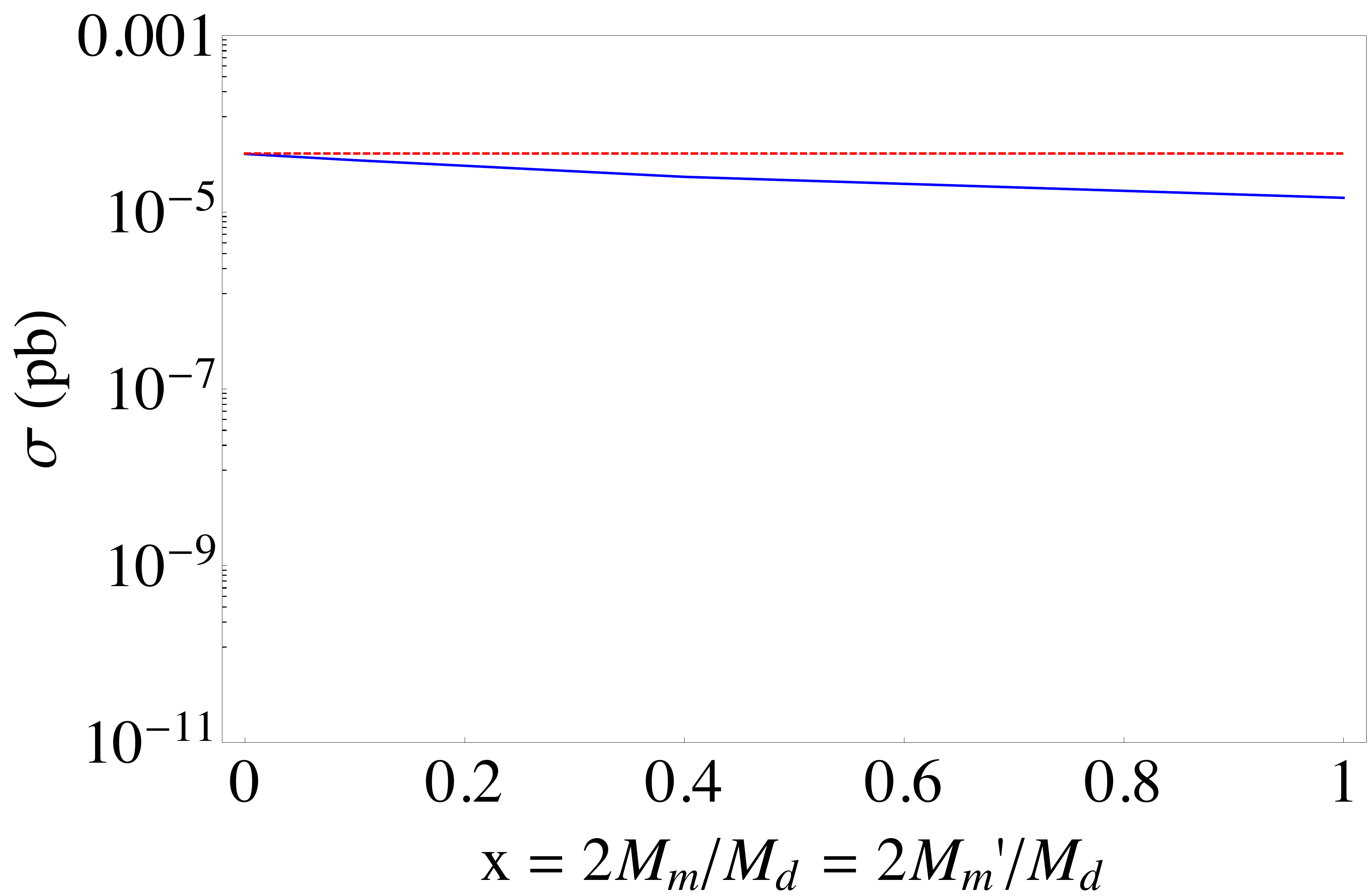}
\caption{$ \tilde{u}_L\tilde{u}_R $}
\label{fig:uLuREqual}
\end{subfigure}
\begin{subfigure}[t]{0.32\textwidth}
\includegraphics[width=5.3cm]{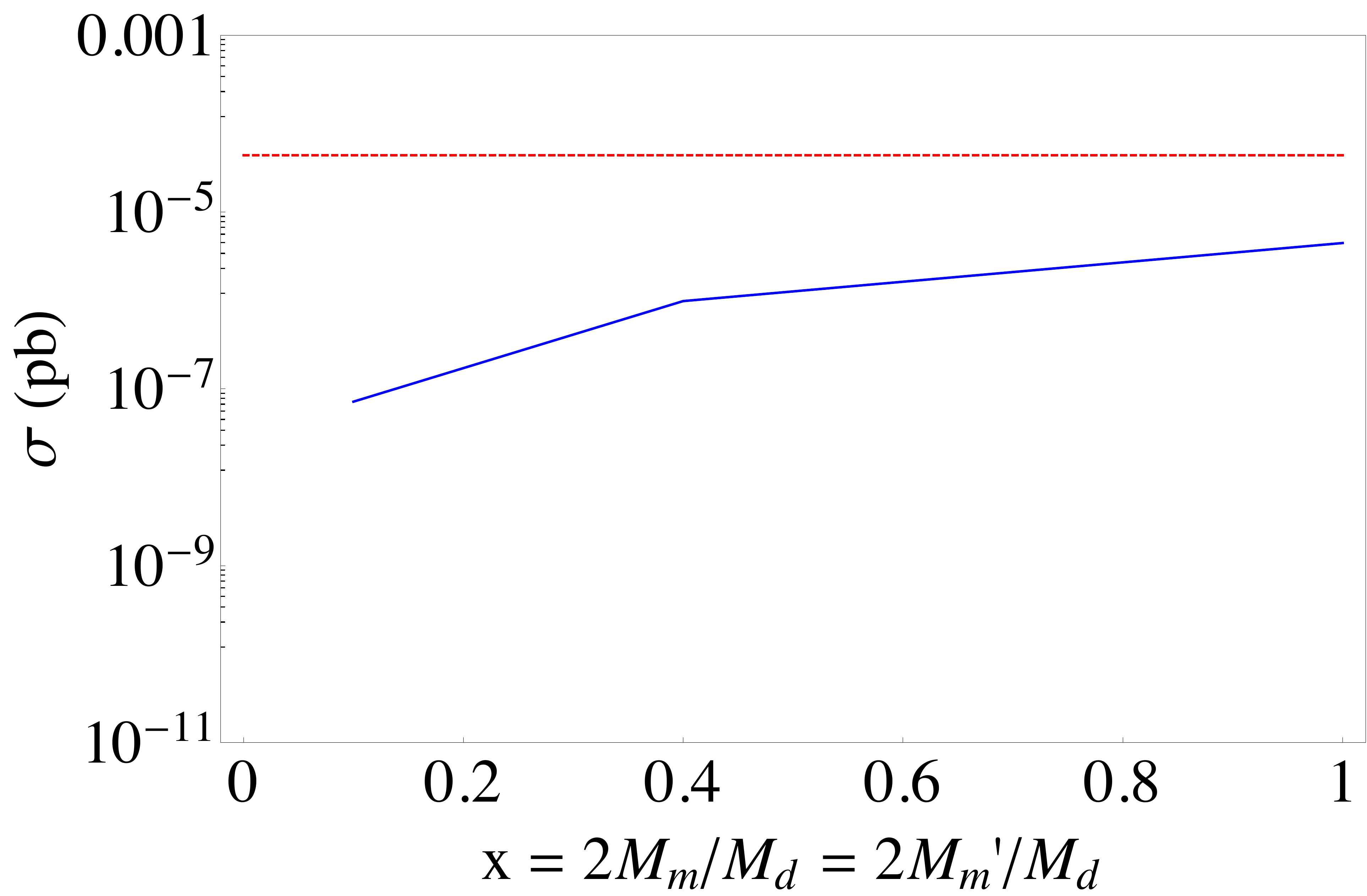}
\caption{$ \tilde{u}_L\tilde{u}^*_R $}
\label{fig:uLuRStarEqual}
\end{subfigure}
\begin{subfigure}[t]{0.32\textwidth}
\includegraphics[width=5.3cm]{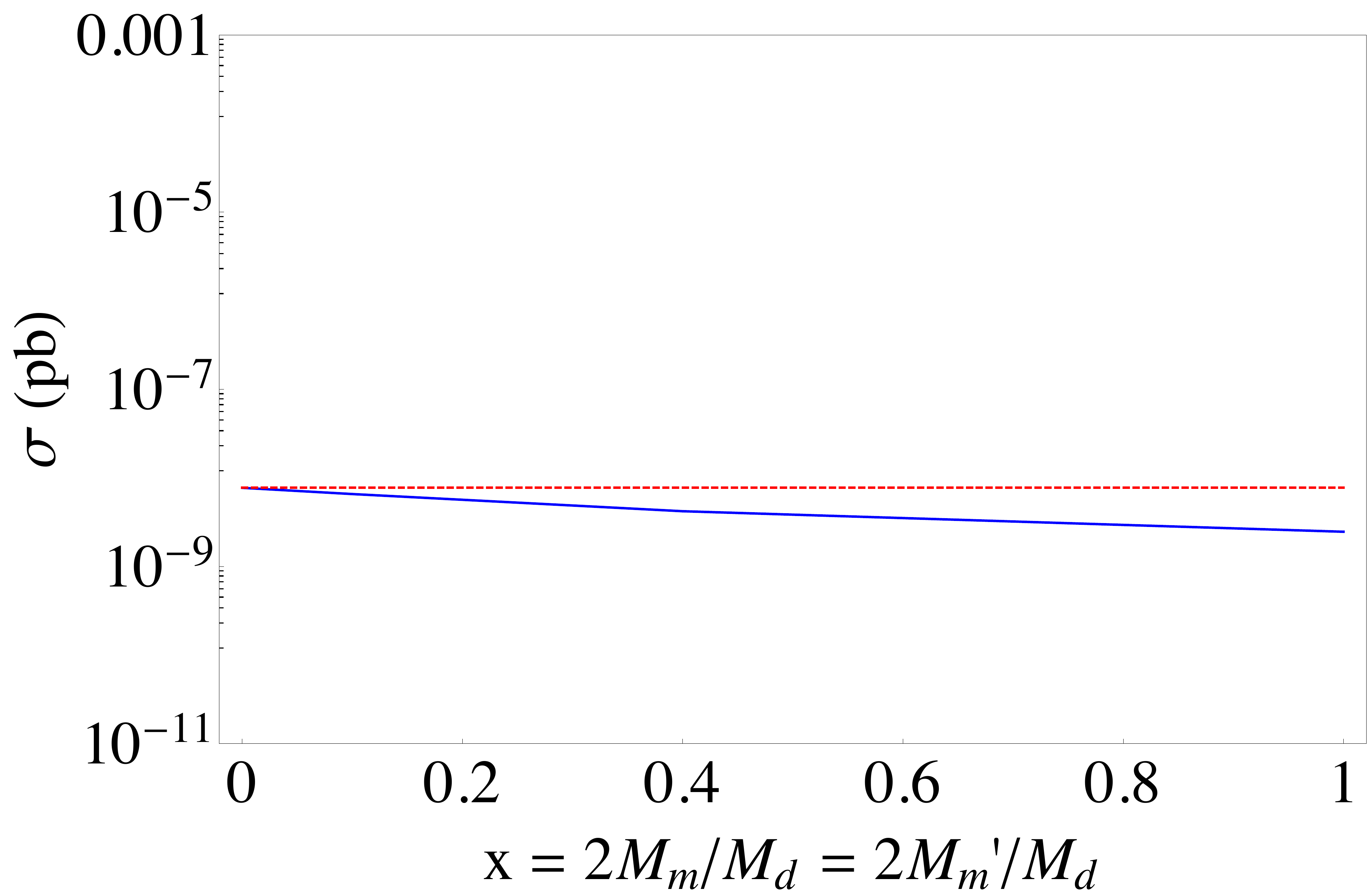}
\caption{$ \tilde{u}^*_L\tilde{u}^*_R $}
\label{fig:uLStaruRStarEqual}
\end{subfigure}
\caption{cross sections of the various unique modes that constitute
up squark production when $M_m'$  and $M_m$ are set equal. The blue curves
show these as a function $x = 2M_m/M_d = 2M_m'/M_d$,
while the dashed red horizontal lines denote the corresponding cross section
for the case of a pure Majorana gluino of the same mass as $\Mone$.
Here the squark mass $M_{\tilde{u}}$ is $1200$~GeV and the mass of the
lighter gluino eigenstate $\Mone$ is $5$~TeV\@.}
\label{ResolvedXSEqual}
\end{figure*}

\begin{figure*}
\begin{centering}
\begin{subfigure}[t]{0.32\textwidth}
\caption{$\Msq=400$~GeV: ratios}
\includegraphics[width=6.0cm]{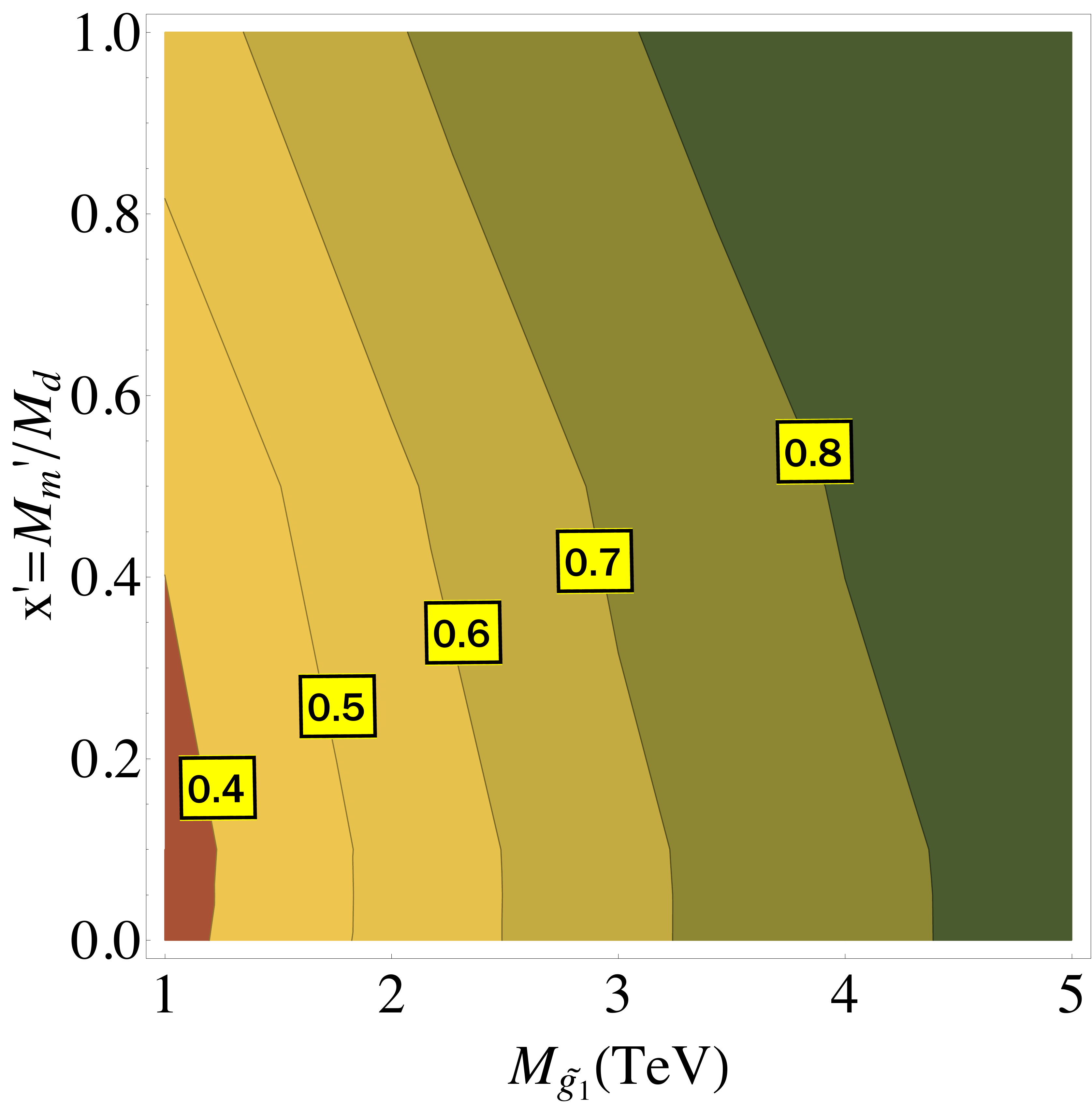}
\label{fig:MPrimebyD400RatioContour}
\end{subfigure} \quad \quad \quad
\begin{subfigure}[t]{0.32\textwidth}
\caption{$\Msq=400$~GeV: cross sections}
\includegraphics[width=6.0cm]{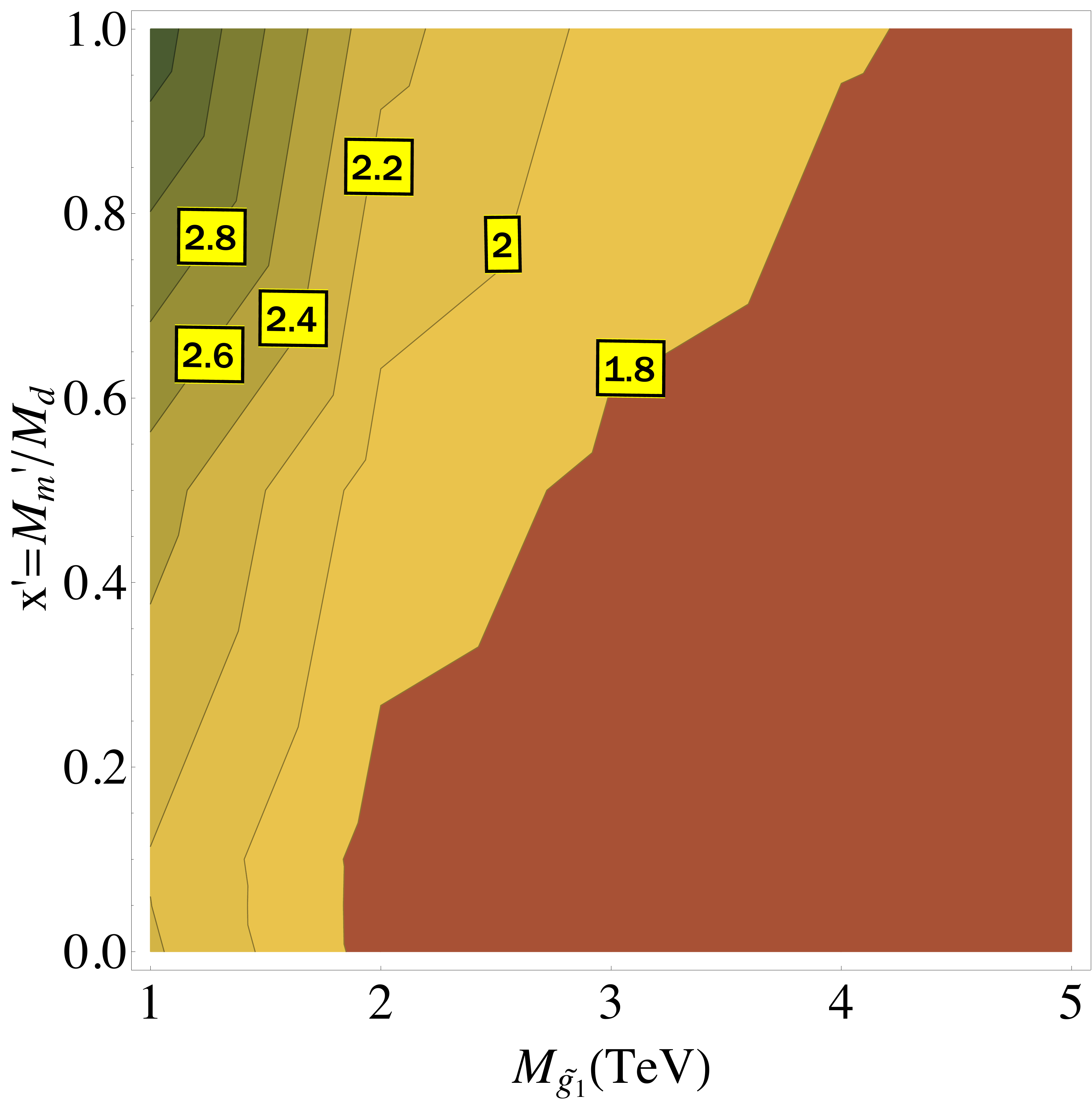}
\label{fig:MPrimebyD400}
\end{subfigure}
\\

\begin{subfigure}[t]{0.32\textwidth}
\caption{$\Msq=800$~GeV: ratios}
\includegraphics[width=6.0cm]{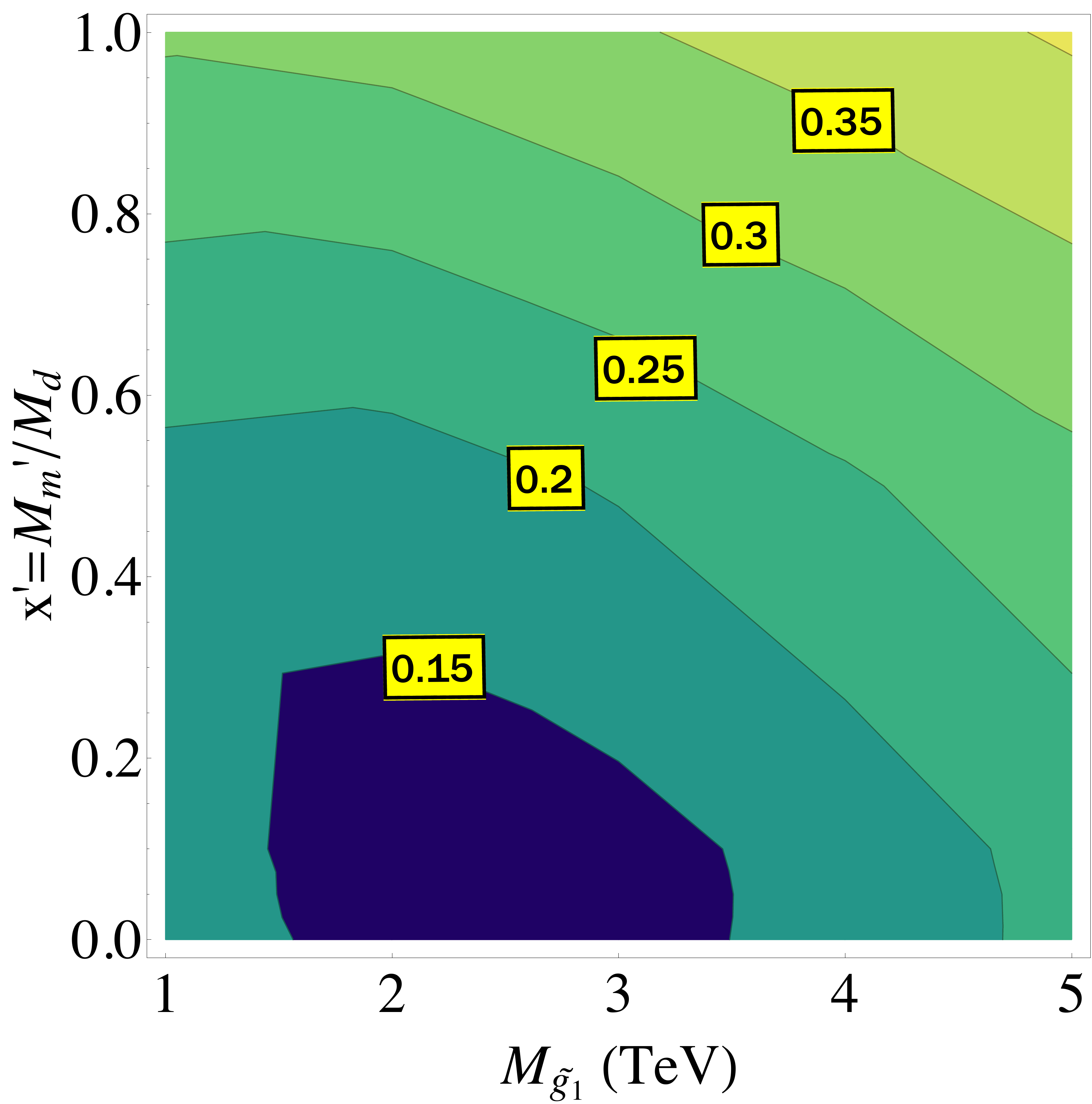}
\label{fig:MPrimebyD800RatioContour}
\end{subfigure} \quad \quad \quad
\begin{subfigure}[t]{0.32\textwidth}
\caption{$\Msq=800$~GeV: cross sections}
\includegraphics[width=6.0cm]{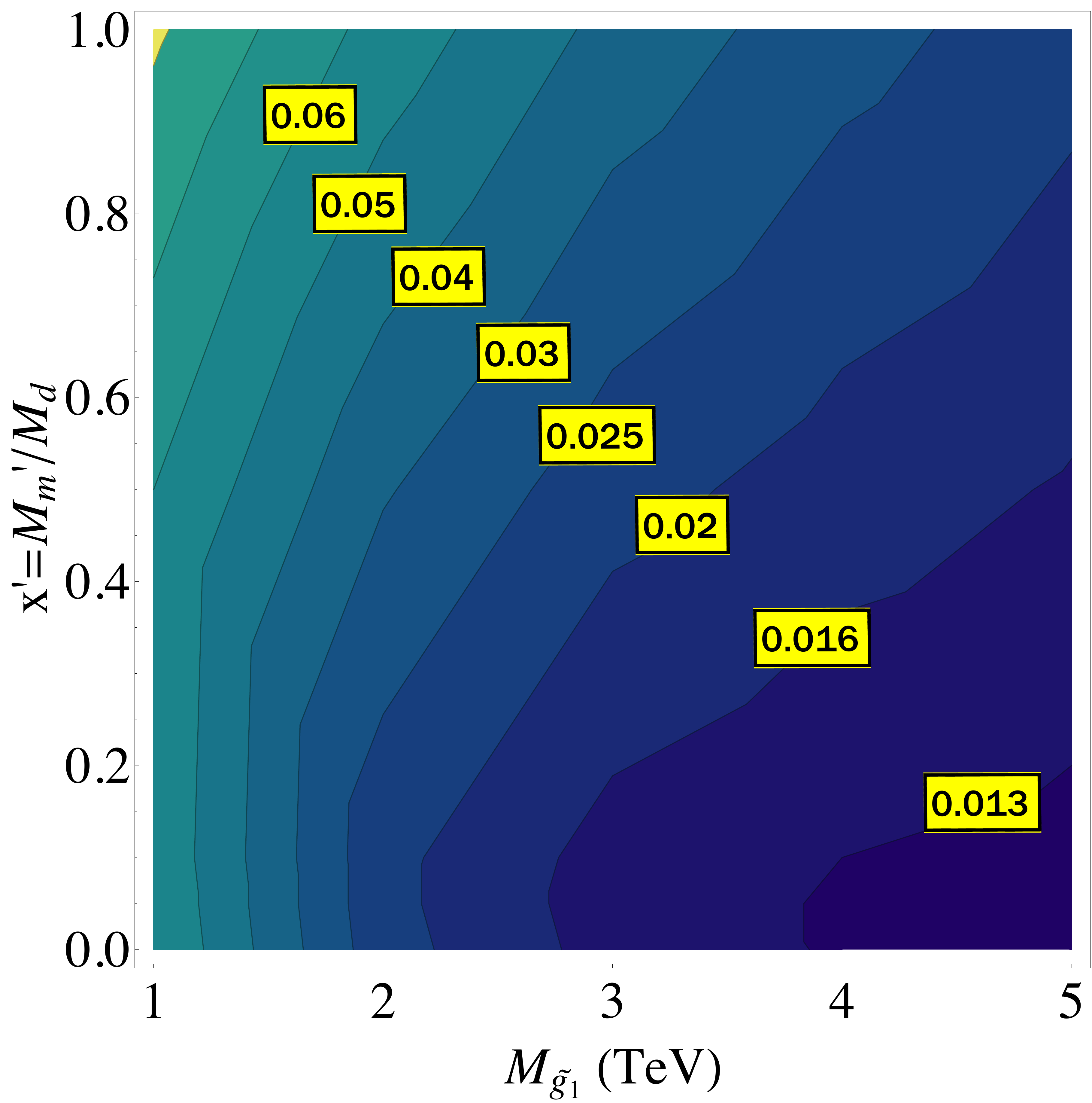}
\label{fig:MPrimebyD800}
\end{subfigure}
\\

\begin{subfigure}[t]{0.32\textwidth}
\caption{$\Msq=1200$~GeV: ratios}
\includegraphics[width=6.0cm]{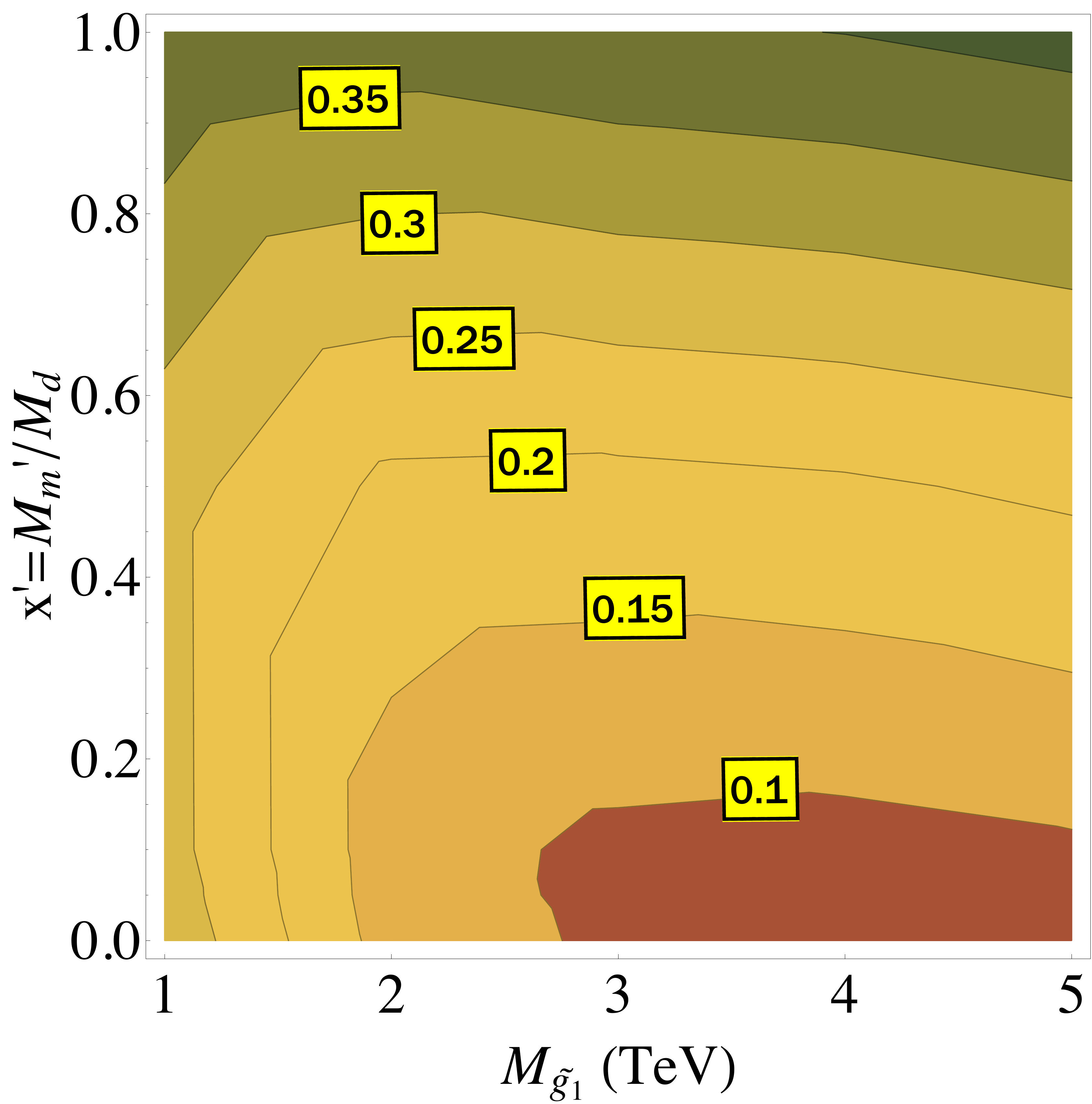}
\label{fig:MPrimebyD1200RatioContour}
\end{subfigure} \quad \quad \quad
\begin{subfigure}[t]{0.32\textwidth}
\caption{$\Msq=1200$~GeV: cross sections}
\includegraphics[width=6.0cm]{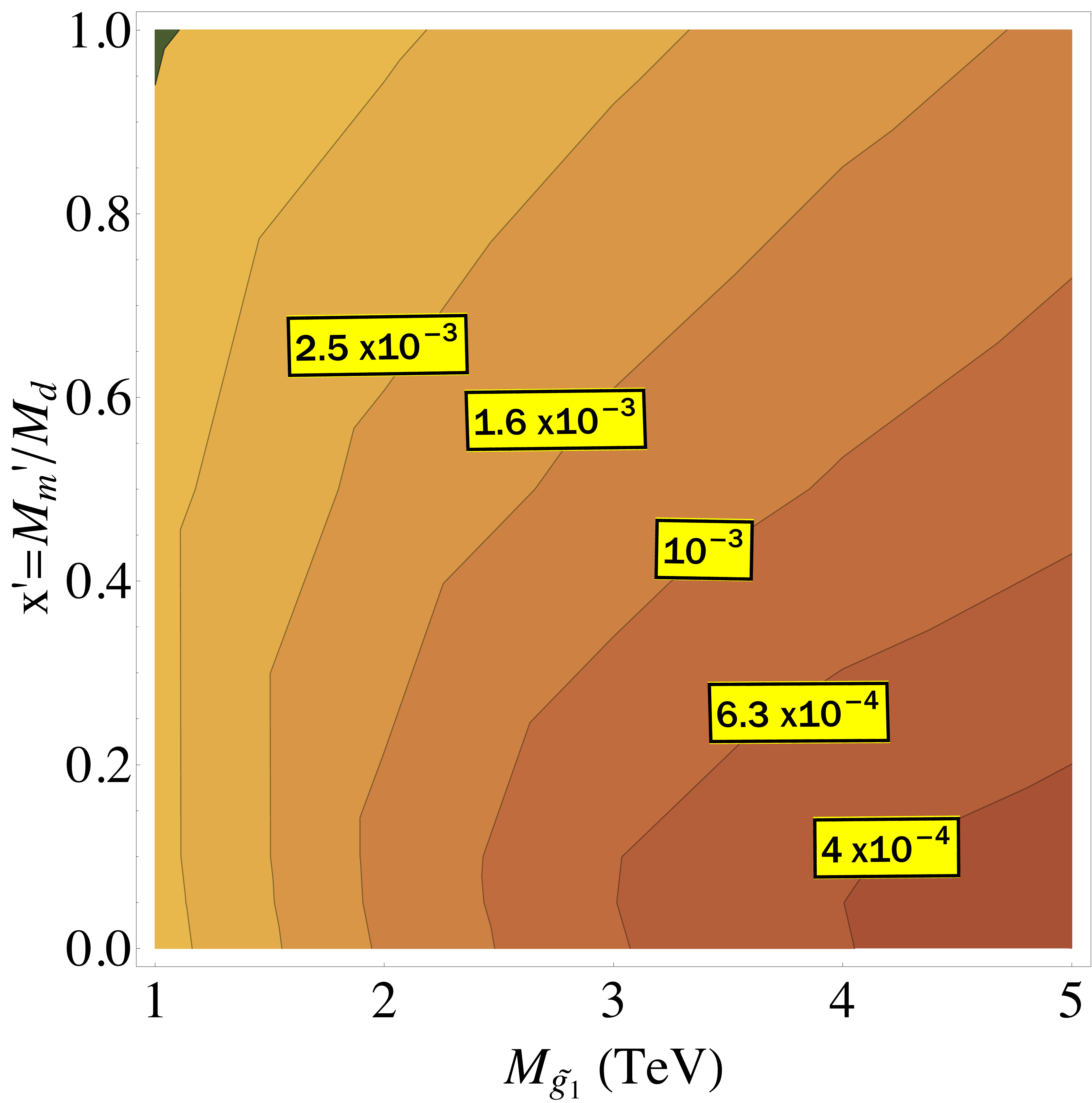}
\label{fig:MPrimebyD1200}
\end{subfigure}
\caption{Plots illustrating Case III. LEFT: Contours of the ratio of the total production cross section of the
first two generations of squarks at LHC with $\sqrt{s} = 8$~TeV\@ in our model
to the cross sections in MSSM\@.  RIGHT: Contours of the cross sections
themselves (at leading order), in pb, at LHC with $\sqrt{s} = 8$~TeV\@.
We show the variation
as the lightest gaugino mass ($\Mone$) is varied simultaneous
with varying $M_m'$ and $M_d$,
parameterized by $x' = M_m'/M_d$. The critical features are explained in the text.}
\label{MajPrimebyDirSqXS}
\end{centering}
\end{figure*}

\begin{figure*}
\begin{subfigure}[t]{0.32\textwidth}
\includegraphics[width=5.3cm]{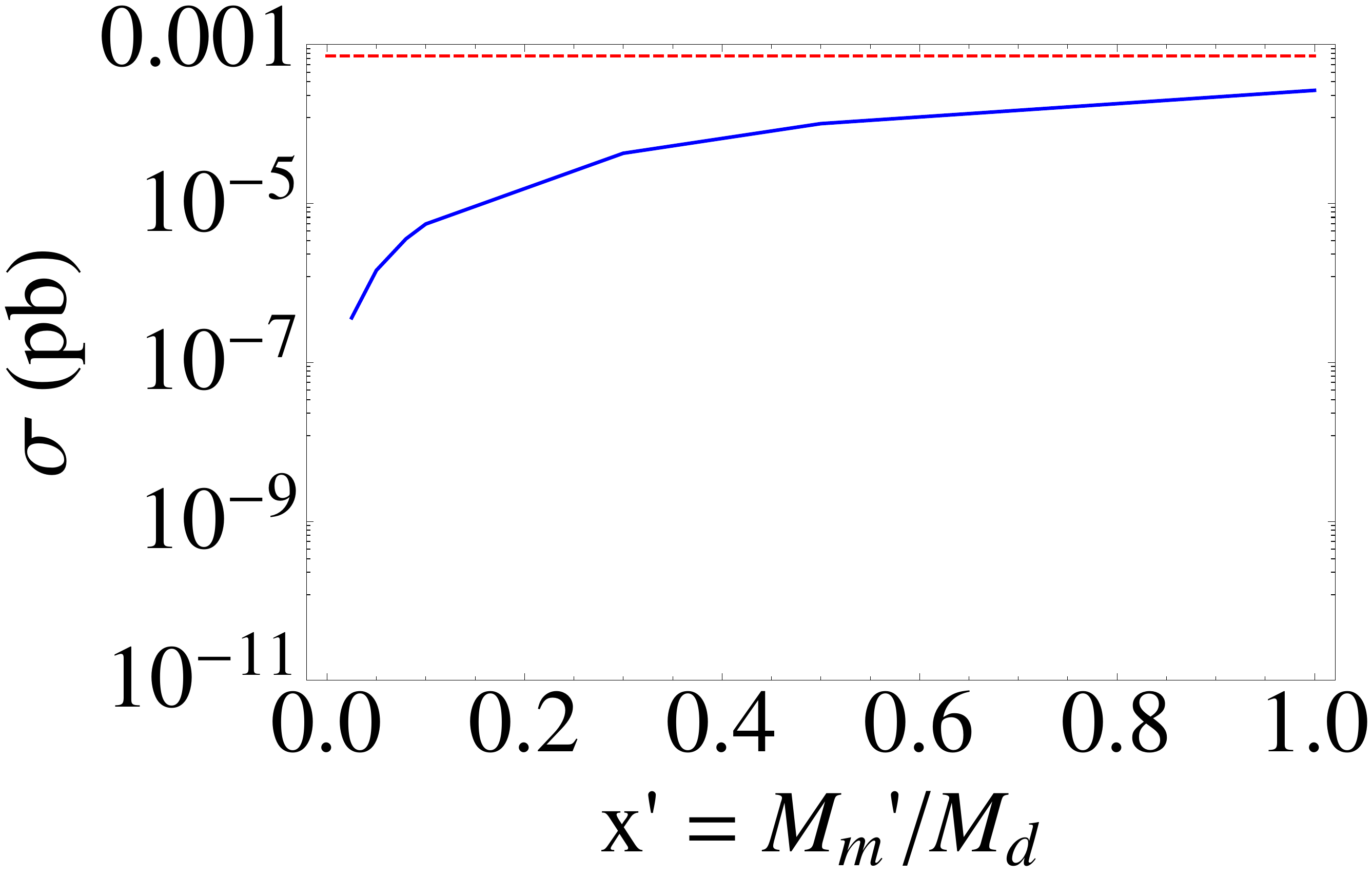}
\caption{$\tilde{u}_L \tilde{u}_L$}
\label{fig:uLuLPrime}
\end{subfigure}
\begin{subfigure}[t]{0.32\textwidth}
\includegraphics[width=5.3cm]{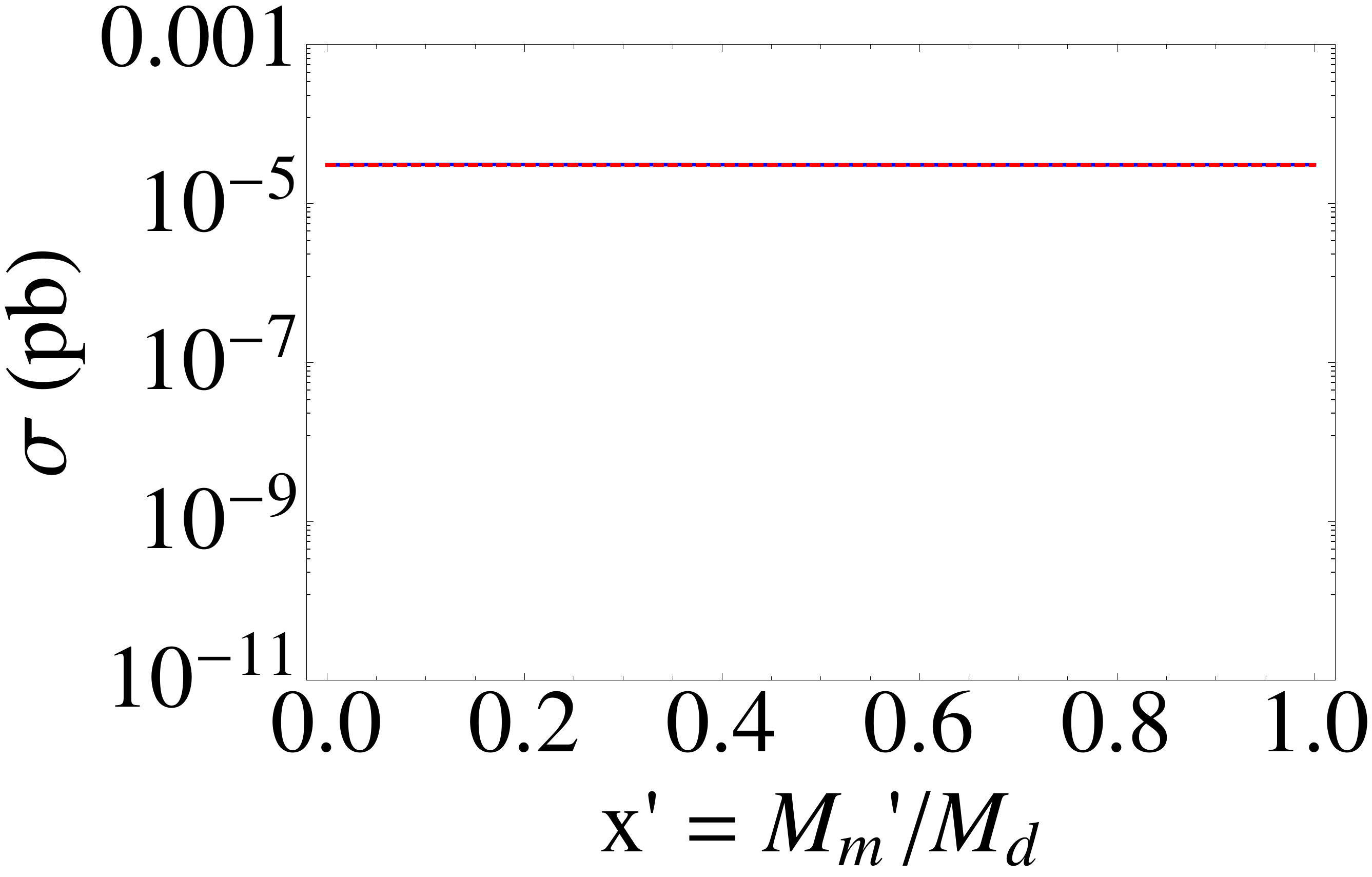}
\caption{$\tilde{u}_L \tilde{u}^*_L$}
\label{fig:uLuLStarPrime}
\end{subfigure}
\begin{subfigure}[t]{0.32\textwidth}
\includegraphics[width=5.3cm]{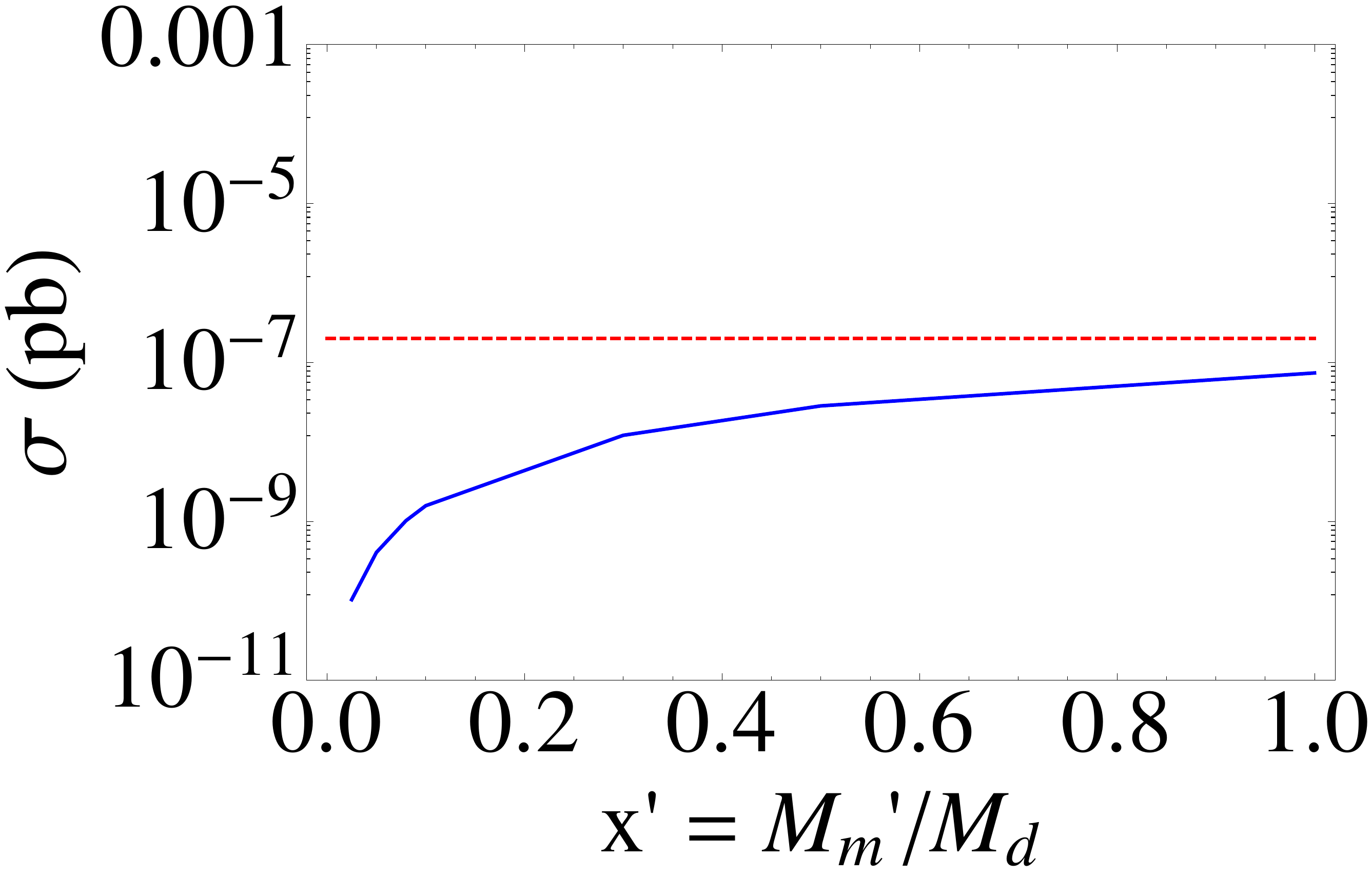}
\caption{$\tilde{u}^*_L \tilde{u}^*_L$}
\label{fig:uLStaruLstarPrime}
\end{subfigure}
\begin{subfigure}[t]{0.32\textwidth}
\includegraphics[width=5.3cm]{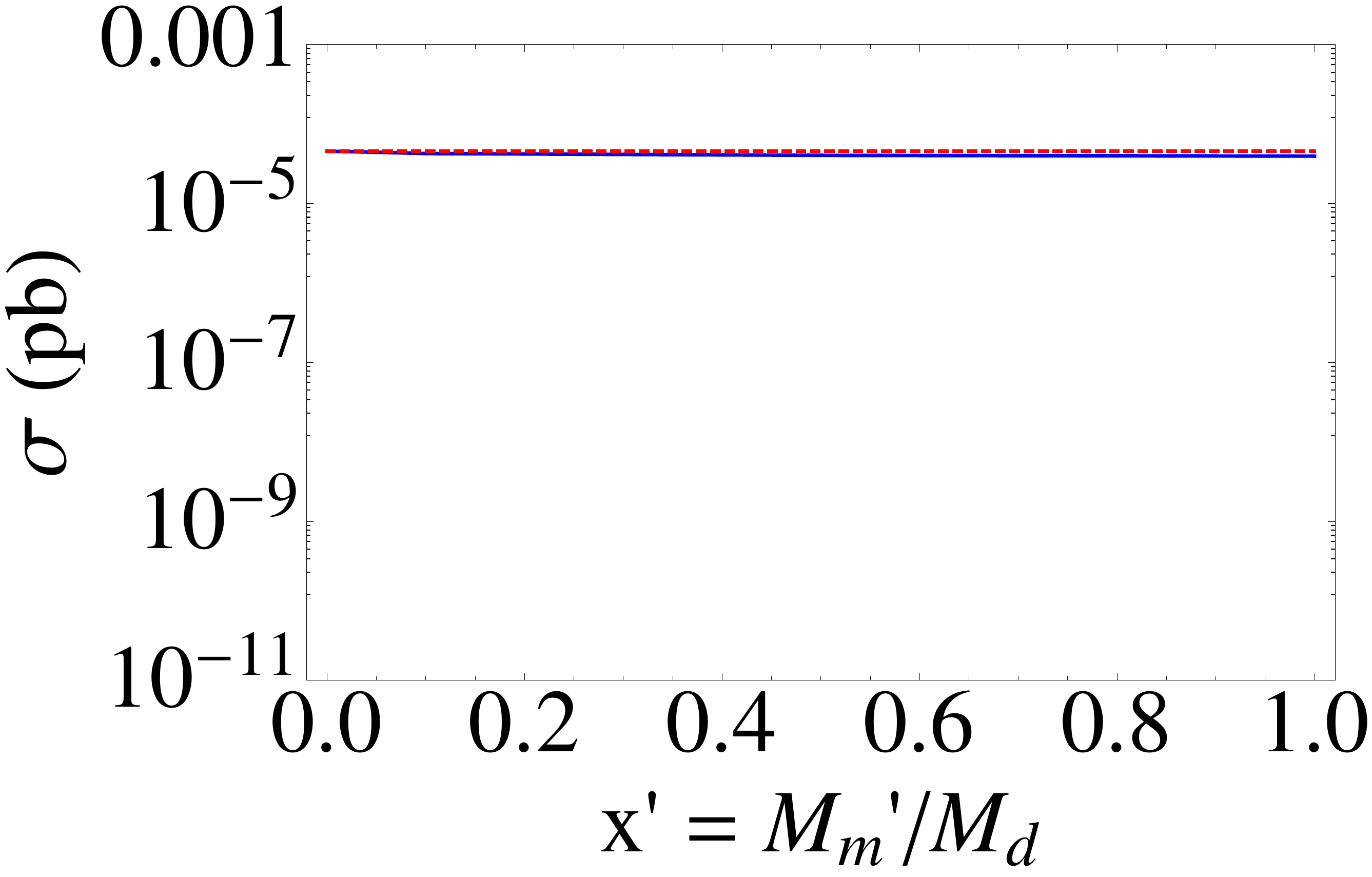}
\caption{$ \tilde{u}_L\tilde{u}_R $}
\label{fig:uLuRPrime}
\end{subfigure}
\begin{subfigure}[t]{0.32\textwidth}
\includegraphics[width=5.3cm]{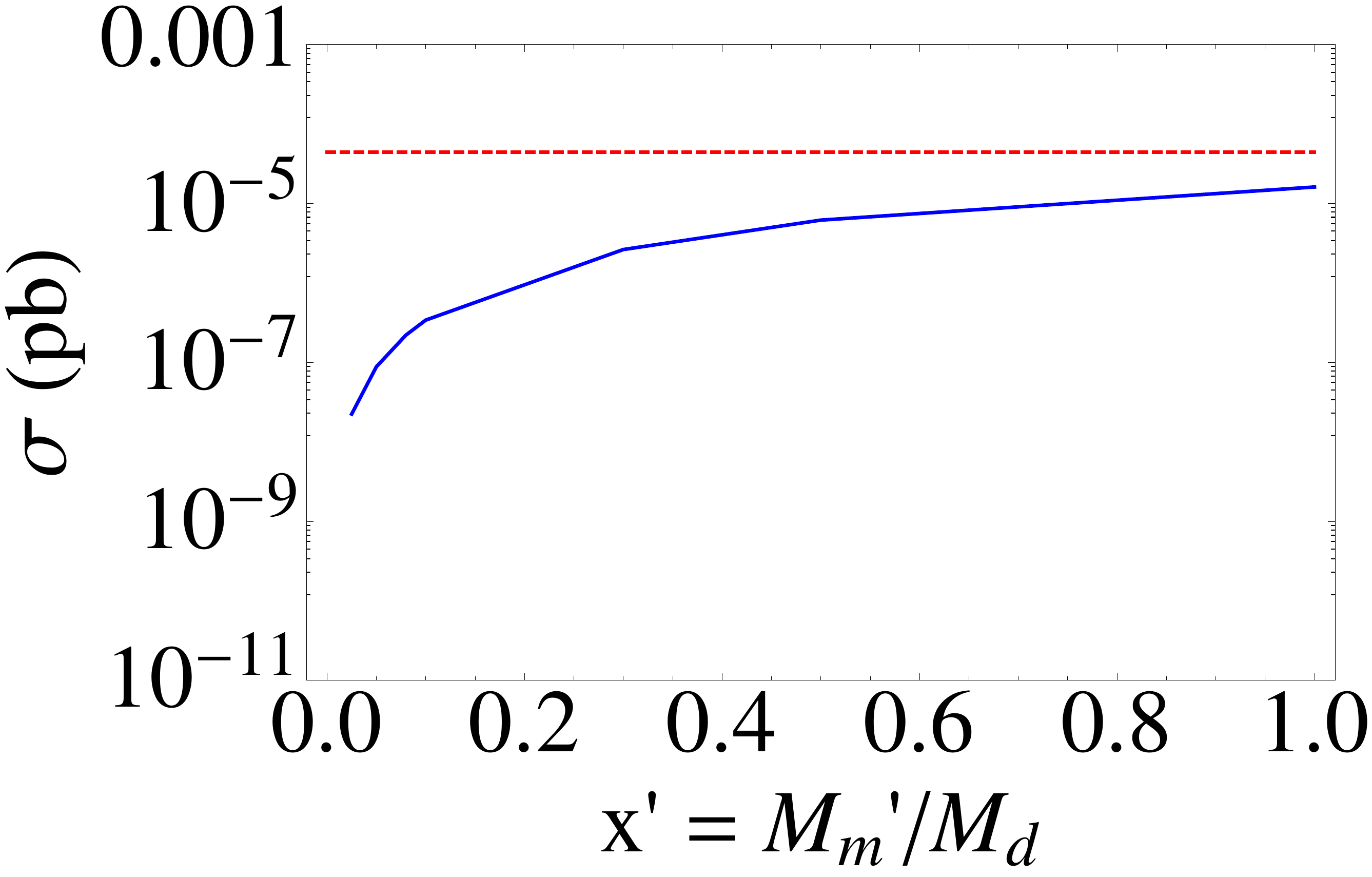}
\caption{$ \tilde{u}_L\tilde{u}^*_R $}
\label{fig:uLuRStarPrime}
\end{subfigure}
\begin{subfigure}[t]{0.32\textwidth}
\includegraphics[width=5.3cm]{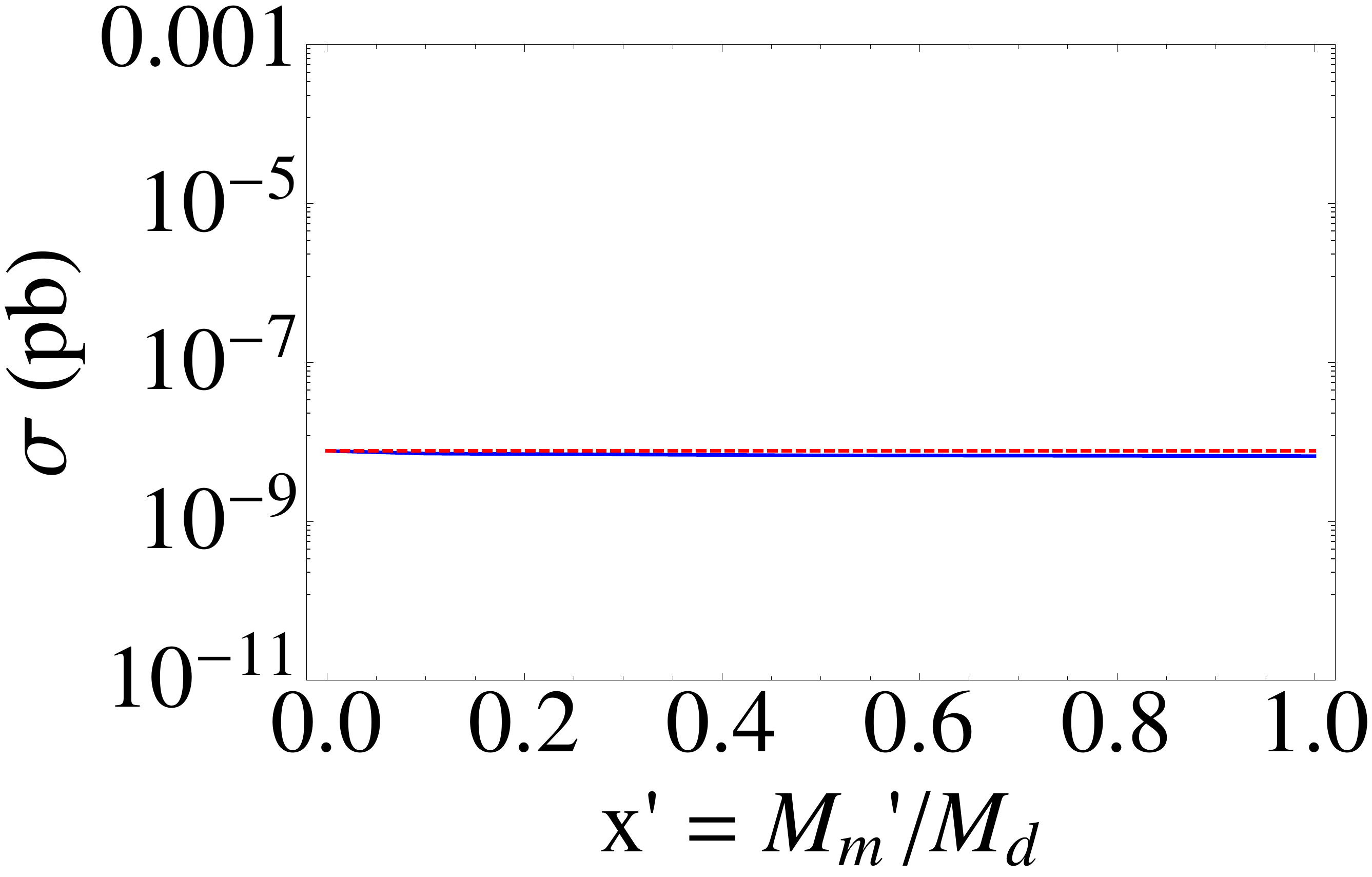}
\caption{$ \tilde{u}^*_L\tilde{u}^*_R $}
\label{fig:uLStaruRStarPrime}
\end{subfigure}
\caption{cross sections of the various unique modes that constitute
up squark production when $M_m$ is set to zero. The blue curves show
these as a function $x = M_m/M_d$,
while the dashed red horizontal lines denote the corresponding cross section
for the case of a pure Majorana gluino of the same mass as $\Mone$.
Here the squark mass $M_{\tilde{u}}$ is $1200$~GeV and the mass of the
lighter gluino eigenstate $\Mone$ is $5$~TeV\@.}
\label{ResolvedXSPrime}
\end{figure*}

\subsubsection{Cross sections across parameter space}

Our first foray into the behavior of the squark cross sections is shown
in Fig.~\ref{MajbyDirSqXS} that contains contour plots  in the
($\Mone$, $x \, (= \, M_m/M_d$)) space.  On the $x$-axis is the eigenvalue of
the lighter of the gluino eigenstates, and on the y-axis is the mixed
nature of the gauginos, parameterized by $M_m/M_d$.  The contours on the right show
the production cross sections summing over all combinations of squarks
and antisquarks of the first and second generations, and the contours
on the left show the ratios of these cross sections to their equivalents
in the scenario of a Majorana gluino with the same mass as $\Mone$.
To illustrate the differences as the squark mass is increased,
the three pairs of plots show three different squark masses:
$400$, $800$ and $1200$~GeV\@.

There are several interesting features shown in Fig.~\ref{MajbyDirSqXS}.
Holding the lightest gluino eigenmass constant, we see that the
squark production cross section \emph{decreases} as a Majorana mass $M_m$
is introduced.  This we explore in detail below.  Next, we see
distinctly different rates of variation in the cross sections across
the three plots.  At $\Msq = 1200$~GeV (Fig.~\ref{fig:MbyD1200})
the cross section falls by an order of magnitude as $\Mone$ goes from
$1$ to $4$~TeV, after which it is roughly constant, whereas for squark masses
$400$~GeV (Fig.~\ref{fig:MbyD400}) and $800$~GeV (Fig.~\ref{fig:MbyD800})
we find much less variation: the cross section drops by a factor of a few
as $\Mone$ is increased from $1$ to $2$~TeV, and then asymptotes to a fixed value.
The larger variation is present because, as we saw earlier, for larger squark masses,
the $s$-channel squark---anti-squark cross section becomes more
competitive with the $t$-channel gluino exchange induced squark-squark
production processes. It is this competition between the two
leading modes for gluino masses
below $\sim 4$~TeV that results in the larger rate of variation of the cross section
in that region in Fig.~\ref{fig:MbyD1200}. The domination of squark-antisquark
production for gluino masses above $4$~TeV results in the constancy of the
cross section observed in the right end of the plot.

We now turn our attention to the plots on the left, depicting contours of
the ratios of the corresponding cross sections on the right to those of a
pure Majorana gluino with a mass the same as $\Mone$. To understand the
features of these plots, we will have to consider the competition between
three different modes: squark--anti-squark production, \emph{same-handed}
squark pair production and \emph{opposite-handed} squark pair production.
Two distinctive features seen here are (i) at a low squark mass of $400$~GeV,
the ratio increases as we move horizontally to the right, as shown in
Fig.~\ref{fig:MbyD400RatioContour}, (ii) at higher squark masses of
$800$ and $1200$~GeV, the ratio first decreases and then increases as
we move in the horizontal direction, with the local minimum shifting to
the right as $\Msq$ is increased, as shown in Figs.~\ref{fig:MbyD800RatioContour}
and \ref{fig:MbyD1200RatioContour}.

The first feature is a result of the same mechanism that results in the
lack of variation in Fig.~\ref{fig:MbyD400}. The squark--anti-squark
production dominates over squark-squark production for a large range of
gluino masses at $\Msq = 400$~GeV, and as $\Mone$ is increased,
this domination increases for both a Majorana and a mixed gluino
(with the domination in the Majorana case weaker) as we saw earlier in
Fig.~\ref{fig:DirGlu400}. Hence we observe a uniform increase in the ratio,
seen to approach unity. The second feature can be understood in terms of
Figs.~\ref{fig:DirGlu800} and \ref{fig:DirGlu1200}. In Fig.~\ref{fig:DirGlu800},
for instance, we notice that near the left extreme ($\Mone \sim 1$~TeV),
the Majorana cross section is dominated by squark pair production and the
Dirac cross section gets nearly equal contributions from both squark--anti-squark
and squark pair production. Near the right extreme ($\Mone \sim 5$~TeV),
the dominant mode of Majorana cross section has fallen and the
total cross section has near-equal contributions from both modes,
while the Dirac cross section, dominated strongly by squark--anti-squark
production, is now comparable to either mode of the Majorana case.
At either extreme, the total Dirac cross section is able to catch up to
an extent with the total Majorana cross section, for different reasons.
In the intermediary mass range, however, the Dirac cross section,
dominated by only squark--anti-squark production, is much smaller than
the Majorana case. This argument can be extended to mixed gluinos as well,
and hence the local minimum observed in Fig.~\ref{fig:MbyD800RatioContour}.
The above discussion applies also to Fig.~\ref{fig:MbyD1200RatioContour},
except that, as seen in Fig.~\ref{fig:DirGlu1200}, the Dirac cross section
catches up with the Majorana at even higher gluino masses. This results
in the rightward shift compared to the $\Msq = 800$~GeV case in the
local minimum.

If we now move vertically anywhere in Fig.~\ref{fig:MbyD1200},
or for gluino masses below $2$~TeV in Figs.~\ref{fig:MbyD400} and \ref{fig:MbyD800},
we observe a drop in cross section. We notice the same for the contours of
the ratios of cross sections, i.e., Figs.~\ref{fig:MbyD400RatioContour},
\ref{fig:MbyD800RatioContour} and \ref{fig:MbyD1200RatioContour}.
This may seem counter to what we would
expect when increasing the Majorana content of the model. The reasons for
the reduction would become clear were we to investigate the physics of
each individual subprocess separately.

\subsubsection{Individual modes}
\label{sec:IndivModes}

\begin{figure}[htbp]
\begin{center}
\includegraphics[width=8cm]{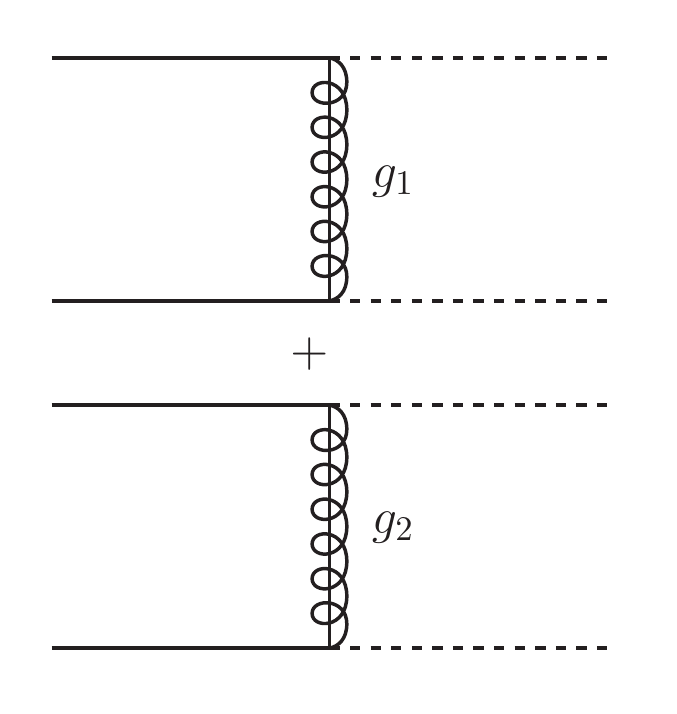}
\caption{General Feynman diagrams (without arrows) for $t$-channel gluino-mediated
squark production. The solid lines (initial state) may be labeled with
all combinations from the quark fields $q_L, q^\dagger_L, q_R, q^\dagger_R$,
and the dashed lines (final state) with the corresponding squark fields
$\tilde{q}_L, \tilde{q}^*_L, \tilde{q}_R, \tilde{q}^*_R.$}
\label{fig:Feyn}
\end{center}
\end{figure}

Let us now consider primarily the gluino $t$-channel pair-production of squarks
with quarks in the initial state. A Feynman diagram depicting this channel
is shown in Fig.~\ref{fig:Feyn}. No arrows and labels are shown, which allows us to
keep the discussion as generic as possible at this point.
Let us first divide pair production into six distinct possibilities:
\begin{eqnarray}
\nonumber
(i)   &~~~&
         \tilde{q}_L, \tilde{q}_L \ \text{or} \ \tilde{q}_R, \tilde{q}_R
         \nonumber \\
(ii)  && \tilde{q}^*_L, \tilde{q}_L \ \text{or} \ \tilde{q}^*_R, \tilde{q}_R
         \nonumber \\
(iii) && \tilde{q}^*_L, \tilde{q}^*_L \  \text{or} \ \tilde{q}^*_R \tilde{q}^*_R
         \nonumber \\
(iv)  && \tilde{q}_L, \tilde{q}_R
         \nonumber \\
(v)   && \tilde{q}_L, \tilde{q}^*_R  \ \text{or} \ \tilde{q}_R, \tilde{q}^*_L
         \nonumber \\
(vi)  && \tilde{q}^*_L, \tilde{q}^*_R
         \nonumber
\end{eqnarray}

In Fig.~\ref{ResolvedXS}, we illustrate the physics behind each of these modes
with a single flavor: up squarks.  Here the squark mass is taken as $1200$~GeV
and the absolute mass of the lighter gluino eigenstate $|\Mone| = 5$~TeV
while the heavier eigenvalue, $\Mtwo$, is varied.  These are illustrative values,
to gain intuition for the effects of varying $x = M_m/M_d$ on the
cross sections of the individual modes.
In this section, we state the results obtained, leaving the detailed behavior
of the analytic expressions of certain amplitudes to App.~\ref{app:Indivmodes}.

$(i) \ \tilde{u}_L, \tilde{u}_L $

The cross section increases from zero and saturates at a value far below
the Majorana cross section as $x = M_m/M_d$ is increased,
as shown in Fig.~\ref{fig:uLuL}.  The amplitude is written in
App.~\ref{app:Indivmodes}, where we find that it is suppressed
by $p^2/\Mone^3$ (times a function of $x$ that becomes just
one power $x$ for small values), considerably smaller than the
naive result of $1/\Mone$.  Moreover, at larger $x \simeq 1$,
the amplitude is not scaling with $x$.
This is due to the lightest gaugino eigenstate
becoming increasingly the adjoint fermion, which does not couple
to quarks and squarks.

$(ii) \  \tilde{u}^*_L, \tilde{u}_L$

The dominant contribution to this diagram is production via an $s$-channel gluon.
In Fig.~\ref{fig:uLuLStar} we see a nearly unvarying cross section as we
increase $x$ as shown by the the blue line.
Since the sub-dominant $t$-channel gluino diagram is negligible,
we find that the cross section values nearly coincides with the
pure Majorana case.

$(iii) \ \tilde{u}^*_L, \tilde{u}^*_L$

The physical principles are the same as $(i)$, hence the similar trends
observed in Fig.~\ref{fig:uLStaruLstar}.  However, the cross section values
are much smaller since the PDF effects of anti-up quarks cause to
suppress this mode.

$(iv) \ \tilde{u}_L, \tilde{u}_R$

The amplitude, and hence the cross section, turns out to be numerically
the same for the cases of pure Majorana and pure Dirac gluinos.
This is reflected in Fig.~\ref{fig:uLuR}, where the blue and red curves
intersect at $x=0$.  As $x$ is increased to $1$, however, the cross section
decreases to roughly $1/13$ of the cross section of the pure Majorana case.
This is evident in the form of the amplitude shown analytically in
App.~\ref{app:Indivmodes}.  As we will see shortly, this is important
in understanding the features of Fig.~\ref{MajbyDirSqXS}.

$(v)  \ \tilde{u}_L, \tilde{u}^*_R$

The physics here is identical to cases $(i)$ and $(iii)$, except for the
suppressing effect of excavating a sea antiquark from one of the protons.
The effect is a decreased cross section as reflected in Fig.~\ref{fig:uLuRStar}.

$(vi) \ \tilde{u}^*_L, \tilde{u}^*_R$

Conceptually similar to case $(iv)$, this production mode suffers from
PDF suppression, resulting in the lowered cross sections seen
in Fig.~\ref{fig:uLStaruRStar}.
\\

We can now answer the question posed at the end of Sec.~\ref{sec:TotalSigmaQCD},
on why the total cross section of squark production declines despite an
addition of Majorana content. We find that an increase in cross section
of the pairs
$(\tilde{q}_L, \tilde{q}_L),
 (\tilde{q}_R, \tilde{q}_R),
 (\tilde{q}_L, \tilde{q}^*_R),
 (\tilde{q}_R, \tilde{q}^*_L),
 (\tilde{q}^*_L, \tilde{q}^*_L),
 (\tilde{q}^*_R, \tilde{q}^*_R)$ -- as expected when departing from a
pure Dirac scenario -- is less relevant in comparison to the decrease
in cross section of the pairs
$(\tilde{q}_L, \tilde{q}_R),
 (\tilde{q}_L, \tilde{q}^*_L),
 (\tilde{q}_R, \tilde{q}^*_R) $ -- due to various kinds of kinematic suppression
as discussed in this section.

The analysis above shows that in addition to the suppression from the
operator dimension (relative dominance of dim-5 or dim-6)
and the kinematics, the third factor that is essential to determine
the cross section trends is the PDFs.
Thus the trends for individual modes would be identical for down squarks
except for the effects of PDF suppression.  As for the second generation
of squarks, the far smaller PDFs of the corresponding second generation quarks
in the proton render most modes negligible, with the only sizeable contribution
coming from $(\tilde{q}_L, \tilde{q}^*_L)$ and $(\tilde{q}_R, \tilde{q}^*_R)$,
which proceed through an $s$-channel gluon.  Therefore we see that the principal
difference between the first and second generations is that $t$-channel
gluino mediation exhibits non-trivial behavior in the former,
while it is practically absent in the latter.

\subsection{Case II: $M_m = M_m', \ x = 2M_m/M_d = 2M_m'/M_d$}

In this scenario, the two gluino mass eigenstates have masses
$-\Mone = M_m - M_d$, $\Mtwo = M_m + M_d$, and the mixing between
the states is maximal ($\cos \theta_{\tilde{g}} = 1/\sqrt{2}$)
independent of $M_m$, $M_m'$ and $M_d$.
We consider the modification
resulting from the Majorana content of gluino in the same way as the
previous section, with the corresponding results shown in
Fig.~\ref{MajEqualbyDirSqXS}.
However, since both Majorana masses are nonzero, the difference
between the eigenvalues $\Mtwo - \Mone = 2 M_m$ (as opposed to just
$\Mtwo - \Mone = M_m$ in Case~I and $\Mtwo - \Mone = M_m'$ in Case~III). 
In order to make an direct comparison of the mixing effects to 
Cases~I and III, while holding the kinematics approximately
equivalent, we define $x$ as $x = 2 M_m/M_d = 2 M_m'/M_d$.

The features shown in Fig.~\ref{MajEqualbyDirSqXS} are in many ways 
similar to those of Case~I\@.
For instance, in the cross section contours
on the right, we see little variation moving horizontally
direction at high $\Mone$ for squark masses $400$ and $800$~GeV
(Figs.~\ref{fig:MEqualbyD400} and \ref{fig:MEqualbyD800}),
for the same reasons as before.
We also notice the local minimum in Figs.~\ref{fig:MEqualbyD800RatioContour}
and \ref{fig:MEqualbyD1200RatioContour} shifts to the right as we
go from $\Msq = 800$~GeV to $\Msq = 1200$~GeV\@.  Notice that,
in all the plots, the values of the cross sections and ratios are
identical to Case~I along $x$=0, since they correspond to a
pure Dirac gluino in either case.

The main differences between Cases~I and II are seen when we move vertically
in the contour plots. Whereas previously the cross section was seen to
uniformly decrease as $x = M_m/M_d$ was increased, we now notice that it
first decreases and then increases, a trend particularly pronounced for
$\Msq = 800$~GeV and $1200$~GeV, as seen in
Figs.~\ref{fig:MEqualbyD800RatioContour}-\ref{fig:MEqualbyD1200}.
This feature can again by understood in terms of the individual subprocesses,
which are given in the plots of Figs.~\ref{ResolvedXSEqual}.

In Case~I, we saw that the subprocess setting the total cross section
was the production mode $\tilde{q}_L\tilde{q}_R$, which decreased by
roughly an order of magnitude as $x$ was increased from $0$ to $1$.
Even though the modes
($\tilde{q}_L \tilde{q}_L$, $\tilde{q}_R \tilde{q}_R$, $\tilde{q}_L \tilde{q}_R^*$)
increased in the same range, their values never caught up with the
opposite-handed squark pair production. This is not the situation here.
Figs.~\ref{fig:uLuLEqual} and \ref{fig:uLuRStarEqual}
show that although the same-handed modes begin at zero cross section,
they overtake opposite-handed modes at around $x=0.2$,
bolstering the total production.

\subsection{Case III: $M_m = 0, \ x' = M_m'/M_d$}

Lastly, we consider the scenario $M_m = 0$, $M_m' \lsim M_d$.
In this Case, the simplified expressions for the masses in Eqs.~(\ref{eq:Monesimp})-(\ref{eq:Mtwosimp}) carry over here with
the replacement $M_m \leftrightarrow M_m'$, while the mixing
angle is $\cos\theta_{\g} = \sqrt{\Mone/(\Mone+\Mtwo)}$.
This means that the relevant mixing angle ranges are switched,
with $\cos \theta_{\g}$ varying from $1/\sqrt{2}$ to $0.53$ and
$\sin \theta_{\g}$ from $1/\sqrt{2}$ to $0.85$ as $x' = M_m'/M_d$
is varied from $0$ to $1$.  Hence the lighter eigenstate is more
of the gluino, and the heavier eigenstate more of the adjoint fermion.
If $x'$ were to be taken to infinity, $\cos \theta_{\g} \ra 0$
and the heavier eigenstate decouples, recovering the MSSM
pure Majorana gluino limit.

Therefore, we expect the cross section to increase as $x'$ is increased
from $0$ to $1$. This is exactly the trend we notice in the plots of
Fig.~\ref{MajPrimebyDirSqXS}, corresponding to this case.
The features of the contours here are very similar to those of Case II
when we move horizontally across the plots, and the physical reasons
are the same. The difference is in the variation in the vertical direction;
the cross sections uniformly increase whereas previously there was
a decrease followed by an increase.

Once again we may understand such a trend by inspecting the
individual production modes, shown in Fig.~\ref{ResolvedXSPrime}.
The same-handed squark pair production modes catch up with
and overtake their opposite-handed equivalents at small $x'$,
while the $\tilde{q}_L\tilde{q}_R$ cross section remains nearly constant.

\section{Mixed Electroweak Gauginos}
\label{sec:SqProdQCED}

We now turn to the effects of electroweak gauginos on squark production.
We assume Higgsino-quark-squark couplings are negligible and thus can ignore
$t$-channel Higgsino mediation of squark production. This leaves us with only
winos and binos, specifically two neutralinos and one chargino.
The particle content and the effects on the
squark cross sections  depend on whether the electroweak gauginos
acquire Dirac, Majorana, or mixed gaugino masses.

There is one new squark production subprocess, $p p \ra \tilde{u}_L \tilde{d}_L$,
that proceeds through
$t$-channel exchange of a chargino.  Regardless
of the wino's Majorana and Dirac content, the chargino
is obviously a Dirac fermion.  However, this subprocess is absent in
pure supersoft models. We have provided a discussion on this in
App.~\ref{sec:DbutnotD}.

For general mixed (Dirac and Majorana) neutralinos and a mixed chargino,
there is a large parameter space that could be considered.
Here we focus on a particular case  --
pure Majorana electroweak gauginos in interference with a purely Dirac gluino.
As pointed out in \cite{Fox:2002bu}, an explicit Majorana mass for
the $U(1)$ and $SU(2)$ gauginos implies there is no suppression of the
quartic coupling of the Higgs potential.  Moreover, we expect that
purely Majorana electroweak masses will yield the largest effects
on first generation squark production cross sections,
and thus bound what can happen in a general model.

We are primarily interested in bino and/or wino masses at which there is a
noticeable departure from the ``QCD-only'' (i.e., mediated by gluons or gluinos)
cross section, $\sigma_{QCD}$.  We characterize this by finding the total cross section for a
given squark mass within a range of bino and wino masses.
In this section we have taken the gluino to have a sufficiently large
purely Dirac mass so that $\sigma_{QCD}$ is comprised dominantly of
just $s$-channel gluon-mediated squark--anti-squark production.
For concreteness, we took $M_{\tilde{g}} = 5$~TeV, though the precise
value is irrelevant here.  Of course at much lower Dirac gluino masses,
$t$-channel squark-squark production eventually dominates, but this merely
weakens the impact of $t$-channel electroweakino interference.

We find that the largest effect of electroweakinos on the total squark
production cross section occurs when the squark mass is near the
Majorana electroweakino mass.  The explanation becomes apparent when
we consider these two observations:
\begin{itemize}
\item[1.]  Previously, when the neutralinos and charginos were turned off,
the cross section at $5$~TeV gluino mass was dominated by
(a) gluon fusion diagrams producing squark-antisquark at low squark masses
($300$ to $\sim 700$~GeV),
(b) both $\tilde{q}^* \tilde{q}$ production and $t$-channel gluino diagrams
producing $\tilde{q}_L \tilde{q}_R$ at high squark masses
($\sim 800$ to $1200$~GeV).
\item[2.]  As we know from the discussion under
(i) in Sec.~\ref{sec:IndivModes}, the coefficient of the Weyl spinors
in the amplitude for a $t$-channel exchange diagram for same-handed
squark production is
\begin{equation*}
g_f^2 \frac{M_f}{t - M_f^2}
\end{equation*}
where $M_f$ and $g_f$ are the mass and chargino-squark-quark coupling of
the fermion (gaugino) respectively. One can see that, as a function of $M_f$,
this expression is at its maximum when $M_f^2 = -t$, where it becomes
\begin{equation*}
g_f^2 \frac{1}{2\sqrt{t}}
\end{equation*}
Moreover, if $\beta$ is an arbitrary real number, both $M_f = \beta \sqrt{t}$
and $M_f = \beta^{-1} \sqrt{t}$ are fermion masses that confer the same value
to the amplitude,
\begin{equation*}
g_f^2 \frac{\beta}{1+\beta^2}\frac{1}{\sqrt{t}}
\end{equation*}
This leads to an effect on the cross section that is symmetric
with respect to $M_f \ra 1/M_f$, as we will see.
\end{itemize}

Opposite-handed squark production, however, has a different expression
for the co-efficient of the spinors in the amplitude:
\begin{equation*}
g_f^2 \frac{p\cdot\sigma}{t - M_f^2}
\end{equation*}
where the spinor indices are suppressed. The maximum of \emph{this}
expression is achieved when $M_f \rightarrow 0$, at which point it
tends to $ g_f^2 \ (p\cdot\sigma)/t$.

One might be concerned about the possible existence of a $t$-channel
pole if $t$ were to approach $M_f^2$.  However, upon integrating
the total cross section between $-1 < \cos\theta < 1$, corresponding to
$t$ over the range
\begin{eqnarray}
t_{-} &<& t \;<\; t_{+} \\
t_{\pm} &=& \frac{1}{2} \left( - s \pm \sqrt{s^2 - 4 s M_{\tilde{q}}^2} \right)
             + M_{\tilde{q}}^2 \, . \nonumber
\end{eqnarray}
It is clear that $t$ is negative definite, and moreover, approaches a
small (negative) value only when $s$ is large.  The required large $s$ means
there is substantial suppression of the integrated cross section in the integration
region where $t \simeq t_{+}$ is small.  Hence, the case of $M_f \ra 0$
does lead to a divergent contribution to the squark production rate.

Now these observations can be put together when reflecting on what happens
when Majorana winos and/or binos are turned on. We define $\sigma_{QCD}$
as the total cross section when squark production is QCD-only and
$\sigma_{QECD}$ as the cross section when it is mediated by winos, binos, and
gluinos.  Table~\ref{winobinomediation} provides information on the
electroweakino mediation of the individual modes.

\begin{figure}[htbt]
\begin{center}
\includegraphics[width=8cm]{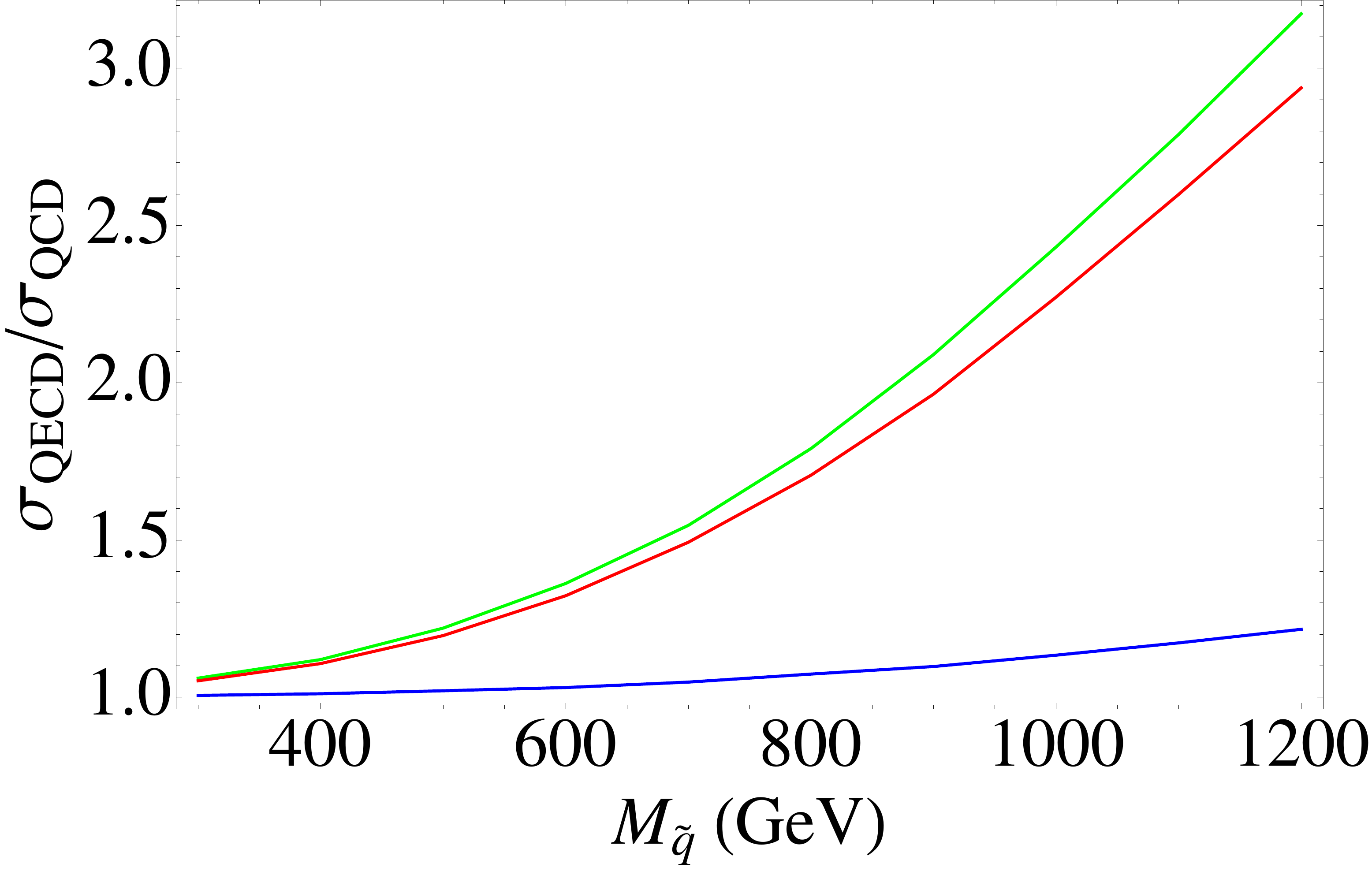}
\caption{Impact of electroweakinos at their maximal electroweakino
interference (MEI) values, where the Majorana wino mass is
equal to the squark mass. The MEI value establishes
the upper bound in cross section on the impact of electroweakino on the
QCD-only pure Dirac gluino scenario. In this plot,
green: both electroweakinos are at their MEI values,
red: the wino is pure Majorana with $\Mwino = \Msq$,
blue: the bino is pure Majorana with $\Mbino = \Msq$,
and the gluino mass is $5$~TeV\@.
These curves show that at the MEI values, the wino is more responsible than
the bino for maximizing the cross section by virtue of its stronger couplings.}
\label{QCEDMaxImpact}
\end{center}
\end{figure}

\subsection{Maximal electroweakino interference (MEI)}

The cross section for $\tilde{q}_L \tilde{q}_L$ production reaches
its maximal value when the wino, which couples only to left-handed squarks,
has a mass $\Mwino = \sqrt{t}$.  Similarly, the cross section of
$\tilde{q}_R \tilde{q}_R$ production reaches its maximal value
when the mass of the bino is also at $\Mbino = \Msq$.
If these sub-processes dominate over the QCD-only squark production,
we may have a significant increase in $\sigma_{QECD}$.
We call this Maximal Electroweak Interference (MEI).
Indeed, these two individually overtake $\tilde{q}_L \tilde{q}_R$ production,
which is the leading sub-process at high squark masses in a QCD-only picture.
The enhancement to $\tilde{q}_L \tilde{q}_L$ is larger than
$\tilde{q}_R \tilde{q}_R$ since the wino couples more strongly
to quarks and squarks than the bino.

In Fig.~\ref{QCEDMaxImpact}, we show the maximum deviation from
$\sigma_{QCD}$, represented by the ratio $\sigma_{QECD}$/$\sigma_{QCD}$,
when both the bino and wino have masses at their $\Msq$-dependent
MEI values.
As expected, the greatest departures are observed
at high squark masses. Two other scenarios are also shown:
$(i)$ a Majorana bino at the MEI value with a Dirac wino (blue),
$(ii)$ a Majorana wino at the MEI value with a Dirac bino (red).
From these we see that the wino, despite coupling only to left-handed squarks,
dominates the increase in the total cross section.

\begin{table}[htdp]
\begin{subtable}{0.5\linewidth}
\centering
\begin{tabular}{c|c|c|}
\multicolumn{1}{c}{} & \multicolumn{1}{c}{wino}  & \multicolumn{1}{c}{bino }  \\
 \cline{2-3}
  $u_{i,L}; \tilde{u}_{i,L}$ & $g/\sqrt{2}$ & $g'/3\sqrt{2}$ \\
  \cline{2-3}
  $d_{i,L}; \tilde{d}_{i,L}$ & $-g/\sqrt{2}$ & $g'/3\sqrt{2}$ \\
  \cline{2-3}
   $u_{i,R}; \tilde{u}_{i,R}$ & 0 & $-4g'/3\sqrt{2}$ \\
  \cline{2-3}
  $d_{i,R}; \tilde{d}_{i,R}$ & 0 & $2g'/3\sqrt{2}$ \\
  \cline{2-3}
\end{tabular}
\end{subtable}
\begin{subtable}{0.5\linewidth}
\centering
\begin{tabular}{|c|c|c|}
\multicolumn{1}{c}{Mode} & \multicolumn{1}{c}{Wino}  & \multicolumn{1}{c}{Bino }  \\
 \cline{1-3}
  $\tilde{q}_L \tilde{q}_L$ & $\checkmark$ & $\checkmark$ \\
  \cline{1-3}
    $\tilde{q}_R \tilde{q}_R$ & \text{\sffamily X}  & $\checkmark$ \\
  \cline{1-3}
 $\tilde{q}_L \tilde{q}^*_L$& $\checkmark$ & $\checkmark$ \\
  \cline{1-3}
   $\tilde{q}_R \tilde{q}^*_R$&\text{\sffamily X}  & $\checkmark$ \\
  \cline{1-3}
  $\tilde{q}^*_L \tilde{q}^*_L$& $\checkmark$ & $\checkmark$ \\
  \cline{1-3}
    $\tilde{q}^*_R \tilde{q}^*_R$& \text{\sffamily X}  & $\checkmark$ \\
  \cline{1-3}
  $\tilde{q}_L \tilde{q}_R$ & \text{\sffamily X}  & $\checkmark$ \\
  \cline{1-3}
   $\tilde{q}_L \tilde{q}^*_R$ &  \text{\sffamily X}  & $\checkmark$ \\
  \cline{1-3}
    $\tilde{q}^*_L \tilde{q}_R$ &  \text{\sffamily X}  & $\checkmark$ \\
  \cline{1-3}
   $\tilde{q}^*_L \tilde{q}^*_R$&  \text{\sffamily X}  & $\checkmark$ \\
  \cline{1-3}
\end{tabular}
\end{subtable}
\caption{(a) Quark-squark-electroweakino couplings of the wino and the bino
for different chiralities.  The index $i$ runs over quark generation;
(b) Categorizing the distinct individual subprocesses of squark production
mediated by the wino and bino. The wino participates in only the
left-handed (anti-)squark production, yet dominates the increase in
the total cross section.}
\label{winobinomediation}
\end{table}

\begin{figure}[htbp]
\begin{center}
\includegraphics[width=0.5\textwidth]{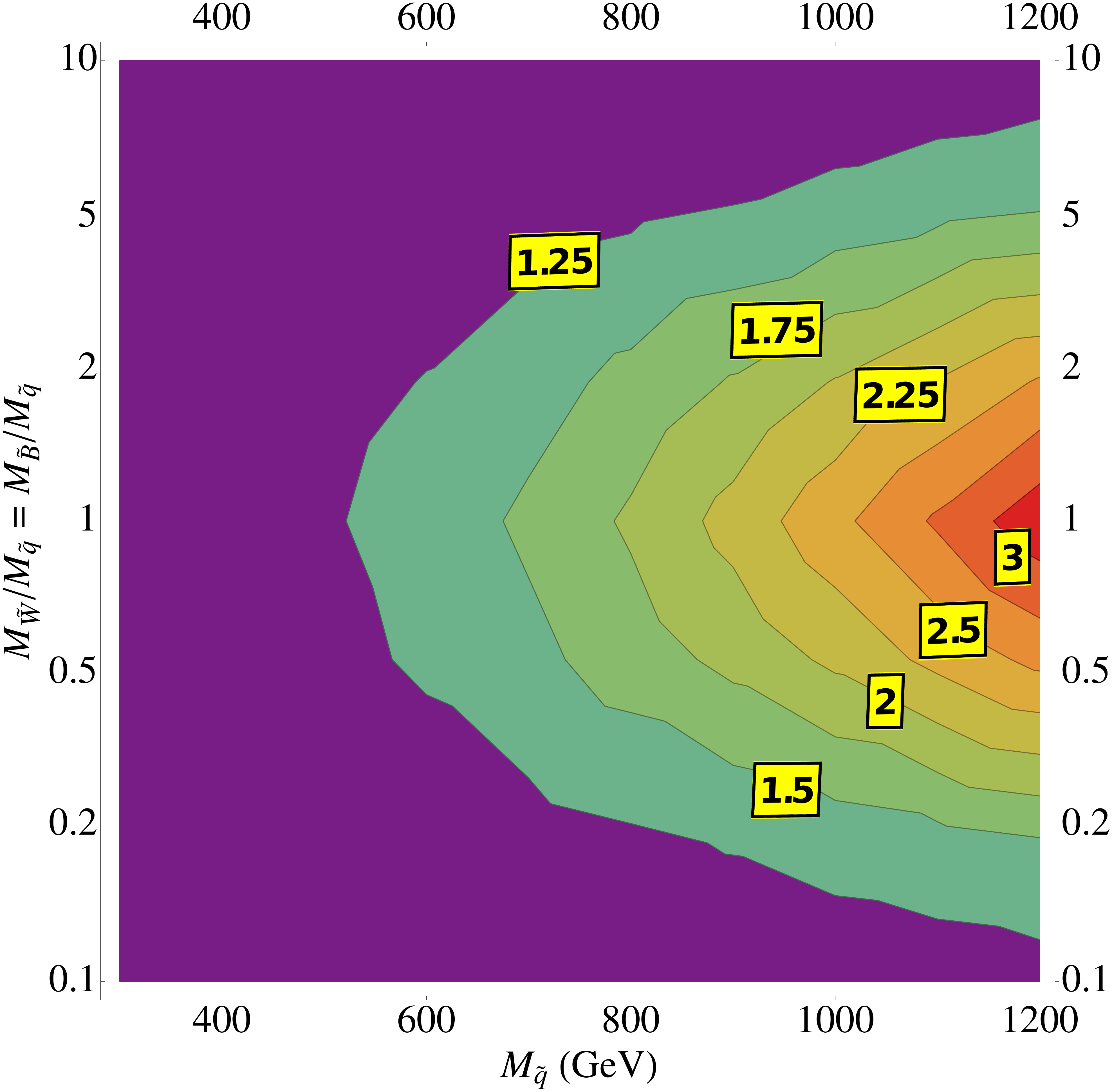}
\end{center}
\caption{Contours showing the impact of electroweak gauginos when both the Majorana wino
and Majorana bino masses are within an order of magnitude of their MEI values.
The peaks are values of $\sigma_{QECD}/\sigma_{QCD}$.
The gluino mass here is $5$~TeV\@.}
\label{contourQCED}
\end{figure}

The contour plot in Fig.~\ref{contourQCED} shows ratios of the cross sections
with and without electroweakino impact, $\sigma_{QECD}/\sigma_{QCD}$,
and spans the parameter space in its most interesting district, that is,
where the masses of the bino and wino are in the neighborhood of
the squark mass. Specifically, we vary the neutralino or chargino mass
in the range $\{0.1 \Msq,10 M_{ \tilde{q}}\}$. The symmetry spoken
of in our second observation, namely, the amplitudes for same-handed
squark production are identical when $M_f/\Msq$ is the same as
$\Msq/M_f$, is reflected in the near-mirror symmetry of the contours
in Fig.~\ref{contourQCED}.

\begin{figure}[htbp]
\includegraphics[width=0.5\textwidth]{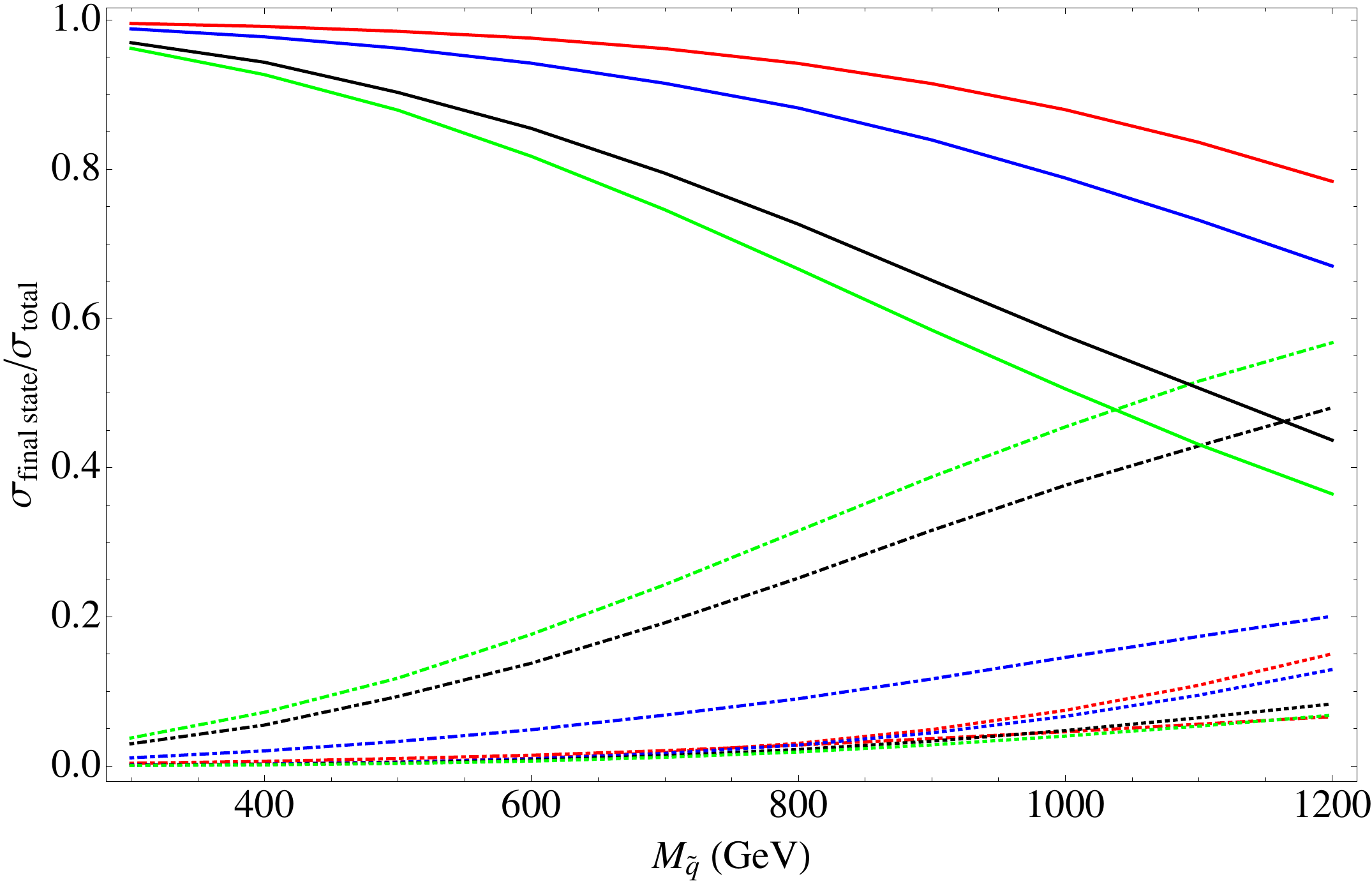}
\caption{Regions of domination: a different look at the plot in
Fig.~\ref{contourQCED}.  The ratio
$\Mwino/\Msq = \Mbino/\Msq$ is represented
by the colors and the code is
(green: $1$; black: $0.5$; blue: $0.2$; red: $0.1$).
The final state of production is given by the constitution of the
line, the code being
(solid: $\tilde{q}_{i,L}, \tilde{q}_{i,L}^*$ and
        $\tilde{q}_{i,R}, \tilde{q}_{i,R}^*$;
dot-dashed: $\tilde{q}_{i,L}, \tilde{q}_{j,L}$ and
            $\tilde{q}_{i,R}, \tilde{q}_{j,R}$;
dashed: $\tilde{q}_{i,L}, \tilde{q}_{j,R}$).
The gluino mass here is $5$~TeV\@.}
\label{DominanceQCED}
\end{figure}

Once again we perceive that the region where the squark mass is high and the
electroweak gaugino masses are close to the squark mass (by a factor of $2$)
is where the colored superpartner production cross section is most enhanced
compared to a pure Dirac gluino.   Different regions of the contour plot
of Fig.~\ref{contourQCED} are dominated in cross section by the production
of different final states. Fig.~\ref{DominanceQCED} is a representation
of these relative effects.
The ratio $\sigma(\text{mode})/\sigma(\text{total})$
is plotted against squark mass for three different kinds of final state modes:
(i) same-handed squark-antisquark (solid lines),
(ii) same-handed squark-squark (dash-dotted), and
(iii) opposite-handed squark-squark (dashed).
The color code is (green, black, blue, red) = ($1$, $0.5$, $0.2$, $0.1$) where
the numbers on the RHS are the ratios of the weak gaugino mass to the
squark mass. The green curves show that as the squark mass exceeds a TeV,
the contribution of the same-handed squark production surpasses the
same-handed squark-antisquark production. As $\Mwino=\Mbino$
is lowered (that is, as red is approached), the final states
$\tilde{q}_L\tilde{q}_L^*$ and $\tilde{q}_R\tilde{q}_R^*$ dominate the
cross section irrespective of the squark mass. These subprocesses,
as seen before, occur chiefly through an $s$-channel gluon with the
initial state as two gluons or a quark and an antiquark.

\section{Recasting LHC Limits}
\label{sec:recast}

\begin{figure}[htpb]
\includegraphics[width=0.5\textwidth]{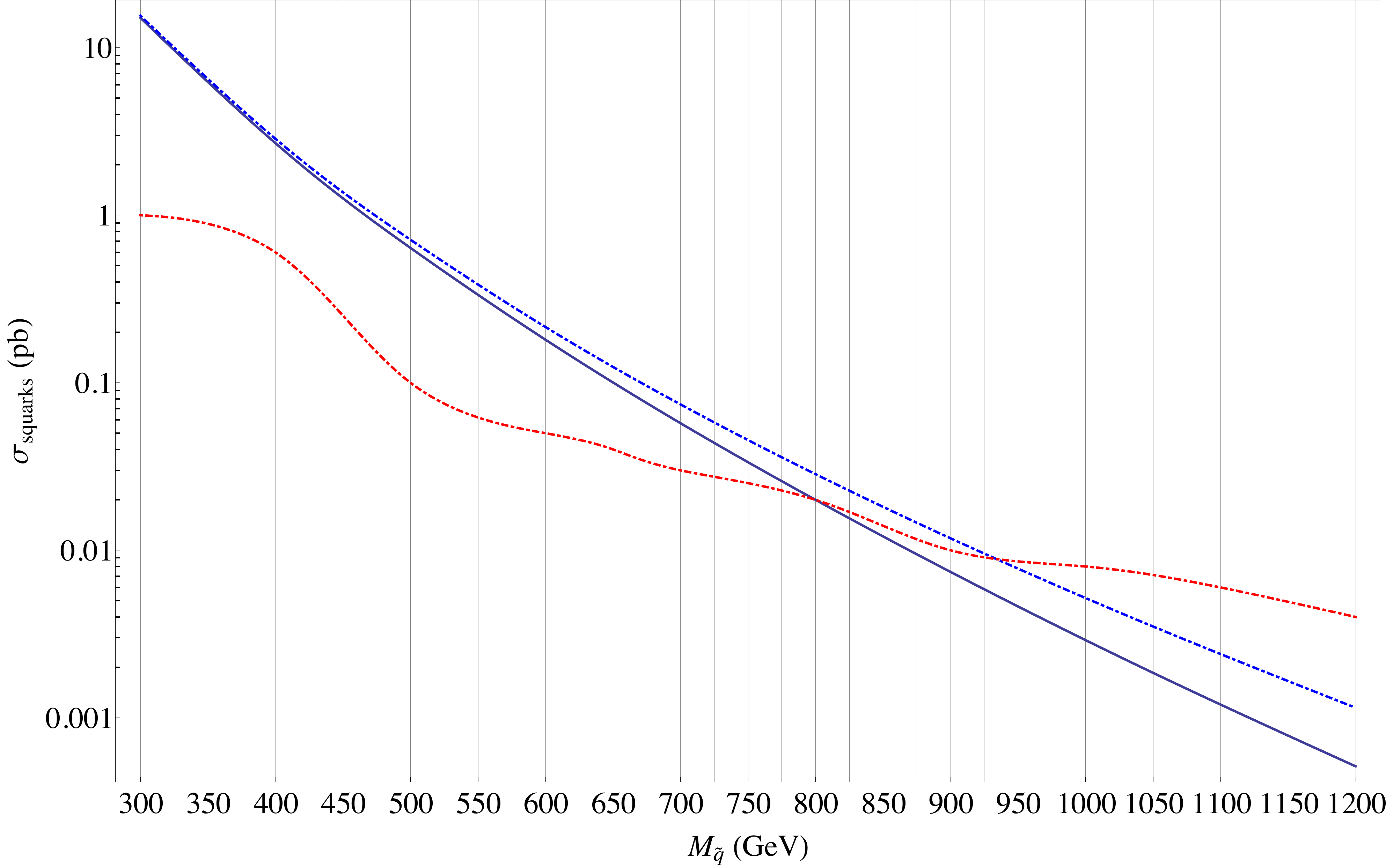}
\caption{The $8$~TeV cross sections at leading order of scenarios with a
pure Dirac gluino (black) and electroweakinos at their MEI values
(blue) intersect with the exclusion cross section set by the
multijet plus missing energy search (red) \cite{CMS2013multijet},
which gives us bounds on the squark mass.}
\label{fig:Bounds}
\end{figure}

\begin{figure}[htbp]
\includegraphics[width=0.5\textwidth]{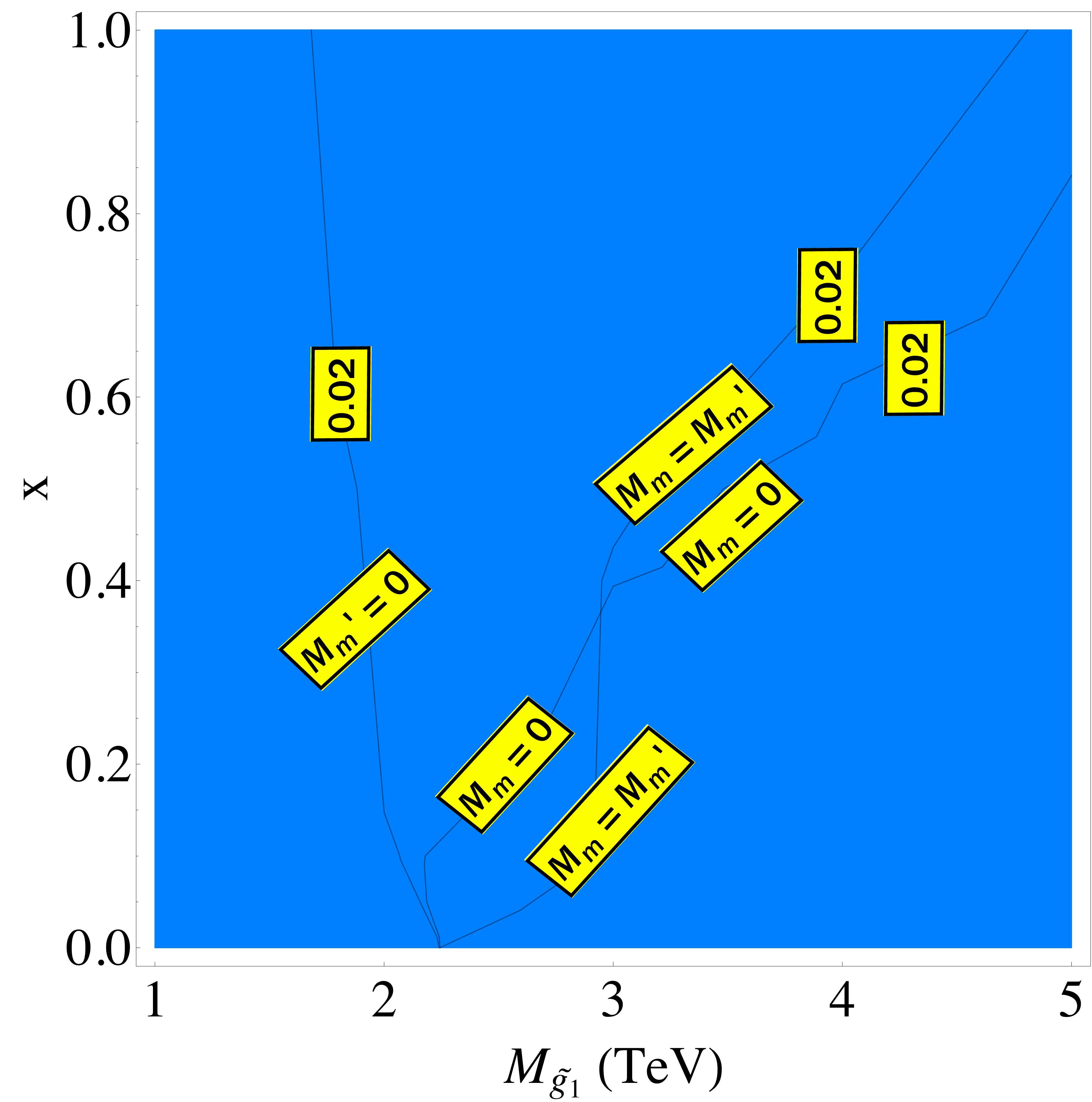}
\caption{Constraints set by the multi-jet plus missing energy search on the parameter space of our model.
Since we find in Fig.~\ref{fig:Bounds} that that the bound is set at
$\Msq \approx 800$~GeV at an exclusion cross section $\approx 0.02$~pb (at leading order),
we have included the contour of that value for that squark mass. 
All three scenarios we have considered are shown, using the appropriate contours from Figs. \ref{fig:MbyD800}, \ref{fig:MEqualbyD800} and \ref{fig:MPrimebyD800}, and the space to the
left of each contour is excluded for the corresponding scenario. Depending on the contour, the y-axis is interpreted as $x = M_m/M_d$ or $x = 2M_m/M_d = 2M'_m/M_d$ or $x' = M'_m/M_d$.}
\label{fig:ContourSq800RECAST}
\end{figure}

We now consider what our results imply for the supersymmetry search strategies
at LHC\@.  The CMS collaboration has provided exclusion cross section limits
on pair-produced first and second generation squarks at 
$\sqrt{s} = 8$~TeV with $19.5$~fb$^{-1}$ of data in their ``T2'' 
simplified model~\cite{CMS2013multijet}.  A similar simplified model,
with the gluino decoupled, has been subjected to a multijet plus missing
energy search analysis by the ATLAS collaboration \cite{ATLASHeavyMSSM:2013}
obtaining similar bounds.  We omit this from our discussion since the CMS
results provided rate bounds throughout the $\Msq$-$M_{\rm LSP}$ plane.

The various cross sections obtained in our model are compared against
the exclusion cross sections of CMS searches that were based on
the search for new physics in multijets and missing momentum final state
at $\sqrt{s} = 8$~TeV and $\mathcal{L} = 19.5$~fb$^{-1}$ \cite{CMS2013multijet}.
In all these analyses the LSP is taken to be massless.

The limits obtained are $\Msq \geq 800$~GeV for a Dirac-gluino-only scenario
and $\Msq \geq 925$~GeV when both the electroweakinos are at their MEI values.
We get these limits by checking where the CMS exclusion cross sections
intersect the cross sections predicted by our models, as plotted in
Fig.~\ref{fig:Bounds}. It deserves to be mentioned that the bound for a
pure Dirac gluino case differs from that found
by the CMS collaboration ($\Msq \geq 840$~GeV) by a small margin.
As a general comment we would like to mention
that such numerical differences in the bounds of simplified models,
particularly when a comparison is made in a plot spanning four orders of magnitude
(like Fig.~\ref{fig:Bounds}), are an inevitable consequence of the nature of
the CMS exclusion plots. The method of reading off cross sections from
color gradients makes it necessarily difficult to pinpoint the values with
great accuracy.

The pure Dirac gluino bound also enables us to set constraints on the
parameter space of mixed gluinos. Since the exclusion cross section at
$\Msq = 800$~GeV is $\sim 0.02$~pb (at leading order), we can overlay the contours of different
mixed gluino scenarios corresponding to that cross section.
Fig.~\ref{fig:ContourSq800RECAST} shows this superimposition, and for each
scenario the parameter space to the left of the corresponding contour
is excluded.

\section{Discussion}

We found that a mixed gluino that acquires both a Dirac mass and a
Majorana mass solely for its gaugino component
($M_m \not= 0$, $M_m' = 0$), is \emph{less} constrained from LHC searches
than a pure Dirac gluino.  This is because the lightest gluino eigenstate
contains more of the adjoint fermion partner that does not couple to
quarks and squarks, and thus further suppresses squark production
through $t$-channel exchange.  This was shown in detail by examining
the individual squark production sub-processes as a function of
the Majorana mass.

A mixed gluino that acquires both a Dirac mass and a
Majorana mass for its adjoint fermion component
($M_m = 0$, $M_m' \not= 0$), or for both of its components
($M_m \not= 0$, $M_m' \not= 0$), is \emph{slightly} more constrained
from LHC searches than a pure Dirac gluino.  This is because the lightest
gluino eigenstate contains more of the gaugino that does couple to
quarks and squarks.  However, the effect is not significant
when the Majorana masses are small compared with the Dirac mass,
roughly $M_m,M_m' \lsim \mathcal{O}(0.1) M_d$.
Again, this was shown in detail by examining
the individual squark production sub-processes as a function of
the Majorana mass(es).

A model with a Dirac gluino and Majorana electroweak gauginos
that both contribute to squark production can have modifications from
the gluino-only cross section by a factor of a few.
The largest effect occurs at the ``maximal electroweakino interference''
mass values of $M_1,M_2 \simeq \Msq$.  As the electroweak gauginos
become larger or smaller than this value, their effect on
squark production becomes suppressed.

New candidates for the LSP are one
of the consequences of finding that light Majorana electroweak
gauginos not significantly affecting cross sections.
In addition to a gravitino LSP, we showed that a Majorana bino
is also perfectly viable since it does not significantly increase
squark production cross sections.  One could also contemplate a
light Majorana wino, however this would introduce new branching
fractions of left-handed squarks to winos.

We conclude by considering several new simplified models
could be studied and constrained
(by the experimental collaborations) that would capture
the essentials of these scenarios with mixed gauginos and
electroweakinos.  It would be particularly insightful to
study the simplified models when not only the squarks but also
the gluino is relatively light, while the LSP mass is allowed
to vary.  Here are several proposals:
\begin{itemize}
\item[1)] \textbf{Dirac gluino, several choices of LSP mass}:
          Cross section bounds in $\Msq - M_D$ plane;
          $M_{\rm LSP} = 0, 200, 400$~GeV\@.
\item[2)] \textbf{Dirac gluino, several choices of gluino mass}:
          Cross section bounds in $\Msq - M_{\rm LSP}$ plane
          $M_D = 1$-$3$~TeV in steps of $0.5$~TeV\@.
\item[3)] \textbf{Mixed gluino, several choices of squark mass}:
          Cross section bounds in $\Mone - x$ plane for Cases I,II,III;
          $\Msq = 500$-$1000$~GeV in steps of $100$-$250$~GeV,
          with a massless LSP.
\item[4)] \textbf{Mixed gluino, several choices of LSP mass}:
          Cross section bounds in $\Mone - x$ plane for Cases I,II,III;
          $\Msq = 500$, $M_{\rm LSP} = 0, 200, 400$~GeV\@.
\end{itemize}

\section*{Acknowledgments}

We thank J.~Alwall, A.~Martin, S.~Martin, and A.~Menon
for several useful discussions.
GDK thanks the Ambrose Monell Foundation for support while at the
Institute for Advanced Study.
GDK and NR are supported in part by the US Department of Energy under
contract number DE-FG02-96ER40969.

\appendix

\section{Individual modes}
\label{app:Indivmodes}

Here we describe the analytic behavior of the individual subprocesses
$\tilde{u}_L \tilde{u}_L$ and $\tilde{u}_L \tilde{u}_R$ that are critical
in understanding the results of Sec.~\ref{sec:MixedGluino}.
\\

(a) $\tilde{u}_L \tilde{u}_L$ \\

This amplitude takes the form
\begin{equation*}
\frac{-i \mathcal{T}}{g^2 C_F} =
  \left( c^2_{\theta_{\tilde{g}}}
  \frac{\Mtwo}{p^2 + \Mtwo^2} + s^2_{\theta_{\tilde{g}}}
         \frac{-\Mone}{p^2 + \Mone^2 } \right)
  u_L u_L
\end{equation*}
where $C_F (=4/3)$ is the appropriate Casimir invariant,
$u_L$ is a 2-component spinor denoting an incoming left-handed up quark with
spinor indices suppressed, and the second term on the RHS has a
minus sign since the mass of $\tilde{g}_1$ is the negative of $\Mone$.

In Case~I ($M_m' = 0$), using the expressions for the mixing angle
in Eq.~(\ref{eq:MixAng}),
expanding the amplitude to leading order in $p^2/M_{\tilde{g}}^2$,
and then writing it in terms of $\Mone$ and $x = M_m/M_d$,
we obtain
\begin{eqnarray}
\frac{c^2_{\theta_{\tilde{g}}} \Mtwo}{p^2 + \Mtwo^2}
- \frac{s^2_{\theta_{\tilde{g}}} \Mone}{p^2 + \Mone^2}
   &=&  \frac{p^2}{\Mone^3} x \left( \sqrt{x^2 + 4} - x \right)^3
        \nonumber \\
   & &{} \qquad + \mathcal{O}(p^4/\Mone^4)
\label{eq:uLuLAmp}
\end{eqnarray}
In Case~II ($M_m = M_m'$), the mixing angle is fixed
$c^2_{\theta_{\tilde{g}}} = 1/2$.  Expanding the amplitude to
leading order in $p^2/M_{\tilde{g}}^2$, and then writing it in terms of
$\Mone$ and $x = 2M_m/M_d = 2M_m'/M_d$, we obtain
\begin{eqnarray}
   &=& - \frac{x}{\Mone (x + 2)}
     + \frac{p^2 x^3 + 12 x}{\Mone^3 (x + 2)^3} \nonumber \\
   & &{} \qquad + \mathcal{O}(p^4/\Mone^4)
\label{eq:uLuLAmpEqual}
\end{eqnarray}
In Case~III ($M_m = 0$), again using Eq.~(\ref{eq:MixAng}),
expanding the amplitude to leading order in $p^2/M_{\tilde{g}}^2$,
and then writing it in terms of $\Mone$ and $x' = M_m'/M_d$,
we obtain
\begin{eqnarray}
  &=& - \frac{x' (x' + \sqrt{x'^2+4})}{2 \Mone}
      + \frac{p^2 x (x'^2+2) (\sqrt{x'^2 + 4} - x')^3}{8 \Mone^3}
      \nonumber \\
  & &{} \qquad + \mathcal{O}(p^4/\Mone^4)
\label{eq:uLuLAmpPrime}
\end{eqnarray}
Clearly, all of these expressions vanish in the Dirac limit, $x \ra 0$.
The key difference is how quickly each expression turns on, and its
asymptotic form as $x$ gets large (by which we mean near $1$).
For example, at small $x$, Case I scales as $p^2 x/\Mone^3$
whereas Case II and III scale as $x/\Mone$.  This illustrates
that Case I is further suppressed as the Majorana mass $M_m$
is turned on.  As a second example, when $x=1$, Case I becomes
$- p^2/\Mone^3$, Case II becomes $-1/(2 \Mone)$, and Case III
becomes $(1 - \sqrt{5})/(2 \Mone)$.  We have checked the the
functional form of the squared amplitudes agrees well with
our results shown in Figs.~\ref{fig:uLuL}, \ref{fig:uLuLEqual}
and \ref{fig:uLuLPrime}.  Finally, we can recover the heavy
pure Majorana case (the MSSM) where $c^2_{\theta} = 1$
and $\Mone = 0, \Mtwo = 5000$~GeV\@.  In this case, the amplitude
becomes $\Mtwo/(p^2 + \Mtwo^2)$ where $\tilde{g}_2$ is interpreted
as the Majorana gluino.  This is obviously suppressed by just one power
of the gluino mass, giving a large cross section as indicated by the
dashed red line in Figs.~\ref{fig:uLuL}, \ref{fig:uLuLEqual}
and \ref{fig:uLuLPrime}.
\\

(b) $\tilde{u}_L \tilde{u}_R$\\

The amplitude for this subprocess is
\begin{equation}
\frac{-i \mathcal{T}}{g^2 C_F} = u_L^\alpha
  \left( c^2_{\theta_{\tilde{g}}}
         \frac{p\cdot\sigma_{\alpha\dot{\beta}}}{p^2 + \Mtwo^2}
         + s^2_{\theta_{\tilde{g}}}
         \frac{p\cdot\sigma_{\alpha\dot{\beta}}}{p^2 + \Mone^2} \right)
  (u^\dagger_R)^{\dot{\beta}}
\end{equation}
where $u^\dagger_R$ denotes an incoming right-handed up quark.
For $|p| \ll M_{\tilde{g}}$, this amplitude is suppressed by $1/M^2$.
In Case~I ($M_m' = 0$), using Eq.~(\ref{eq:MixAng}),
expanding the amplitude to leading order in $p^2/M_{\tilde{g}}^2$,
and then writing in terms of $\Mone$ and $x = M_m/M_d$,
we obtain
\begin{eqnarray}
  \frac{c^2_{\theta_{\tilde{g}}}}{p^2 + \Mtwo^2}
  &+& \frac{s^2_{\theta_{\tilde{g}}}}{p^2 + \Mone^2} \nonumber \\
  &=& \frac{(x - \sqrt{x^2 + 4})^2}{4 \Mone^2}
      + \mathcal{O}(p^2/\Mone^2)
\end{eqnarray}
and in Case~II ($M_m = M_m'$), writing in terms of
$x = 2M_m/M_d = 2M_m'/M_d$ we obtain
\begin{eqnarray}
  \frac{c^2_{\theta_{\tilde{g}}}}{p^2 + \Mtwo^2}
  &+& \frac{s^2_{\theta_{\tilde{g}}}}{p^2 + \Mone^2} \nonumber \\
  &=& \frac{x^2 + 4}{\Mone^2 (x + 2)^2}
      + \mathcal{O}(p^2/\Mone^2)
\end{eqnarray}
and in Case~III ($M_m = 0$), writing in terms of
$x' = M_m'/M_d$ we obtain
\begin{eqnarray}
  & & \frac{c^2_{\theta_{\tilde{g}}}}{p^2 + \Mtwo^2}
  + \frac{s^2_{\theta_{\tilde{g}}}}{p^2 + \Mone^2} \nonumber \\
  &=&
      \frac{(x'^2 + 1) (x' - \sqrt{x'^2 + 4})^2}{4 \Mone^2}
      + \mathcal{O}(p^2/\Mone^2) \, .
\end{eqnarray}
These analytic expressions agree well with our results shown in
Figs.~\ref{fig:uLuR}, \ref{fig:uLuREqual}, and \ref{fig:uLuRPrime}.

We observe in Fig.~\ref{fig:uLuR} that the cross sections for $x=0$
and for the pure Majorana gluino are identical in this mode.
This is because in the pure Dirac case, $s^2_{\theta} = c^2_{\theta} = 0.5$
and $\Mtwo = \Mone = M$ (say), rendering the co-efficient of the spinors
in the amplitude $p\cdot\sigma_{\alpha\dot{\beta}}/(p^2 + M^2)$, and in the
pure Majorana limit, $c^2_{\theta} = 1$ and we once again have
$p\cdot\sigma_{\alpha\dot{\beta}}/(p^2 + M^2)$ in the amplitude.

By inspecting the expressions in Eqs.~(\ref{eq:uLuLAmp}), (\ref{eq:uLuLAmpEqual}),
(\ref{eq:uLuLAmpPrime}) and comparing with their $\tilde{u}_L\tilde{u}_R$ counterparts,
one can also see that (i) in Case~I, $\tilde{u}_L\tilde{u}_L$ never catches up with
$\tilde{u}_L\tilde{u}_R$ as $x$ goes from 0 to 1, (ii) in Case~II, it catches up at
about $x=0.2$, and (iii) in Case~III, it catches up at a very small value of $x$.
This is reflected in Figs.~\ref{ResolvedXS}, \ref{ResolvedXSEqual} and
\ref{ResolvedXSPrime} and hence in the respective contour plots.


\section{``Dirac'' Charginos}
\label{sec:DbutnotD}

\begin{figure}
\centering
\includegraphics[width=8cm]{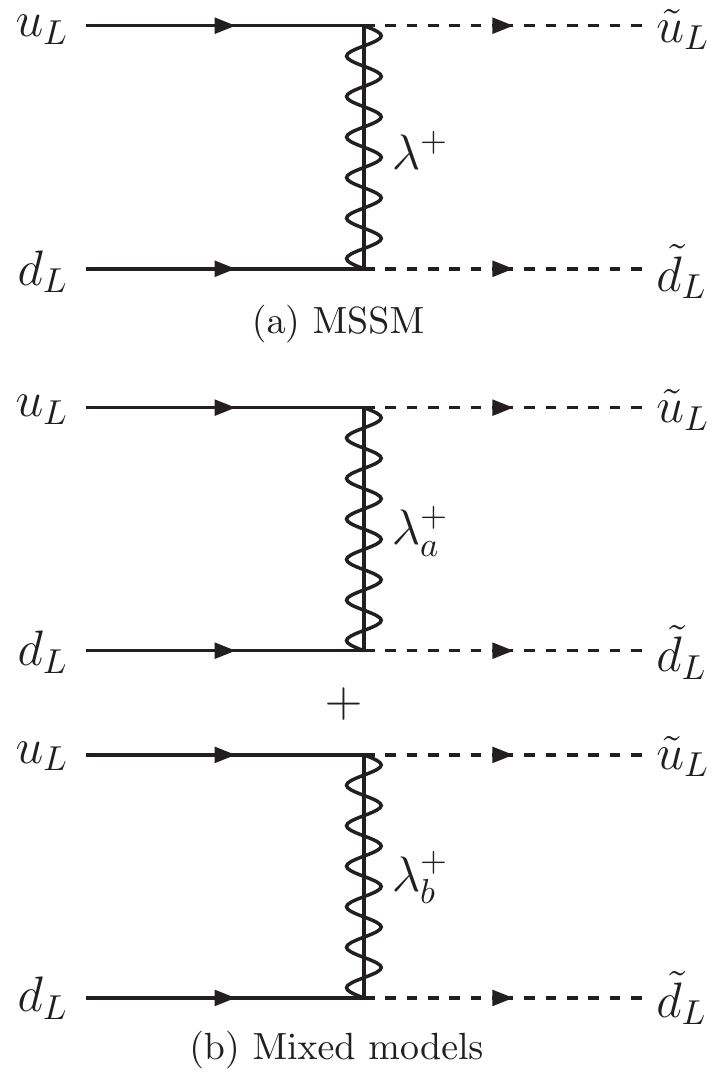}
\caption{Feynman diagrams for the process $p p \ra \tilde{u}_L \tilde{d}_L$
in MSSM and models with both Dirac and Majorana gaugino masses.}
\label{fig:DbutnotD}
\end{figure}

In this section we discuss the differences in the process
$p p \ra \tilde{u}_L \tilde{d}_L$ (and its equivalents for other generations)
for winos that acquire a Majorana mass versus winos that acquire a Dirac mass.
We note that some aspects of ``Dirac'' charginos have been
discussed previously in \cite{Choi:2010gc}.
We are specifically interested in the mediation of this process by
$t$-channel charginos.
In MSSM, this process is shown in the Feynman diagram in
Fig.~\ref{fig:DbutnotD}(a).
For mixed models with both Dirac and Majorana wino masses,
the Feynman diagrams are given in Fig.~\ref{fig:DbutnotD}(b).

The presence of the extra chargino can be understood by studying the
relevant mass terms in the Lagrangian, given in Weyl notation by
\begin{equation}
\mathcal{L}_{\tilde{w} \, \mathrm{mass}} \; = \frac{1}{2} \;
\left( \begin{array}{cc}
       w & \psi
       \end{array} \right)
\left( \begin{array}{cc}
       \hat{M}_m & \hat{M}_d \\
       \hat{M}_d & \hat{M}_m'   	
       \end{array} \right)
\left( \begin{array}{c}
       w \\ \psi
       \end{array} \right)
+ \text{h.c.}
\label{eq:basicmasswino}
\end{equation}
where $w$ is the wino, $\psi$ is the triplet fermion partner,
and the hatted quantities are to distinguish from the analogous
parameters for the gluino.  The notation is somewhat an abuse of
notation, since the eigenvectors on the left- and right-hand sides
of the mass matrix are identical for neutral components
of the wino and triplet, whereas the eigenvectors for the charged
fields must involve opposite electric charge components that pair
$w^+,\psi^+$ with $w^-,\psi^-$.  Also we have neglected the
wino-Higgsino mixings that arise after electroweak symmetry breaking
in order to simply understand the differences between a pure Dirac
wino and a mixed wino with regard to squark production.

A mixed (Majorana and Dirac mass) neutral wino interacts in a way
completely analogous with the gluino.  The charged wino is distinct,
since of course a chargino is always a Dirac fermion.  In the MSSM,
the chargino acquires a Dirac mass by pairing the two charged winos
$\lambda^\pm$ with the ``Majorana'' mass term
$M_2 (\lambda^+\lambda^- + c.c.)$.  In models with a Dirac mass
for the chargino, the charged wino $\lambda^\pm$ acquires mass
with a charged fermion partner $\psi^\mp$.  The mixing is analogous
to the mixed gluino, where now
\begin{equation}
\left( \begin{array}{c} \lambda_a^{\pm} \\
                        \lambda_b^{\pm} \end{array} \right)
 \;=\; \left( \begin{array}{cc}
              \cos \theta_{\tilde{w}}  & \sin \theta_{\tilde{w}} \\
              -\sin \theta_{\tilde{w}} & \cos \theta_{\tilde{w}} \end{array} \right)
\left( \begin{array}{c} w^{\pm} \\
                        \psi^{\pm} \end{array} \right)
\label{eq:charginomix}
\end{equation}
with the same form of the mass eigenvalues and mixing angles as
Eqs.~(\ref{eq:EigMass}) and (\ref{eq:MixAng}).  Since the
wino couples to quarks and squarks, while the triplet partner
does not, the usual wino interaction terms
\begin{equation}
\mathcal{L} = - g_2 \ (\tilde{u}^*_{L,i} \lambda^+  d_{L,i}
              + \tilde{d}^*_{L,i}\lambda^-  u_{R,i} ) + \text{h.c.}
\end{equation}
become
\begin{eqnarray}
\mathcal{L} = - g_2 \ (\tilde{u}^*_{L,i} \lambda_a^+ \cos \theta_{\tilde{w}} \ d_{L,i}
              + \tilde{u}^*_{L,i} \lambda_b^+\sin  \theta_{\tilde{w}} \ d_{L,i}
              \nonumber \\
              + \tilde{d}^*_{L,i} \lambda_a^- \cos \theta_{\tilde{w}} \ u_{R,i}
              + \tilde{d}^*_{L,i} \lambda_b^- \sin  \theta_{\tilde{w}} \ u_{R,i} )
              + \text{h.c.}
              \nonumber \\
\label{eq:2compchargLagAlloyed}
\end{eqnarray}
Interestingly, in the pure Dirac mass limit where $\hat{M}_m,\hat{M}_m' = 0$,
the mixing angles become maximal, and then for the same reasons that
$qq \ra \tilde{q}_L \tilde{q}_L$ vanishes for a Dirac gluino, one can
show that $qq' \ra \tilde{q}_L \tilde{q'}_L$ vanishes for a Dirac wino.
We did not utilize this observation in our studies, since our main focus
was the interference between Majorana wino and bino with a Dirac gluino.

\section{14 TeV extrapolation}

\begin{figure*}
\begin{centering}
\begin{subfigure}[t]{0.32\textwidth}
\caption{$M_m'=0$: ratios}
\includegraphics[width=6.0cm]{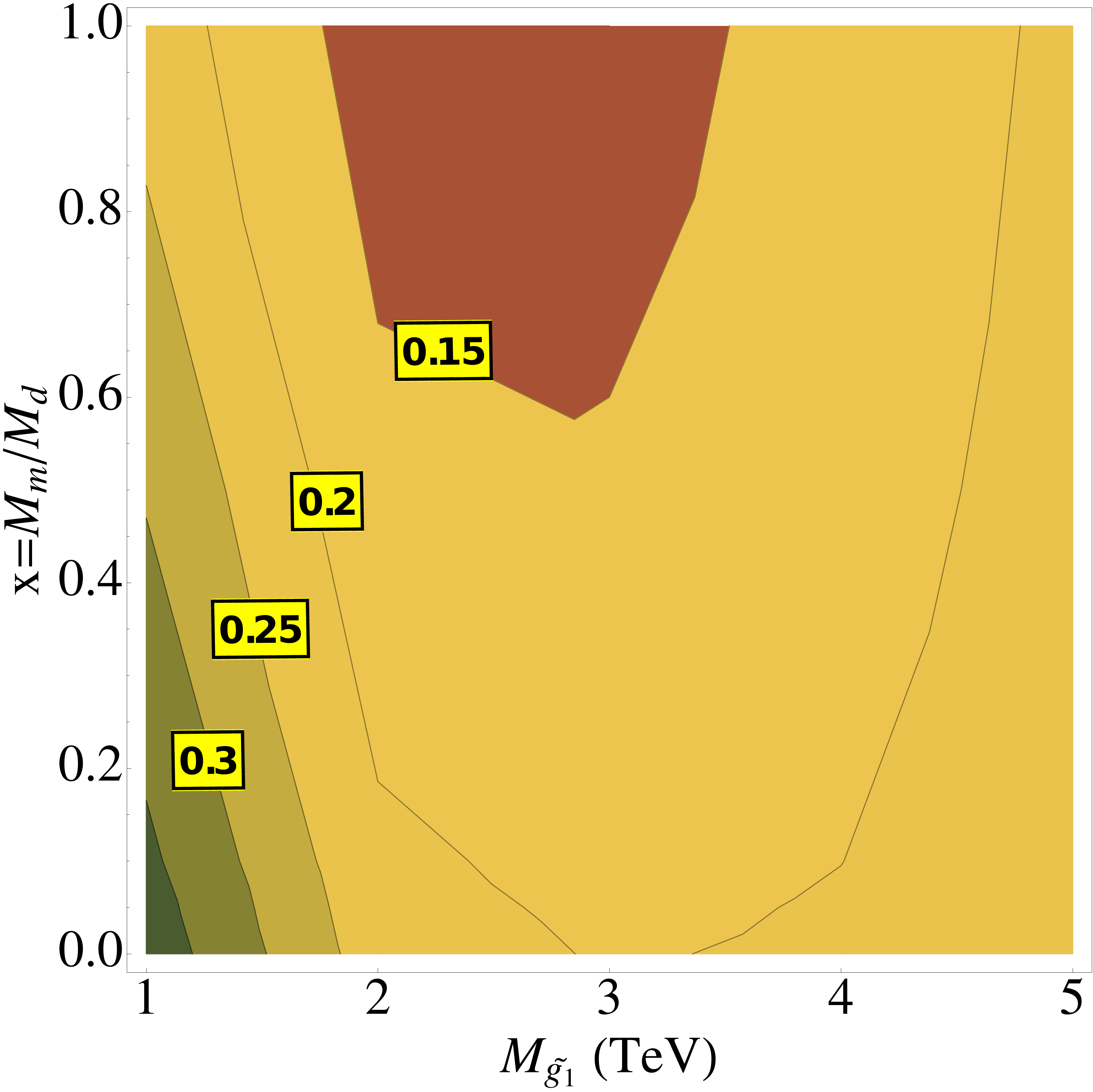}
\label{fig:MbyD1200RatioContour14TeV}
\end{subfigure} \quad \quad \quad
\begin{subfigure}[t]{0.32\textwidth}
\caption{$M_m'=0$: cross sections}
\includegraphics[width=6.0cm]{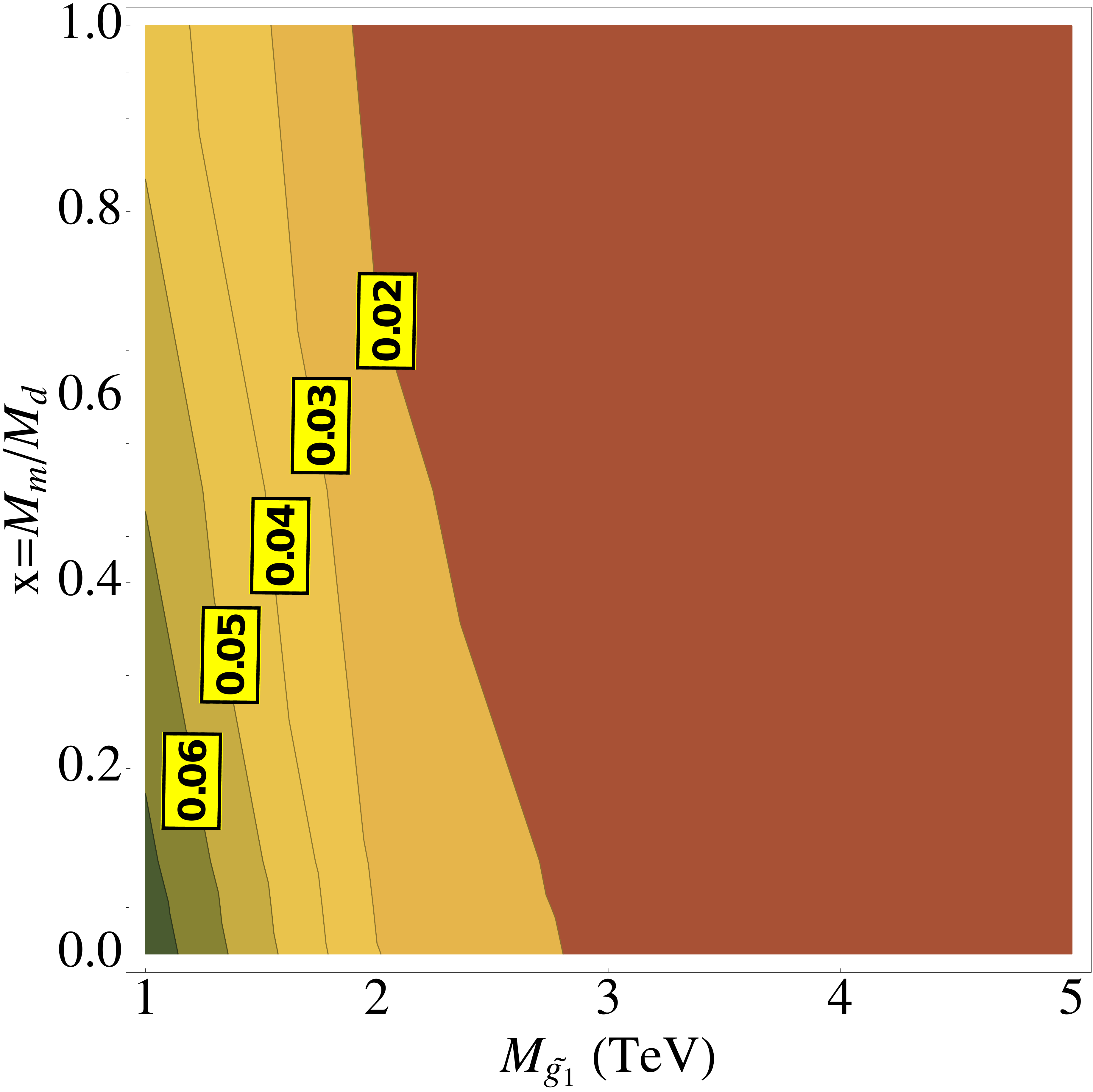}
\label{fig:MbyD120014TeV}
\end{subfigure}
\\
\begin{subfigure}[t]{0.32\textwidth}
\caption{ $M_m=M_m'$: ratios}
\includegraphics[width=6.0cm]{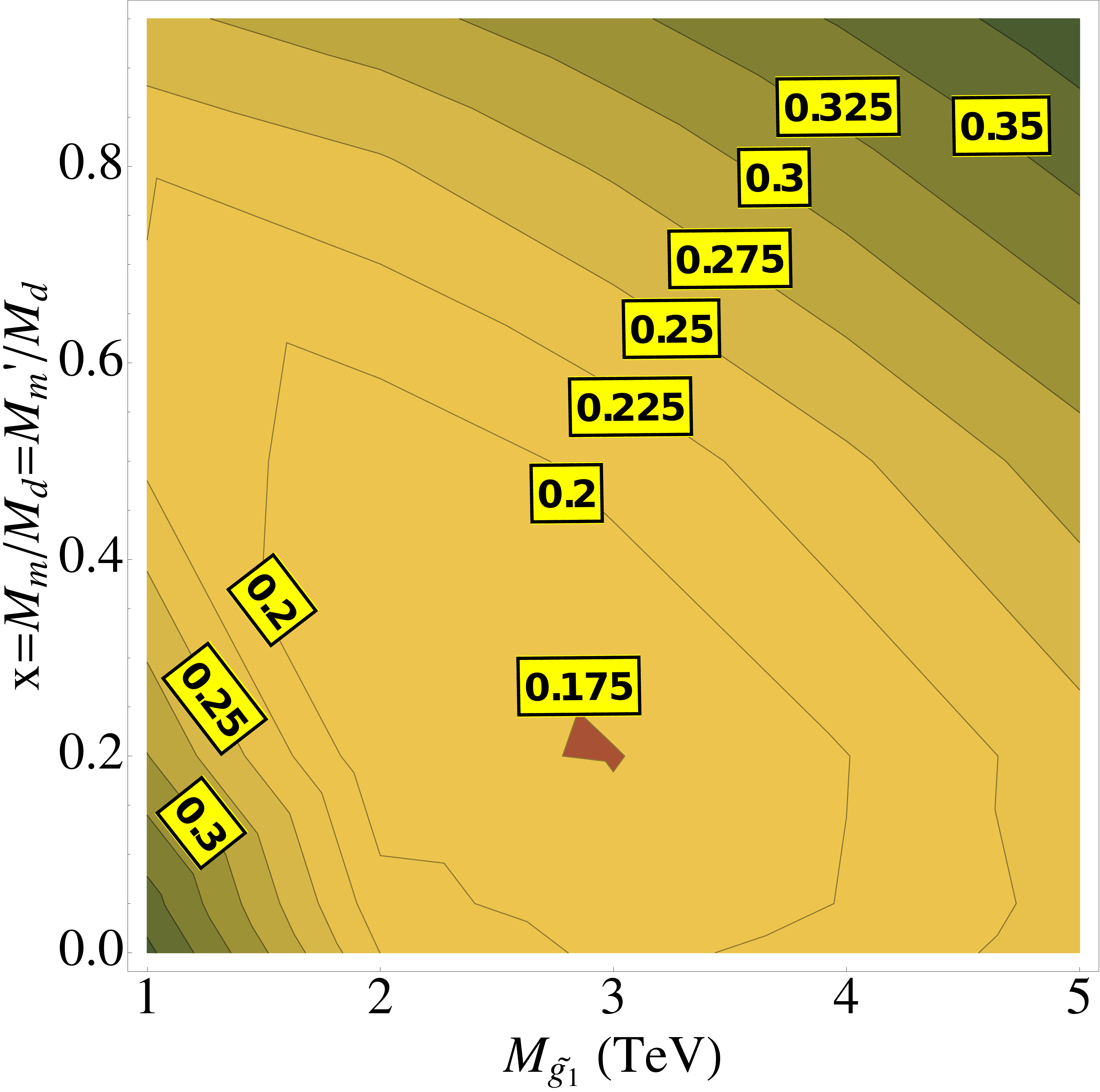}
\label{fig:MEqualbyD1200RatioContour14TeV}
\end{subfigure} \quad \quad \quad
\begin{subfigure}[t]{0.32\textwidth}
\caption{ $M_m=M_m'$: cross sections}
\includegraphics[width=6.0cm]{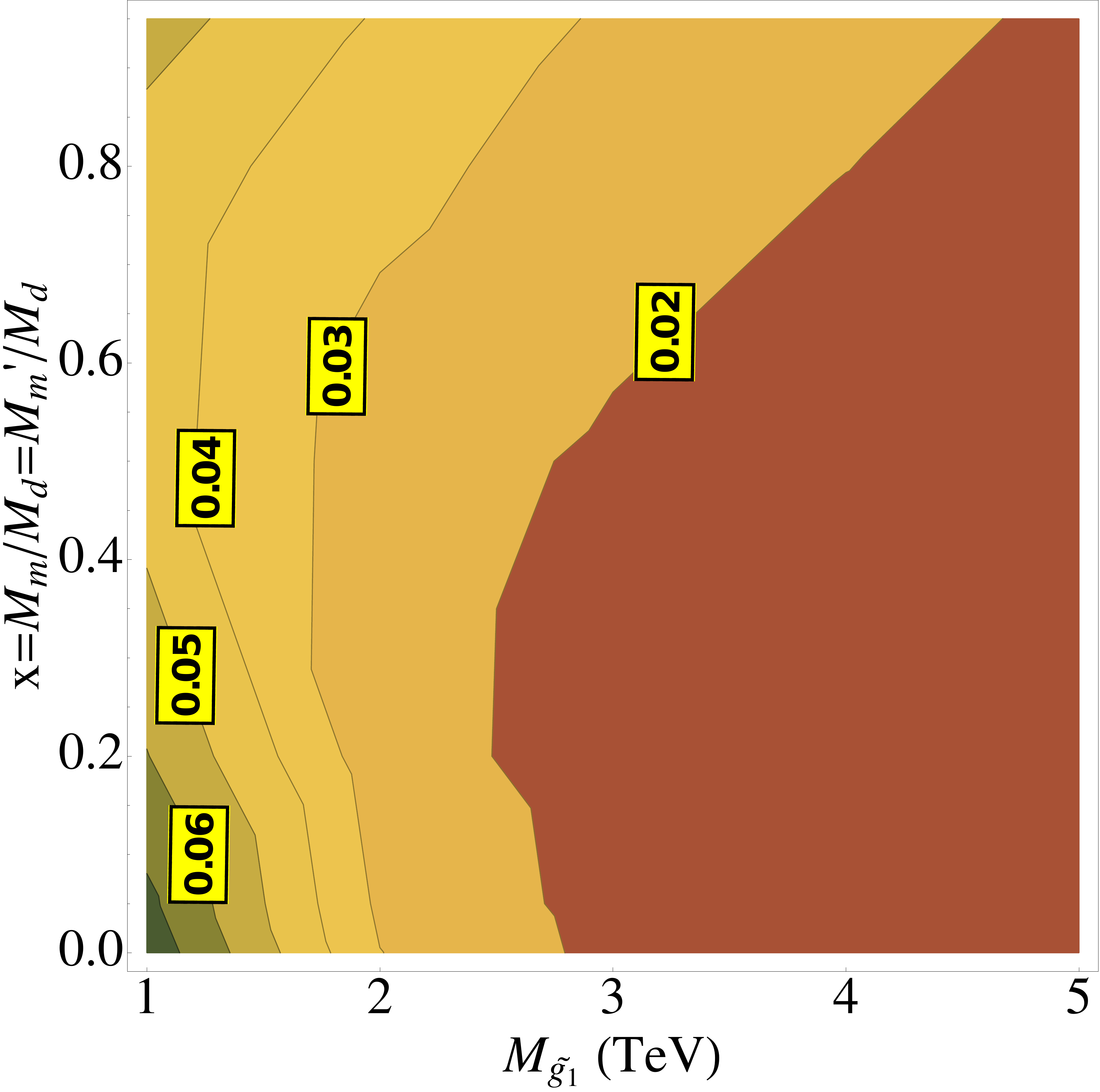}
\label{fig:MEqualbyD120014TeV}
\end{subfigure}
\\
\begin{subfigure}[t]{0.32\textwidth}
\caption{ $M_m=0$: ratios}
\includegraphics[width=6.0cm]{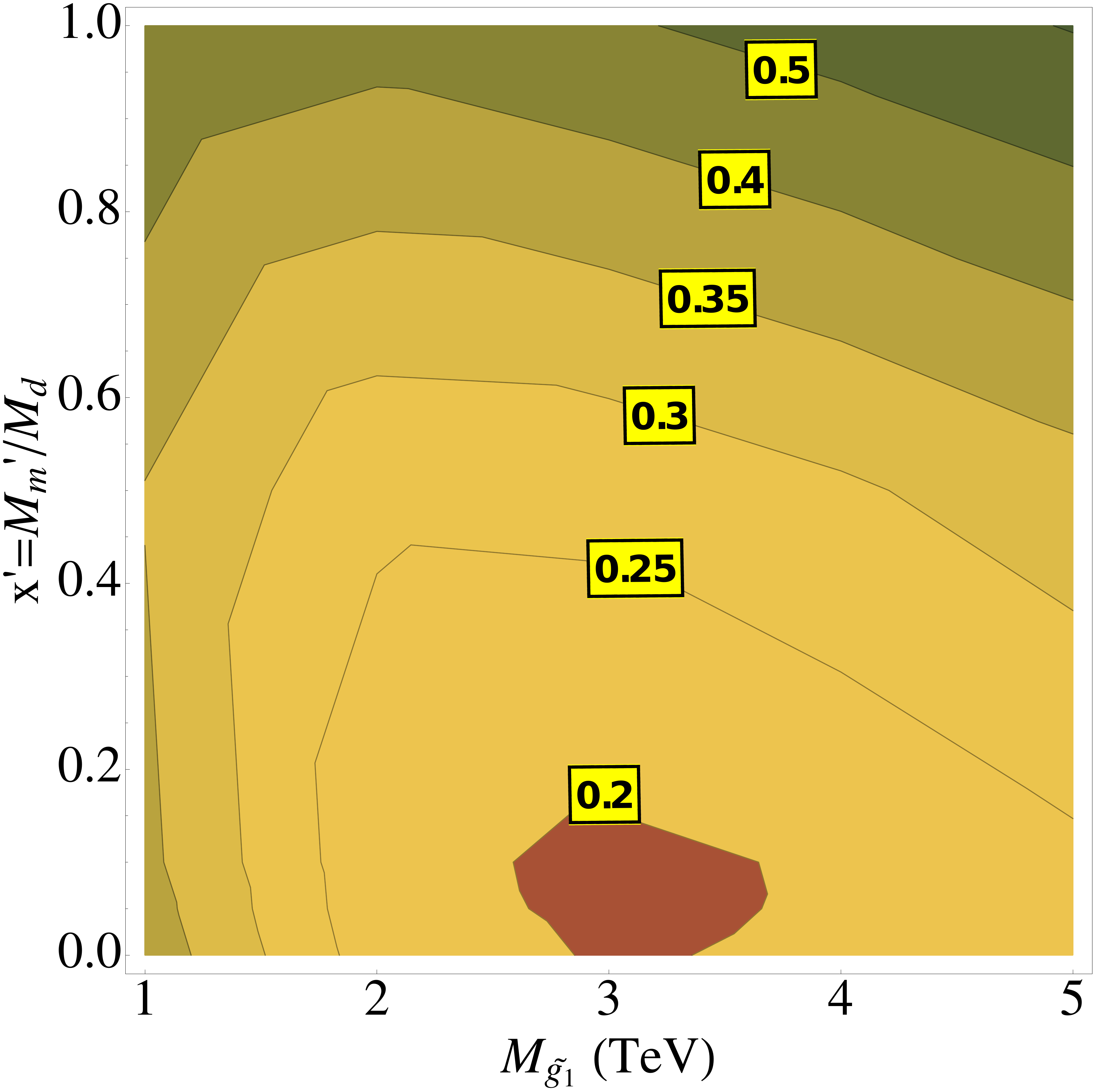}
\label{fig:MPrimebyD1200RatioContour14TeV}
\end{subfigure} \quad \quad \quad
\begin{subfigure}[t]{0.32\textwidth}
\caption{$M_m=0$: cross sections}
\includegraphics[width=6.0cm]{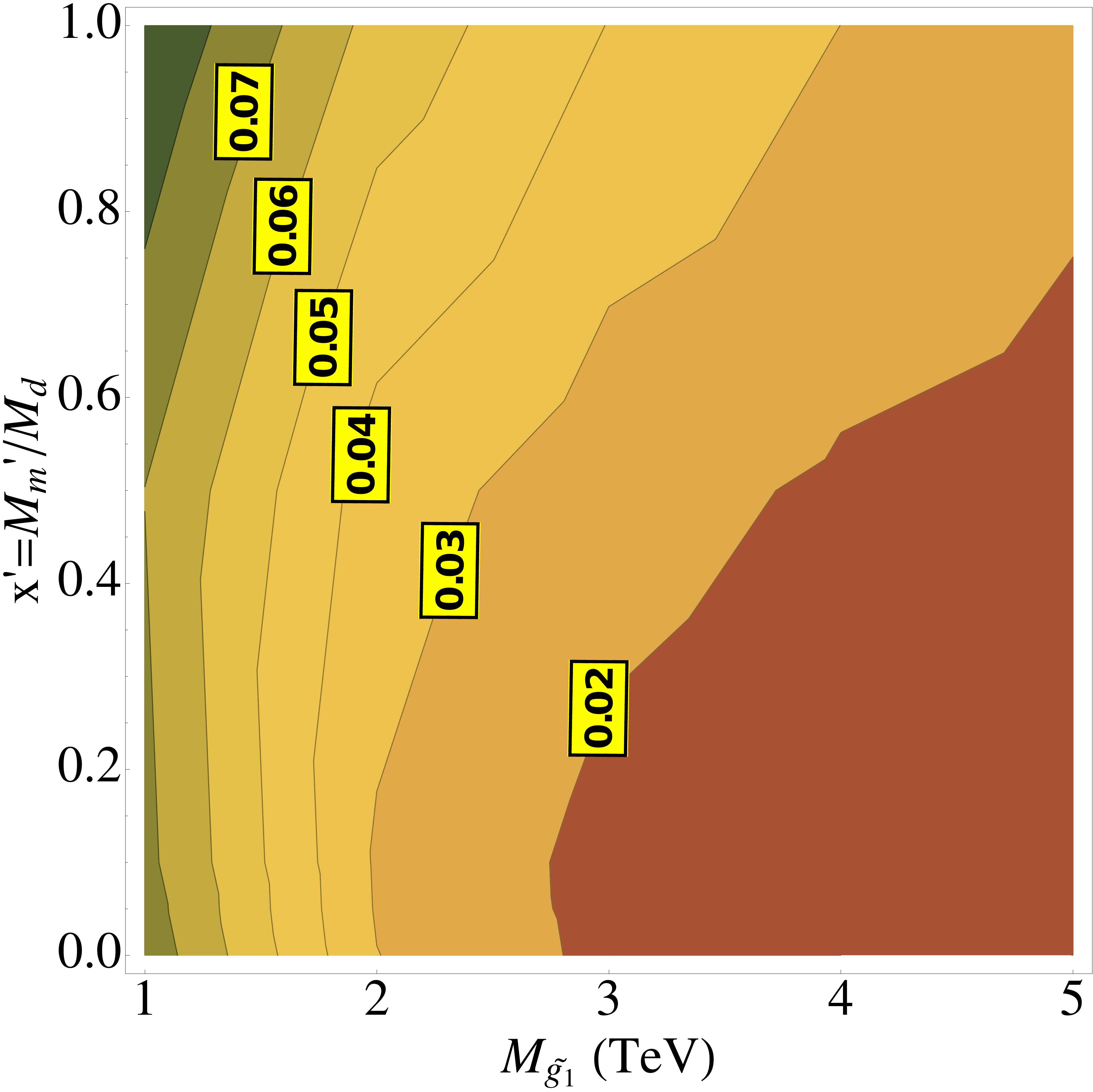}
\label{fig:MPrimebyD120014TeV}
\end{subfigure}
\caption{LEFT: Contours of the ratio of the total production cross section of the
first two generations of squarks at LHC with $\sqrt{s} = 14$~TeV\@ (extrapolated)
in our model to the cross sections in MSSM\@.  RIGHT: Contours of the cross sections
themselves (at leading order), in pb, at LHC with $\sqrt{s} = 14$~TeV\@.
The squark mass here is $1200$~GeV and the parameterization of the axes is similar to
Figs. ~\ref{MajbyDirSqXS}, \ref{MajEqualbyDirSqXS} and \ref{MajPrimebyDirSqXS}.
The details of the critical features are explained in the text.}
\label{MajAllbyDirSqXS14TeV}
\end{centering}
\end{figure*}

\begin{figure}[htbp]
\centering
\includegraphics[width=8cm]{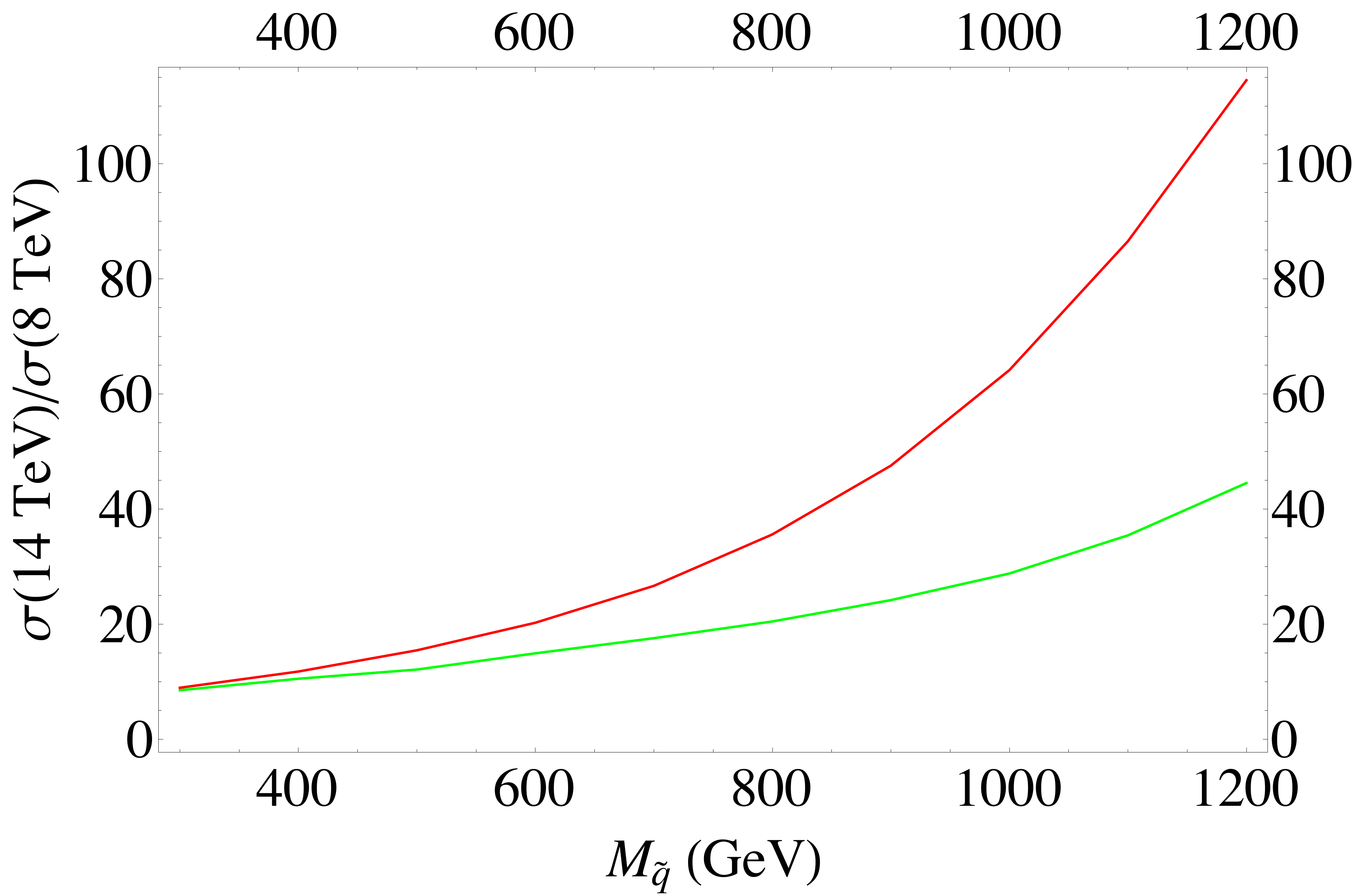}
\caption{Ratios of squark production cross sections
at $\sqrt{s} = 14$~TeV to those at $\sqrt{s} = 8$~TeV\@. Here,
red: QCD-only Dirac gluino case,
green: electroweakinos at their MEI values,
which provides an upper bound on the impact of electroweakinos
in the presence of a Dirac gluino.}
\label{fig:14TeV}
\end{figure}

In this appendix we extend our results to $\sqrt{s} = 14$~TeV at the LHC\@.
Given that the current LHC bound on the squark mass is roughly $800$~GeV
(with a massless LSP), we illustrate the $\sqrt{s} = 14$~TeV results
for $\Msq = 1200$~GeV\@.
The contour plots in Fig.~\ref{MajAllbyDirSqXS14TeV},
parameterized analogously to those of Sec.~\ref{sec:TotalSigmaQCD},
show the changes one would observe for this squark mass.
Specifically, when compared to
Figs.~\ref{MajbyDirSqXS}, \ref{MajEqualbyDirSqXS} and \ref{MajPrimebyDirSqXS},
we find that the cross sections and ratios increase for all three scenarios.
Moreover, at $14$~TeV the $s$-channel gluon-mediated diagrams producing
squark--anti-squark dominate over squark-squark production at all gluino masses
shown in the plots, which was not the case at $\sqrt{s} = 8$~TeV\@.
This implies that the features of the ratio and cross section contours
for $\Msq = 1200$~GeV at $\sqrt{s} = 14$~TeV resemble their
equivalents for, say, $\Msq = 800$~GeV at $\sqrt{s} = 8$~TeV,
and this is the trend observed in all of the plots
in Fig.~\ref{MajAllbyDirSqXS14TeV}.

As for the impact of the mixed electroweak gauginos, a comparison with the
$\sqrt{s} = 8$~TeV LHC results is presented in Fig.~\ref{fig:14TeV},
where the ratios of squark production cross section at $\sqrt{s} = 14$~TeV
to those at $\sqrt{s} = 8$~TeV have been plotted. The green curve indicates
electroweakinos at their MEI values while the red curve shows the QCD-only
Dirac gluino case.  The gluino mass is again taken to be $5$~TeV\@.
Here again, we emphasize that the MEI value is not a special point.
It merely sets an upper bound on the impact of electroweakinos on a
pure Dirac gluino scenario.  We note two features:
(a) The ratios increase as the squark mass increases. This happens because
at $\sqrt{s} = 14$~TeV, the cross section is dominated by
squark--anti-squark production, unlike the case at $\sqrt{s} = 8$~TeV,
where at high squark masses there is competition between squark--anti-squark
and squark-squark modes;
(b) The green curve increases at a slower rate with respect to squark mass
than the red curve. The impact of the electroweakinos on the total
cross section is by affecting $t$-channel (mainly left-handed) squark-pair
production, and such an impact would weaken as $\sqrt{s}$ is increased.
This causes squark--anti-squark production through gluon fusion diagrams
and $s$-channel gluon-mediated subprocesses to dominate.
These features show that the impact of the electroweakinos at their MEI values
are expected to be much less for LHC operating at $14$~TeV\@.



\end{document}